\documentclass[a4paper,10pt]{article}

\usepackage{theorem}
\usepackage{amsmath,amssymb}

\newtheorem{theorem}{Theorem}[section]
\newtheorem{proposition}[theorem]{Proposition}

\newtheorem{lemma}[theorem]{Lemma}

\numberwithin{equation}{section} \allowdisplaybreaks[1]

\allowdisplaybreaks

\newcount\driver
\newcount\bozza

\font\msytw=msbm10 scaled\magstep1
\font\msytww=msbm8 scaled\magstep1

{\count255=\time\divide\count255 by 60
\xdef\hourmin{\number\count255}
        \multiply\count255 by-60\advance\count255 by\time
   \xdef\hourmin{\hourmin:\ifnum\count255<10 0\fi\the\count255}}

\let\a=\alpha \let\b=\beta    \let\g=\gamma     \let\d=\delta     \let\e=\varepsilon
\let\z=\zeta  \let\h=\eta     \let\th=\vartheta \let\k=\kappa     \let\l=\lambda
\let\m=\mu    \let\n=\nu      \let\x=\xi        \let\p=\pi        \let\r=\rho
\let\s=\sigma \let\t=\tau            \let\c=\chi
\let\ps=\psi   \let\o=\omega     
 \let\D=\Delta       \let\L=\Lambda    
             
\let\O=\Omega 

\def\PP{{\cal P}}\def\MM{{\cal M}}\def\VV{{\cal V}}
\def\CC{{\cal C}}\def\WW{{\cal W}}
\def\TT{{\cal T}}\def\NN{{\cal N}}\def\BB{{\cal B}}
\def\LL{{\cal L}}
\def\DD{{\cal D}}

\def\RRR{\mathbb{R}}   

       \def\CCC{\mathbb{C}} 

       \def\ZZZ{\hbox{\msytw Z}}
\def\zzzz{\hbox{\msytww Z}}       
\def\TTT{\mathbb{T}}  

\def\pp{{\bf p}}\def\qq{{\bf q}}\def\xx{{\bf x}}
 
\def\yy{{\bf y}}\def\kk{{\bf k}}\def\nn{{\bf n}}
\def\zz{{\bf z}}\def\uu{{\bf u}}\def\vv{{\bf v}}
\def\hh{{\bf h}}
\def\bP{{\bf P}}

\def\oo{{\underline \omega}}

          \def\ux{{\underline\xx}}
\def\uk{{\underline \kk}}

       \def\hv{{\widehat v}}

\def\hh{{\widehat \h}}      
       
\def\hp{{\widehat \ps}}     \def\hF{{\widehat F}}
         
           \def\hJ{{\widehat \jmath}}
\def\hJ{{\widehat J}}       \def\hg{{\widehat g}}

\def\hG{{\widehat G}}

\def\be#1\ee{\begin{equation}#1\end{equation}}
\def\bsp#1\esp{\begin{split}#1\end{split}}
\def\bal#1\eal{\begin{align}#1\end{align}}
\def\ba#1#2\ea{\begin{array}{#1}#2\end{array}}

\def\bea{\begin{eqnarray}}   \def\eea{\end{eqnarray}}
\def\bean{\begin{eqnarray*}} \def\eean{\end{eqnarray*}}
\def\bfr{\begin{flushright}} \def\efr{\end{flushright}}
\def\bc{\begin{center}}      \def\ec{\end{center}}
\def\bd{\begin{description}} \def\ed{\end{description}}
\def\bv{\begin{verbatim}}

\def\lft{\left}                  \def\rgt{\right}
\def\la{{\langle}}               \def\ra{{\rangle}}

\def\erp{{\perp}}
\def\arp{{\parallel}}

\def\mp#1{\marginpar{\tiny\bf#1}}

\def\nn{\nonumber}
\def\Halmos{\hfill\vrule height10pt width4pt depth2pt \par\hbox to \hsize{}}
\def\pref#1{(\ref{#1})}
\def\virg{\;,\quad}
\def\qed{\raise1pt\hbox{\vrule height5pt width5pt depth0pt}}
\let\dpr=\partial

\let\==\equiv

\let\io=\infty
\let\0=\noindent

\def\*{{\hfill\break\null\hfill\break}}

\def\eg{\hbox{\it e.g.\ }}
\def\tilde#1{{\widetilde #1}}

\def\Tr{\rm Tr}

\def\lb#1{\label{#1}}
\def\ins#1#2#3{\vbox to0pt{\kern-#2 \hbox{\kern#1 #3}\vss}\nointerlineskip}

\newdimen\xshift \newdimen\xwidth \newdimen\yshift
\newcount\griglia

\def\insertplot#1#2#3#4#5#6{%
\xwidth=#1pt \xshift=\hsize \advance\xshift by-\xwidth \divide\xshift by 2%
\begin{figure}[ht]
\vspace{#2pt} \hspace{\xshift}
\begin{minipage}{#1pt}
#3 \ifnum\driver=1 \griglia=#6
\ifnum\griglia=1 \openout13=griglia.ps \write13{gsave .2
setlinewidth} \write13{0 10 #1 {dup 0 moveto #2 lineto } for}
\write13{0 10 #2 {dup 0 exch moveto #1 exch lineto } for}
\write13{stroke} \write13{.5 setlinewidth} \write13{0 50 #1 {dup 0
moveto #2 lineto } for} \write13{0 50 #2 {dup 0 exch moveto #1
exch lineto } for} \write13{stroke grestore} \closeout13
\includegraphics{griglia.ps} \fi
\includegraphics{#4.ps}\fi%
\ifnum\driver=2 \fi
\end{minipage}
\caption{#5}
\end{figure}
}

\ifnum\bozza=1 
\usepackage[notref,notcite]{showkeys}\fi

\begin{document}

\title{Universal Luttinger Liquid Relations in the\\
1D Hubbard Model}%

\author{G. Benfatto$^1$ \and P. Falco$^2$ \and V. Mastropietro$^1$}

\maketitle

\footnotetext[1]{\normalsize 
Dipartimento di Matematica, Universit\`a di Roma ``Tor Vergata",
00133 Roma, Italy.}
\footnotetext[2]{\normalsize School of Mathematics,
Institute for Advanced Study, Princeton, New Jersey 08540
}
\begin{abstract}
We study the 1D extended Hubbard model with a weak repulsive
short-range interaction in the non-half-filled band case, using
non-perturbative Renormalization Group methods and Ward
Identities. At the critical temperature,  $T=0$, the response
functions  have anomalous power-law decay with multiplicative
logarithmic corrections. The critical exponents, the
susceptibility and the Drude weight verify the universal Luttinger
liquid relations. Borel summability and (a weak form of)
Spin-Charge separation is established.
\end{abstract}

\tableofcontents

\section{Main Results}

\subsection{Introduction}\lb{sec1.1}
 The Hubbard model, see {\it e.g.} \cite{L95}, describing
interacting spinning fermions on a lattice, plays the same role in
quantum many body theory as the Ising model in classical
statistical mechanics, that is it is the simplest model displaying
many real world features: it is however much more difficult to
analyze. While our understanding of the Hubbard model in higher
dimensions at zero temperature is really poor (except for special
choices of the lattice as in \cite{FKT04} and \cite{GM010_1}), the
situation is better in $d=1$, when the model furnishes an accurate
description of real systems, like quantum wires or carbon
nanotubes \cite{Gi004}.

The one dimensional Hubbard model (from now on the Hubbard model
tout court) can be exactly solved by {\it Bethe ansatz}, as shown
by Lieb and Wu \cite{LW068}: the system is insulating in the half
filled band case while it is a metal otherwise and the elementary
excitations are {\it not} electronlike, a phenomenon which is
nowadays called {\it electron fractionalization} \cite{A097_1}.
Recently in \cite{LW03} a strategy for a proof that the lowest
energy state of Bethe ansatz form is really the ground state
has been outlined (see also \cite{G05}). This method is however of
little utility for understanding the asymptotic behavior of
correlations; and does not apply in studying the ground state of a
slight generalization of the model, the extended Hubbard model,
that consists in replacing the local quartic interaction with a
short-ranged one. Other approaches has been therefore developed to
get more insights into the physical properties of the Hubbard model.

Under certain {\it drastic approximations}, like replacing the
sinusoidal dispersion relation with a linear relativistic one and
neglecting certain terms called {\it backward} and {\it umklapp}
interactions (see after \pref{ff11} for their definition), one
obtains the {\it spinning Luttinger Model}, which is exactly
solvable in a stronger sense,  \cite{ML065}, \cite{Ma064}: all its
Schwinger functions, at distinct points, can be explicitly
computed. This model, describing interacting fermions, can be
exactly mapped in a model of two kinds of {\it free bosons},
describing the propagation of charge or spin degrees of freedom
and with {\it different} velocities ({\it spin-charge
separation}); again, a phenomenon of electron fractionalization
which has received a considerable attention in the context of high
$T_c$ superconductors \cite{A097}. Moreover, as in spinless
Luttinger model, the correlations have a power law decay rate {\it
with interaction dependent exponents}.

However, neglecting the lattice effects and backscattering or
umklapp interactions is not safe, and indeed the mapping to free
bosons is not expected to be true in the Hubbard model. A somewhat
more realistic effective description 
can be obtained by including the backward interaction in the
spinning Luttinger Model, so obtaining the  {\it g-ology model}.
This system is no more solvable; however, a perturbative
Renormalization Group (RG) analysis,  \cite{So079}, shows that, in
the {\it repulsive case}, such extra coupling is {\it marginally
irrelevant}, i.e. becomes negligible over large space-time scales.
In \cite{MD091} the necessity of implementing Ward Identities in a
RG approach was emphasized, in order to go beyond purely
perturbative results, but the analysis was limited to the
Luttinger model and no attempt was done to include the effects of
nonlinear bands. In \cite{GS089} it was observed that the
correlations of the repulsive g-ology model would qualitatively
differ from the Luttinger model ones for the presence of
multiplicative {\it logarithmic corrections}.

A new point of view, that extended previous ideas
of Kadanoff, \cite{KW071}, and Luther and
Peschel, \cite{LP075}, was proposed by Haldane, \cite{Ha080},
and  is nowadays known as {\it Luttinger
Liquid Conjecture}. The idea is to exploit the concept of {\it
universality}, a basic notion in statistical physics saying that
the critical properties are largely independent from the details
of the model, at least inside a certain class of models. In the
present case, as the exponents are non trivial functions of the
coupling, universality has a meaning more subtle than usual; it
does not mean that the exponents are the same (the exponents {\it
do depend} on the details of the model), but that the exponents
and certain thermodynamic quantities verify a set of {\it
universal relations} which are {\it identical} to the Luttinger
model ones. Such relations give an exact determination of
physical quantities in terms of a few measurable parameters.
The validity of such relations has been checked in
certain special solvable spinless fermionic lattice models
\cite{Ha080}, but a proof of their validity in the Hubbard model
(or in the non solvable extended Hubbbard model) is an open
problem. It should be remarked that, even thought the Hubbard
model differs from the spinning Luttinger model for irrelevant or
marginally irrelevant terms in the Renormalization Group sense (in
the weak non half filled band case), this would not imply at all
the validity of the same relations as in the Luttinger model;
irrelevant terms {\it do renormalize} the exponents and the
thermodynamic quantities.

Starting from the 90's, the methods developed in constructive
Quantum Field Theory (QFT) for the analysis of QFT models in
$d=1+1$ \cite{GK085_1,Le087} were applied to interacting non
relativistic spinless fermionic systems in the continuum
\cite{BGPS994}, so establishing the anomalous dimension in a non
solvable model, by combining Renormalization Group methods with non
perturbative information coming from the {\it exact solution of
the Luttinger model}; the extension to spinning fermions was done
in \cite{BoM995}. An important technical advance was achieved in
\cite{BM002,BM005}, by implementing Ward Identities based on local
symmetries with Renormalization Group methods. A well known
difficulty in any Wilsonian Rermalization Group approach is that
the momentum cut-offs break the local symmetries on which Ward
Identities are based; in \cite{BM002,BM005} it was developed a
technique allowing to rigorously take into account the extra terms
produced in the Ward Identities by the cut-offs, so that
interacting non relativistic fermions in $d=1$ were constructed
{\it without any use of exact solutions} \cite{BM005,M005,FM008}.
The main outcome, with respect to the early Renormalization Group
analysis \cite{So079}, is that the results are exact (the lattice and non
linear bands are fully taking into account) and non-perturbative;
and physical observables are written in terms of convergent
expansions so that they can be computed with arbitrary precision.
The complexity of such expansions made however impossible to
verify explicitly the universal Luttinger Liquid relations in
\cite{Ha081}; 
they  have finally been proven for
interacting {\it spinless} fermions on the lattice (the $XXZ$
chain and extended versions) in \cite{BM010,BFM010} 
through Ward identities; this fact appears  to be related to a
non-perturbative version of the Adler-Bardeen theorem of the non
renormalization of the anomalies, \cite{M007}. In this paper we
will extend such ideas to {\it spinning} fermions in the Hubbard
model; as we will see, the extension is rather non
trivial and new phenomena take place. 

\subsection{Extended Hubbard Model and Physical Observables}

Let $\b>0$ be the {\it inverse temperature}, $-\m$ the {\it chemical potential}
and $\CC=\{-[L/2],\ldots,|[(L-1)/2]L\}$ a one dimensional lattice of $L$ sites.
The extended Hubbard model \cite{Hu064} describes fermions hopping on $\CC$
with a short-range density-density interaction; the Hamiltonian plus the
chemical potential term is
\be\lb{1.1}  H=-{1\over 2}\sum_{x\in \CC\atop s=\pm}
(a^+_{x,s}a^-_{x+1,s}+a^+_{x,s}a^-_{x-1,s}) + \m \sum_{x\in \CC\atop s=\pm 1}
a^+_{x,s}a^-_{x,s} +\l \sum_{x,y\in \CC\atop s,s'=\pm 1}
v(x-y)a^+_{x,s}a^-_{x,s} a^+_{y,s'}a^-_{y,s'} \ee
where $a^\pm_{x,s}$ are fermionic creation and annihilation operators at site
$x$ with 'spin' $s$, and $v(x)$ is a smooth, even  potential such that
$|v(x)|\le C e^{-\k |x|}$ (short range condition); periodic boundary conditions
are assumed: $a^\e_{L+1,s}=a^\e_{1,s}$.

The set-up of the Grand Canonical Ensemble is standard, and we remind it
concisely; more details are, for example, in \cite{Mah000}. If $O_x$ is a
monomial in the operators $a^\e_{x,s}$, $\e, s=\pm$, or in the density
operators $a^{\e}_{x,s} a^{\e'}_{x,s'}$, $\e,\e',s,s'=\pm$, given $x_0\in
[0,\b]$, define $\xx=(x,x_0)$ and $O_\xx:=e^{H x_0} O_x e^{-H x_0}$ (so that $x_0$ has the
meaning of {\it imaginary} time); then, given the observables
$O_{\xx_1},\ldots, O_{\xx_n}$, their Grand Canonical correlation is
\be\lb{trc}
\la O_{\xx_1}\cdots O_{\xx_n}\ra_{L,\b}
:= {{\rm Tr}[e^{-\b H}{\bf T}(O_{\xx_1}\cdots O_{\xx_n})]
\over {\rm Tr}[e^{-\b H}]}
\ee
where ${\bf T}$ is the time order product.
%
Similarly, $\la O_{\xx_1};\cdots; O_{\xx_n}\ra_{T;L,\b}$ denotes
the corresponding truncated correlations. We are interested in the
correlations in the thermodynamic limit $L\to\io$ and at the
critical temperature $\b^{-1}=0$; the limit $L,\b\to\io$
will be indicated by  dropping the labels $L,\b$.

Define $\bar p_F\in [0,\p]$, the {\it free Fermi momentum}, and $\bar v_F$, the
{\it free Fermi velocity}, such that
$$
\cos\bar p_F=\m\qquad \bar v_F=\sin \bar p_F\;.
$$
In this paper we have three main assumptions on the parameters:
%
\be\lb{ma} \bar p_F \not=0,\p/2,\p \virg \hat v(2\bar p_F) > 0\virg \l \ge 0\ee
The condition $\bar p_F \not=0,\p$ means that the {\em empty band} and the
{\em filled band} cases are {\em not} included; the reason of such exclusion
is that, if $\bar v_F=0$, the scaling of the model would be very different
and would depend in a critical way on the interaction. The condition $\bar
p_F \not= \p/2$ excludes the {\em half-filled band} case; it will have the
effect to make the Umklapp interaction (see \S\ref{ss2.5a}) irrelevant (in
the RG language). The two other conditions can be loosely called the {\it
repulsive condition} on the interaction; they indeed imply that one of the
contribution to the effective interaction (in the RG language) is strictly
positive at all scales.

The model is $SU(2)$ symmetric, as the Hamiltonian is invariant under
transformation $a^\pm_{x,s}\to \sum_{s'}M_{s,s'}a^\pm_{x,s'}$ with $M\in
SU(2)$; and includes the standard and the U-V Hubbard models, corresponding to
the interactions $\l v(x-y)=U \d_{x,y}$ and $\l v(x-y)=U \d_{x,y} + \frac12
V\d_{|x-y|,1}$, respectively: in the former case the repulsive condition is
$U\ge 0$.

By definition, the  $T=0$ {\it free energy} is
\be\lb{fe}
E(\l):=-\lim_{L,\b\to \io} (L\b)^{-1}\log {\rm Tr}[e^{-\b H}]\;,
\ee
and the   2-point {\it Schwinger function} is
\be
S_{2,\b,L}(\xx-\yy):=  \la
a^{-}_{\xx,+}a^{+}_{\yy,+}\ra_{\b,L} =\la
a^{-}_{\xx,-}a^{+}_{\yy,-}\ra_{\b,L}\;.
\ee
The connection with experimental physics is through the {\it response
functions}, defined as Fourier transforms of the following truncated
correlations:
\be \O_{\a,\b,L}(\xx-\yy):= \la \r^\a_{\xx}
\r^\a_{\yy}\ra_{T;\b,L}:=\la \r^\a_{\xx} \r^\a_{\yy}\ra_{\b,L}- \la
\r^\a_{\xx}\ra_{\b,L} \la\r^\a_{\yy}\ra_{\b,L}
\ee
where $\r^\a_{\xx}$ is one of the following densities
(see pagg. 54, 55 of \cite{Gi004}):
\bal
\r^C_{\xx}&= \sum_{s=\pm }a^+_{\xx,s}a^-_{\xx,s} &{\rm (charge\
density)}\nn\\
\r^{S_i}_{\xx}&= \sum_{s,s'=\pm } a^+_{\xx,s}\s^{(i)}_{s,s'} a^-_{\xx,s'}
&{\rm (spin\ densities)}\nn\\
\lb{rho}\nn\\[-30pt]
\\
\r^{SC}_{\xx}&= \frac12 \sum_{s=\pm \atop \e=\pm} s\, a^\e_{\xx,s}
a^\e_{\xx,-s} & {\rm (singlet\ Cooper\ density)}\nn\\
\r^{TC_i}_\xx&= \frac12 \sum_{s,s'=\pm \atop \e=\pm} a^\e_{\xx,s}
\tilde \s^{(i)}_{s,s'} a^\e_{\xx+{\bf e},s'}\virg {\bf e}=(1,0)
&{\rm (triplet\ Cooper\ densities)}\nn
\eal
where $i=1,2,3$ and
\bal \s^{(1)} &= \begin{pmatrix} 0 &1\\ 1& 0 \end{pmatrix}
&\s^{(2)} &= \begin{pmatrix} 0 &-i\\ i& 0 \end{pmatrix}
&\s^{(3)} &= \begin{pmatrix} 1 &0\\ 0& -1 \end{pmatrix}\nn\\
\tilde\s^{(1)} &= \begin{pmatrix} 1 &0\\ 0& 0 \end{pmatrix}
&\tilde\s^{(2)} &= \begin{pmatrix} 0 &1\\ 1& 0 \end{pmatrix}
&\tilde\s^{(3)} &= \begin{pmatrix} 0 &0\\ 0& 1 \end{pmatrix}\nn
\eal
When the interaction is off ($\l=0$), all  functions
$\O_{\a}(\xx-\yy)$
have power law decay for large $|\xx-\yy|$ with the same exponent;
as we will see, turning on the interaction removes such
degeneracy: some correlations are enhanced by the presence of
interaction and others are depressed, so that the exponents become
different.

Define the Fourier transform of the response functions as
\be \hat\O_\a(\pp):=\lim_{\b,L\to\io}\int_{-\b/2}^{\b/2}
dx_0\sum_{x\in\CC} e^{i\pp\xx}\;\O_{\a,\b,L}(\xx)\ee
where $\pp=(p, p_0)$, with  $p\in {2\pi \over L}\CC$ and $p_0\in {2 \pi\over
\b}\ZZZ$. The {\it susceptibility} is given by
\footnote{in a fermion system, $\k=\k_c\r^2$, where $\k_c$ is
the {\it 
compressibility} and $\r$ the {\it 
density}, see \eg (2.83) of \cite{PN066}}
\be\lb{kk} \k:=\lim_{p\to 0}\hat\O_C(p,0)\;. \ee
The paramagnetic part of the current $J_x$ is defined as
\be J_x={1\over 2i}\sum_{s=\pm} [a^+_{x+1,s} a^-_{x,s}- a_{x,s}^+
a^-_{x+1,s}] \ee
while the diamagnetic part is
\be\t_x=-{1\over 2} \sum_{s=\pm}[a^+_{x,s} a^-_{x+1,s}+a_{x+1,s}^+
a^-_{x,s}] \ee

The {\it Drude weight} is defined as
%
\be D=-\la \t_x\ra -\lim_{p_0\to 0}\lim_{p\to 0} \lim_{\b,L\to\io}
\int_{-\b/2}^{\b/2} dx_0\sum_{x\in\L} e^{i\pp\xx} \la J_\xx J_{\bf
0}\ra_{T;L,\b}\equiv \lim_{p_0\to 0}\lim_{p\to 0} \hat
D(\pp)\label{cc}\ee
where the first term is a constant independent from $x$. If one
assumes analytic continuation in $p_0$ around $p_0=0$, one can
compute the conductivity in the linear response approximation by
the Kubo formula, that is $\s = \lim_{\o\to 0}\lim_{\d\to 0}{\hat
D(-i \o+\d,0)\over -i \o+\d}$. Therefore, a nonvanishing $D$
indicates infinite conductivity.

The conservation law
\be\lb{eqm} {\partial \r^C_\xx\over \partial x_0}= e^{H x_0}
[H,\r_x] e^{-H x_0} =-i\dpr^{(1)}_x J_{\xx} \= -i
[J_{x,x_0}-J_{x-1,x_0}]\;, \ee
where $\dpr^{(1)}_x$ denotes the lattice derivative, implies exact
relations, called {\it Ward identities} (WI), between the
Schwinger functions, the density correlations and the {\it vertex
functions}, defined as $G^{2,1}_\r(\xx,\yy,\zz) = \la \r^{(C)}_\xx
a^-_\yy a^+_\zz\ra_{T}$ and $G^{2,1}_j(\xx,\yy,\zz) = \la J_\xx
a^-_\yy a^+_\zz\ra_{T}$. Some Ward Identities, which will play an
important role in the following, are
\bal -ip_0\hat G^{2,1}_\r(\kk,\kk+\pp)-i(1-e^{-ip}) \hat
G^{2,1}_j(\kk,\kk+\pp)
&= \hat S_2(\kk) - S_2(\kk+\pp)\label{ref1}\\
-i p_0  \hat\O_{C}(\pp) -i(1-e^{-ip})
\hat\O_{j,\r}(\pp) &=0\label{ref2}\\
-i p_0  \hat\O_{\r,j}(\pp) -i (1-e^{-ip}) \hat D(\pp)
&=0\label{ref3} \eal
where $\O_{j,\r}(\xx,\yy)=\la \r^C_{\xx} J_{\yy}\ra_{T,\b,L}$.


%
%
%

%
%
%
%
%
%

\subsection{Anomalous exponents and logarithmic corrections}
In the free case, $\l=0$, $\bar p_F$ determines the position of the two
singularities of the Fourier transform of $S_2$, which are at $\kk=(\pm \bar
p_F,0)$; whereas $\bar v_F$ is the velocity of the large-distance leading term
of  $S_2$,
$$
S_2(\xx)\sim \sum_{\o=\pm }
\frac{e^{-i\o \bar p_F x}}
{\bar v_F x_0+i\o x}\;.
$$
In the following theorem we see that, when the interaction is turned on,
$\l>0$, the singularities of the Fourier transform of $S_2$ are moved into  new
positions, $\kk=(\pm p_F,0)$, with $p_F=\bar p_F+O(\l)$. Moreover, the power
law decay of $S_2(\xx)$ and the response functions is strongly modified: the
decay exponent is changed ({\em anomalous dimension}) and there are logarithmic
corrections.

\begin{theorem}\lb{th1.1}

If the conditions \pref{ma} are satisfied, there exist $\l_0>0$
such that, if $0<\l<\l_0$, there exist continuous functions
$$
p_F\=p_F(\l)=\bar p_F +O(\l)\qquad
v_F\=v_F(\l)=\sin p_F(\l)+O(\l)
$$
%
(depending also of the other parameters of the model, like $v(x)$ and $\m$),
such that, setting
\be
\bsp
\tilde\xx:=(x,v_F x_0)&\virg L(\xx)=1+ b\l\hv(2\bar p_F) \log |\xx| \virg
b=2(\pi \sin \bar p_F)^{-1}\\
\bar\O_0(\xx) := \frac{x_0^2 - x^2}{x_0^2+ x^2}&\virg \bar
S_0(\xx):=\frac{v_F x_0\cos p_F -  x\sin p_F} {|\tilde\xx|} \esp
\ee
the large $|\xx|$ asymptotics of the two-points Schwinger function is
\be S_2(\xx) \sim \lft[\bar S_0(\xx) + R_2(\xx)\rgt] {L(\xx)^{\z_z}\over
|\tilde \xx|^{1+\h}} \ee
where $R_2(\xx)$ is a continuous function of $\l$ and $\xx$, such that
$|R_2(\xx)|\le C_\th \l^{1-\th}$, for some positive constants $C_\th$ and
$\th<1$.
 Besides, the large $|\xx|$ asymptotics of the correlations are
\bal {\rm for}\quad \a=C,S_i\qquad &\O_{\a}(\xx)\sim {\bar\O_0(\tilde\xx) +
R_\a(\xx)\over \p^2|\tilde\xx|^2} + \cos[2 p_F x] {L(\xx)^{\z_\a}\over
\pi^2|\tilde\xx|^{2 X_\a}} \lft[1 + \tilde R_\a(\xx)\rgt]\nn\\
{\rm for}\quad \a=SC \qquad &\O_{\a}(\xx) \sim - \lft[ \bar\O_0(\tilde\xx) +
\tilde R_\a(\xx) \rgt] \cos(2 p_F x) {L(\xx)^{\tilde \z_\a} \over
\pi^2|\tilde\xx|^{2 \tilde X_\a}}  - {1\over \pi^2}{L(\xx)^{\z_\a} \over
|\tilde\xx|^{2 X_\a}} \lft[1 + R_\a(\xx)\rgt]\nn\\
\lb{asymp} {\rm for}\quad \a=TC_i \qquad &\O_{\a}(\xx) \sim -  {v_F^2\over
\pi^2}{L(\xx)^{\z_\a}\over |\tilde\xx|^{2 X_\a}} \lft[1 + R_\a(\xx)\rgt]
\eal
with the functions $R_\a(\xx)$ and $\tilde R_\a(\xx)$ having the
same properties of $R_2(\xx)$. Moreover, the critical exponents $\h$ and $X_\a$,
are continuous functions of $\l$, while the exponents $\tilde\z_{SC}$ and $\z_\a$
of the logarithmic corrections could also depend on $\xx$ (we can not exclude it), but
satisfy the bounds $|\tilde \z_{SC}|\le C\l$ and $|\z_\a-\bar\z_\a|\le C\l$,
for a suitable constant $C$, with
\be\lb{zalfa} \bar\z_z = 0 \virg \bar\z_C=-{3\over 2}\virg
\bar\z_{S_i}={1\over 2}\virg \bar\z_{SC}=-{3\over 2}\virg
\bar\z_{TC_i}= \frac12\ee
\end{theorem}

\medskip

In the free $\l=0$ case the response functions decay for large
distance with power laws of exponent equal to $2$. The interaction
partially removes such degeneracy by producing {\it anomalous
exponents} which are (in general) non trivial functions of the
coupling (see Theorem 1.2); in particular the response to charge
and spin densities are {\it enhanced} by the interaction, while
the response to triplet Cooper densities are depressed. While the
presence of non universal exponents is a common feature with the
Luttinger model, the presence of {\it logarithmic corrections} is
a striking difference. Such corrections remove the degeneracy in
the response of charge and spin densities: the response to spin
density is dominating. Note on the other hand that the exponents
of the non oscillating part of charge or spin density correlations
are the same as in the free case; also logarithmic corrections are
excluded.

Similar formulas have been derived in the physical literature
\cite{GS089} under the {\it g-ology approximation}, that is
replacing the Hubbard model with the continuum g-ology model
describing fermions with linear dispersion relation. The existence
of anomalous exponents was proved previously in Theorem 1 of
\cite{M005} (see also \cite{BoM995}), and the absence of
logarithmic corrections in the non oscillating part of charge or
spin density correlations was also previously proved in
\cite{M007_3}. The above theorem improves
such results, as it
proves for the first time the existence of logarithmic corrections
and universal relations.

\subsection{The Luttinger liquid relations}

\begin{theorem}\lb{th1.2}
Under the same conditions of Theorem \ref{th1.1}, there exist
continuous functions
\be\lb{g1} K\=K(\l)= 1-c\l + O(\l^{3/2}),\quad\quad\bar K\=\bar K(\l)=
1-c\l + O(\l^{3/2})\ee
with $c=[\hat v(0)-\hat v(2p_F)](\pi \sin \bar p_F)^{-1}$,
such that the critical exponents satisfy the extended scaling formulas
\be\bsp\lb{1.8}
4\h = K + K^{-1}-2\;,\qquad& 2X_{C} = 2X_{S_i}=K+1\;,\\
2X_{TC_i} = 2X_{SC}=K^{-1}+1\;,\qquad& 2\tilde X_{SC} = K +
K^{-1}\;. \esp \ee
Moreover
%
%
%
\be\lb{den}
\bsp \hat \O_{C}(\pp) &= {\bar K\over \pi v}{v^2p^2\over p_0^2+v^2
p^2}+A(\pp)\\
\hat D(\pp) &= {v\over \pi}\bar K {p_0^2\over p_0^2+v^2 p^2}
+B(\pp) \esp \ee
with $A(\pp)$, $B(\pp)$ continuous and vanishing at $\pp=0$, $v=\sin \bar
p_F+O(\l)$ and $\bar K=1+O(\l)$; therefore the Drude weight $D$ \pref{cc} and
the susceptibility $\k$ \pref{kk} are $O(\l)$ close to their free values and
verify the Luttinger liquid relation
\be\lb{hh1} v^2=D/\k\ee
\end{theorem}

The above Theorem says that, even if the logarithmic corrections
alter the power law decay of the spinning Luttinger model, the
exponents verify the same {\it universal relations} \pref{1.8}, in
agreement with {\it the Luttinger liquid conjecture}
\cite{KW071,LP075,Ha080}. Such relations say that the knowledge of
a single exponents implies the determination of all the others.

The Fourier transform of the density correlation is similar to the
free one, the interaction producing a renormalization of the
velocity and of the amplitude $\bar K$. The susceptibility and the
Drude weight are {\it finite}, saying that the system has a
metallic behavior (contrary to what happens in the half-filled
band case).

Besides the universal relations involving the critical exponents, there
is also the {\it universal relation} \pref{hh1}, which relates
the susceptibility and the Drude weight to the charge velocity $v$
appearing in \pref{den}; this relation was conjectured in \cite{Ha080}
(in the spinless case, but the extension to the spinning case is
straightforward, see {\it e.g.} \cite{Gi004}). In the notation of
\cite{Ha080}, $v_N=\pi\k^{-1}, v_J={D\over\pi}$ so that
\pref{hh1} takes the form
\be v_N v_J=v^2 \ee
The validity of \pref{1.8}, \pref{hh1} is a rather remarkable
universality property following from the combination of
conservation laws of the Hubbard model and Ward Identities coming
from the asymptotic gauge invariance of the effective theory. Wether a
similar relation holds also for the spin conductivity is an interesting
open problem.

\subsection{
Spin Charge separation}

\begin{theorem}\lb{th1.3}
Under the same condition of Theorem \ref{th1.1}, the Fourier transform of the
2-point Schwinger function is given by \be \hat S_2(\kk+{\bf
p}_F^\o)=Z(\kk)\hat S_{M,\o}(\kk)[1+R(\kk)]\virg {\bf p}_F^\o = (\o p_F,0)\ee
where
\be |R(\kk)|\le C {\l^2\over 1+ a |\l \log |\kk||}\virg a\ge 0\;, \ee
\be Z(\kk)=L(|\kk|^{-1})^{\z_z}[1+R'(\kk)]\virg |R'(\kk)|\le C|\l| \ee
$L(t)$, $t\ge 1$, is the function defined in Theorem \ref{th1.1} and $\hat
S_{M,\o}(\kk)$ is a function whose Fourier transform is of the form
\be\lb{hh}
S_{M,\o}(\xx)= \frac1{2\p v_F} { [v_\r^2 x_0^2 +(x_1/v_F)^2]^{-\h_\r/2}\over
(v_\r x_0+i\o x_1/v_F)^{1/2} (v_\s x_0+i\o x_1/v_F)^{1/2}} e^{C +O(1/|\xx|)}
\ee
with $v_{\r,\s}=1 +O(\l)$, $\h_\r=O(\l^2)$, $v_\r-v_\s=c_v\l+O(\l^2)$, with
$c_v\not=0$.
\end{theorem}


The above theorem says that the two point function can be written,
up to a logarithmic correction, as the 2-point function of the
spinning Luttinger model, a model which shows the phenomenon of
{\it spin-charge separation} (see also \cite{FM008}). A
manifestation of spin charge separation is that the 2-point
function is factorized in the product of two functions, similar to
Schwinger functions of particles with different velocities. In
this sense, the above theorem says that the spin-charge separation
occurs approximately also in the Hubbard model, but is valid only
at large distances and up to logarithmic corrections. Similar
expressions are true also for the density correlations (the
explicit formulae are in \S 5 and are not reported here for
brevity).
In the spinning Luttinger model $v_\r=v$, where $v$ is the
velocity appearing in \pref{den}; in the present case we can
verify this identity only at the lowest order in $\l$, and whether
this identity holds or not in the Hubbard model is an interesting
open problem.
%
%

\subsection{Borel summability}

In \S\ref{sec2.6} we shall prove the following Theorem.

\begin{theorem}\lb{th1.4}
Given $\d\in (0,\p/2)$, there exists $\e\=\e(\d)>0$, such that the
free energy, the Schwinger functions and the density correlations
are analytic in the set
\be\lb{dom}
D_{\e,\d}=\{\l\in\CCC: |\l|< \e, |\text{Arg } \l|< \p-\d \}
\ee
and continuous in the closure, $\bar D_{\e,\d}$. Moreover,
if $f(\l)$ is one of these
functions, there exist three constants $c_0$, $c_1$, $\k$, and a family of
functions $f_h(\l)$, $h\le 0$, analytic in the set
\be\lb{domh} D^{(h)}_{\e,\d}:=
D_{\e,\d} \bigcup \lft\{l\in\CCC:|\l| < \frac{c_0} {1+|h|}\rgt\}\ee
such that
\be\lb{Lesn}
|f_h(\l)| \le c_1 e^{-\k|h|} \virg f(\l) = \sum_{h=-\io}^0 f_h(\l)
\ee
\end{theorem}

By using the Lemma in \cite{Le087}, see pag. 466, this Theorem
implies that all the functions satisfy the Watson Theorem, see
pag. 192 of \cite{Hardy049}. Hence they are Borel summable in the
usual meaning.

\subsection{Contents of the paper}
The paper is organized in the following way.

\begin{enumerate}

\item In \S\ref{ss2.5a}--\ref{sec2.2} we resume the RG analysis of the
    extended Hubbard model given in \cite{BM001,M005}. The fermionic field
    is decomposed as a sum of fields $\psi^{(h)}$, $h$ integer and $\le 1$.
    The field $\psi^{(1)}$ is associated to the momenta far from the Fermi
    points, while the fields $\psi^{(h)}$ with $h\le 0$ are supported
    closer and closer to the two Fermi points, hence are more and more
    singular in the infrared region. The iterative integration of such
    fields, accompanied by a {\em free measure renormalization} (the field
    strength renormalization), leads to a sequence of {\it effective
    potentials} $V^{(h)}$, expressed as renormalized expansions in a set of
    $5$ {\it running couplings} $\vec v_h=(\n_h, \d_h, g_{1,h}, g_{2,h},
    g_{4,h})$, whose flow is driven by a recursive relation called {\it
    beta function}. The coupling $\n_h$ is associated to the only relevant
    term (in the usual RG language) present in the effective potentials
    (the others are marginal) and describes the change of the Fermi
    momentum due to the interaction; in order to control its flow, we
    change the chemical potential $\m$ in $\m-\n$ and we compensate this
    operation by adding to the interaction a {\em counterterm} $\n
    \sum_{x,s} a^+_{x,s} a^-_{x,s}$. The value of $\n$ is then chosen, by
    an iterative argument, so that $\n_h\to 0$ as $h\to -\io$; this will
    implicitly determine the interacting Fermi momentum. This procedure
    works since one can prove, under the conditions of Theorem \ref{th1.1},
    the convergence of expansions in the running couplings, by using two
    crucial technical tools: the determinant bounds for fermionic truncated
    expectations and the {\em partial vanishing of the beta function} (see
    \pref{beta23} below for the definition), which implies
    the convergence of $\vec v_h$ to $\vec v_{-\io} = (0, \d_{-\io}, 0,
    g_{2,-\io}, g_{4,\io})$. This limit can be seen as characterizing a
    point in a set of {\em fixed points}, depending on $3$ parameters, of
    the RG transformation, suitably scaled; of course, the chosen fixed
    point depends on all details of the model.

    \item In Appendix \ref{appA} a property of the effective couplings flow is
    proved, implying the convergence of our expansions in the set
    $D_{\e,\d}$, defined in \pref{dom}. At this point, as explained in
    \S\ref{sec2.6}, simple dimensional arguments allow us to prove Theorem
    \ref{th1.4} and then the Borel summability of the free energy and all
    the correlation functions.

\item In \S\ref{sec2.4}--\ref{sec2.5} the analysis is extended to the
    2-point function and to the response functions, which are expressed in
    terms also of renormalizations of the density operators. In particular,
    from the flow of these renormalizations one obtains the critical
    exponents and the logarithmic corrections appearing in \pref{asymp}.
    The critical exponents only depend on $v_F$ and $\vec v_{-\io}$, what will
    play a crucial role in the subsequent analysis. This is apparently not
    true for the logarithmic corrections; such corrections, absent in the
    spinless case, are due to the weaker convergence of the effective
    couplings to their limiting values.

\item The Luttinger liquid relations \pref{1.8} or \pref{den} can
be
    checked directly by the expansions at lowest orders, but the complexity
    of such expansions makes essentially impossible their proof at any
    order. Similarly, the partial vanishing of the beta function, on which
    the RG analysis is based, cannot be proven directly from the
    expansions. Such properties are related to the {\it asymptotic}
    validity of certain symmetries, and our strategy consists in the
    introduction of a suitable {\it effective model}, for which such
    symmetries are {\it exact}, and in showing that certain quantities
    computed in the effective model coincides, with a proper choice of its
    parameters, with analogues quantities in the extended Hubbard model.
    The {\it effective model} is introduced in \S\ref{ss2.5c} and is
    expressed directly in terms of functional integrals with linear
    dispersion relation; the model has an infrared and an ultraviolet
    momentum cut-off, and several interactions are present, non local and
    short ranged (both in space and time). The model can be consider a
    variation of the {\it g-ology} models introduced in the physical
    literature, the main difference being that the cut-offs are on space
    and time momentum components, what is of advantage for our approach
    (but makes the model not accessible to bosonization techniques).

\item The non locality of the interaction allows us the removal of the
    ultraviolet cut-off and a Renormalizaton Group analysis in the infrared
    region can be performed (similar to that performed in the Hubbard
    model), leading to convergent expansions in the effective couplings,
    see Appendix \ref{appB}. A first use of the effective model is in the
    proof of the partial vanishing of the beta function of the extended
    Hubbard model; a proof of this property was already given in
    \cite{M005}, but we present here a simplified version of it (the main
    novelty is that the ultraviolet cut-off in the effective model is
    removed), see Appendix \ref{appC}. Ward Identities and Schwinger-Dyson
    equations, with corrections due to the infrared cut-off, can be
    combined in the absence of back-scattering interactions to get
    relations implying the partial vanishing of the Beta function of the
    effective model; from this fact and using the symmetry properties in
    Appendix \ref{appB}, we can derive the partial vanishing of the beta
    function of the extended Hubbard model. Note the remarkable fact that a
    model with no back-scattering interaction and not spin-symmetric is
    used to prove properties of the Hubbard model, which is spin symmetric
    and in which the back-scattering interaction is present.

\item If the back-scattering coupling is set equal to zero and both the
    infrared and ultraviolet cut-offs are removed, the model becomes {\it
    exactly solvable}, in the sense that the Schwinger functions verify a
    set of {\it closed equations}, obtained combining Ward Identities and
    Schwinger-Dyson equations, derived in \S\ref{sec4.1}--\ref{ss4.4}. The
    non locality of the interaction has the effect that the anomalies in
    the Ward Identities verify the Adler-Bardeen non renormalization
    property, see \cite{M007_2}, \cite{BFM010}, so that they can be {\it
    exactly computed}; therefore also the critical exponents, which are
    expressed in terms of such anomalies, can be exactly computed and the
    analogue of the relations \pref{1.8} for the effective model are
    obtained.

\item In \S\ref{ss2.5d} we prove that the limiting values $\vec v_{-\io}$ of the
    effective couplings of the effective model with no back-scattering
    coincide with that of the Hubbard model, if a suitable fine tuning of
    the {\em bare} couplings of the effective model is done. Since the
    critical exponents only depend on $\vec v_{-\io}$ and have the same
    functional dependence on it as in the Hubbard model, this implies that
    also the critical exponents coincide. Therefore we can prove that
    \pref{asymp} is satisfied for the extended Hubbard model, with the
    critical exponents verifying the relations \pref{1.8}.

\item In \S\ref{ss2.5dhy} the proof of Theorem \ref{th1.3} is presented. By
    using the closed equation of the effective model in the limit of
    removed cut-offs, we can prove for it the exact Spin-Charge separation.
    We show that this implies the {\it approximate} Spin-Charge separation
    for the extended Hubbard model.

\item Finally, in \S\ref{ss2.5e} we compute the Drude weight and the
    susceptibility and we prove the Luttinger liquid relation \pref{hh1}.
    The proof is based on the fact that these quantities are related to the
    Fourier transform of some density correlations. However, the bounds
    obtained in Theorem \ref{th1.1} in coordinate space do not allow us to
    exclude logarithmic singularities. In order to prove their finiteness,
    we use the Ward Identities of the Hubbard model \pref{ref1},
    \pref{ref2}, \pref{ref3}, combined with the information coming from the
    effective model, keeping in this case also the backward interactions;
    this works since, even if the effective model is not completely
    solvable in that case, the Fourier transforms of the density
    correlations can be still exactly computed from the Ward Identities. We
    can prove that, by a suitable fine tuning of the couplings of the
    effective model, the Fourier transforms of the current and density
    operators coincide up to a constant and a renormalization, whose values
    are fixed by the Ward Identities \pref{ref1}, \pref{ref2},
    \pref{ref3}: this implies \pref{hh1}.

\end{enumerate}

\section{RG  Analysis for the Hubbard Model}\lb{sec2}

\subsection{Functional integral representation}\lb{ss2.5a}

The analysis of the Hubbard model correlations is done by a rigorous
implementation of the RG techniques. To begin with,  we need a {\it functional
integral representation} of the model, because the RG techniques are optimized
for that. We give here a concise description of it; a thorough discussion is in
Sec 2 of \cite{M005}. The main object to study is the functional
$\WW(J,\h)\=\WW_{M;L,\b}(J,\h)$, defined by
\be\lb{1z}
e^{\WW(J,\h)}=\int P(d\psi) e^{-\VV(\psi)+ \sum_\a \int d\xx J^{\a}_\xx
\r^{\a}_\xx  + \sum_s\int d\xx [\h^+_{\xx,s} \psi^-_{\xx,s} + \ps^+_{\xx,s}
\h^-_{\xx,s}]} \ee
where $\psi^\pm_{\xx,s}$ and $\h^\pm_{\xx,s}$ are Grassmann variables and the
fermionic density operators $\r^{\a}_\xx$ are defined as in \pref{rho}, with
$\psi^\pm_{\xx,s}$ in place of $a^\pm_{\xx,s}$, $J^{\a}_\xx$ are commuting
variables, $\int d\xx$ is a short form for $\sum_{x\in\CC}\int_{-\b/2}^{\b/2}
dx_0$, $P(d\psi)$ is a Grassmann Gaussian measure in the field variables
$\psi^\pm_{\xx,s}$ with covariance (the free propagator) given by
\bal\lb{cv0} &\int P(d\ps)\; \ps^\e_{\xx,s}\ps^\e_{\yy,s'}=0\;,
\qquad \int P(d\ps)\; \ps^-_{\xx,s}\ps^+_{\yy,-s}=0\;, \cr &\int
P(d\ps)\; \ps^-_{\xx,s}\ps^+_{\yy,s}= \bar g_M(\xx-\yy) :=
{1\over\b L}\sum_{\kk\in\DD_{L,\b}} {\chi(\g^{-M} k_0) e^{i
k_0\d_M} e^{-i\kk(\xx-\yy)} \over -i k_0+ (\cos \bar p_F -\cos
k)}\;. \eal
In the above formulae, $\chi(t)$ is a smooth compact support function equal to
$1$ for $|t|<1$ and equal to $0$ if $|t|\ge \g$, for a given {\sl scaling
parameter} $\g>1$, fixed throughout the paper; $M$ is a positive integer;
${\cal D}_{L,\b}:={\cal D}_L \times {\cal D}_\b$, ${\cal
D}_L:=\frac{2\p}{L}\CC$, ${\cal D}_\b:=\frac{2\p}{\b}(\zzzz+\frac12)$;
\be
\VV(\psi)=\l \sum_{s,s'=\pm}\int d\xx d\yy\; \psi_{\xx,s}^+\psi_{\xx,s}^-
v(\xx-\yy) \psi_{\yy,s'}^+ \psi_{\yy,s'}^-
\ee
with $v(\xx-\yy)=\d(x_0-y_0)v(x-y)$. Due to the presence of the
ultraviolet cut-off $\g^M$, the Grassmann integral has a finite
number of degree of freedom, hence it is well defined. The time
shift in \pref{cv0}, $\d_M:=\b/\sqrt{M}$, is introduced in order
to take correctly into account the discontinuity of $g(\xx)$ at
$\xx=0$: our definition guarantees that, fixed $L$ and $\b$,
$\lim_{M\to\io} \bar g_M(\xx)=g(\xx)$ for $\xx\not=0$, while
$\lim_{M\to\io} \bar g_M(0,0)=g(0,0^-)$, as it is to be for
Proposition \ref{p2.1} below.

If $\l=0$, the Hubbard model correlations can be easily calculated
by using \pref{cv0}, hence they are singular at momenta $(\o \bar
p_F,0)$, $\o=\pm1$. Since in the interacting theory, $\l\neq0$,
the position of the singularity is expected to change by O$(\l)$,
when the first of conditions \pref{ma} is satisfied, we add to the
interaction a {\it counterterm}
$$
\n \NN(\ps)=\n \sum_{s=\pm}\int d\xx\;
\psi_{\xx,s}^+\psi_{\xx,s}^-
$$
and, to leave unchanged $\WW(J,\h)$ in \pref{1z}, we subtract the
same term from the free measure, that has then a covariance:
%
\be\lb{cutM}
g_M(\xx)= {1\over\b L}\sum_{\kk\in\DD_{L,\b}} {\chi(\g^{-M} k_0) e^{i
k_0\d_M} e^{-i\kk \xx} \over -i k_0+ (\cos p_F -\cos k)}\;, \qquad
\ee
where $p_F$ is the {\it interacting  Fermi momentum}
defined such that
$$
\cos p_F=\m-\n\;.
$$
We introduce the following Grassmann integrals:
\be\lb{f2}
S_n^{M,\b,L}(\xx_1,s_1,\e_1;....;\xx_n,s_n,\e_n)
={\partial^n\over\partial\h^{-\e_1}_{\xx_1,s_1}...
\partial\h^{-\e_n}_{\xx_n,s_n}} \WW(J,\h)\Big|_{0,0}
\ee
It is well known that such Grassmann integrals, called {\em Schwinger
functions}, can be used to compute the thermodynamical properties of the model
with Hamiltonian \pref{1.1}. This follows from the following proposition.

\begin{proposition}\lb{p2.1}
For any finite $\b$ and $L$, there exists a complex disc, centered in
the origin, $D_{L,\b}$,
such that, if $\l\in D_{L,\b}$,
\be\lb{trace}
{\Tr[e^{-\b H}{\bf T} a^{\e_1}_{\xx_1,s_1}\cdots a^{\e_n}_{\xx_n,s_n} ]\over
\Tr[e^{-\b H}]}|_T=\lim_{M\to\io}
S_n^{M,\b,L}(\xx_1,s_1,\e_1;....;\xx_n,s_n,\e_n)\;;
\ee
besides both members are analytic in $\l$ in the same disc.
A similar statement holds for the density correlations.
\end{proposition}

The proof of this theorem can be done exactly as in the spinless case, see
\cite{BM001}. The main point, strictly related with the fact that we are
treating a fermionic problem, is that, for $L$ and $M$ finite, the l.h.s. of
\pref{trace} is the ratio of the traces of two matrices whose coefficients are
entire functions of $\l$, hence it is the ratio of two entire functions of
$\l$. Then, it may have a singularity only if $\Tr[e^{-\b H}]$ vanishes, which
certainly does not happen in a neighborhood of $\l=0$ small enough. On the
other hand, it is rather easy to prove that also the r.h.s. of \pref{trace} is
analytic in a small neighborhood of $\l=0$ and that its Taylor coefficients
coincide with those of the l.h.s.. This follows from the fact that the UV
singularity of the free propagator is very mild and can be controlled with a
trivial resummation, in the RG expansion, of the tadpole terms (see pag. 1383
of \cite{BM001}). In this resummation, the only important thing to check is
that $\lim_{M\to\io} g_M(0,0)= g(0,0^-)$ (otherwise the perturbative expansions
of the two sides of \pref{trace} would not coincide).

The RG analysis will allow to prove \mp{...}
that the analyticity domain is indeed of the form $D_{\b,L}=\{\l, |\l|\le c\e_0
\min \{(\log\b)^{-1}, (\log L)^{-1}\}\} \bigcup \{|\l| \le \e_0, |\arg
\l|<{\pi\over 2}+\d\}$, with $c, \e_0>0$, $0<\d<\pi/2$ independent of $\b$ and
$L$.

\subsection{Multiscale analysis for the effective potential}

We will briefly recall here the RG analysis for interacting fermionic systems
on the lattice as developed in \cite{BM001} and \cite{M005} in the spinless and
the spinning case, respectively. Note that the proof of many technical points
do not depend on the spin, hence we shall refer to \cite{BM001} for the
corresponding details.

Let $\TTT$ be the one dimensional torus, $\|k-k'\|_\TTT$ the usual distance
between $k$ and $k'$ in $\TTT $ and $\|k\|_\TTT=\|k-0\|_\TTT$. We introduce a
positive function $\c(\kk') \in C^{\io}(\TTT \times \RRR)$, $\kk'=(k_0,k')$,
such that $ \c(\kk') = \c(-\kk') = 1$ if $|\kk'| <t_0 = a_0
v_F/\g$ and $=0$ if $|\kk'|>a_0 v_F$, where $v_F=\sin p_F$, $a_0=
\min \{{p_F\over 2}, {\p- p_F\over 2}\}$ and
$|\kk'|=\sqrt{k_0^2+v^2_F \|k'\|_{\TTT }^2}$. The above definition
is such that the supports of $\c(k-p_F,k_0)$ and $\c(k+p_F,k_0)$
are disjoint and the $C^\io$ function on $\TTT \times R$
\be
\lb{f1} \hat f_1(\kk) := 1- \c(k-p_F,k_0) - \c(k+p_F,k_0) \ee
is equal  to $0$, if $v_F^2\|\big[|k|-p_F\big]\|_{\TTT }^2 +k_0^2<t_0^2$. We
define also, for any integer $h\le 0$,
\be f_h(\kk')= \c(\g^{-h}\kk')-\c(\g^{-h+1}\kk') \ee
which has support $t_0 \g^{h-1}\le |\kk'|\le t_0 \g^{h+1}$ and equals
1 at $|\kk'| =t_0\g^h$; then
\be \c(\kk') = \sum_{h=h_{L,\b}}^0 f_h(\kk') \ee
where
\be
h_{L,\b} :=\min \lft\{h:t_0\g^{h+1} > |\kk_{\rm m}|\rgt\}\qquad
{\rm for}\quad\kk_{\rm m}=(\p/\b,\p/L)\;.
\ee
%
For $h\le 0$ we also define
%
\be \hat f_h(\kk) = f_h(k-p_F,k_0) +f_h(k+p_F,k_0) \ee
(for $h=1$ the definition is \pref{f1}). This definition implies that, if $h\le
0$, the support of $\hat f_h(\kk)$ is the union of two disjoint sets, $A_h^+$
and $A_h^-$. In $A_h^+$, $k$ is strictly positive and $\|k-p_F\|_{\TTT }\le
t_0\g^h \le t_0$, while, in $A_h^-$, $k$ is strictly negative and
$\|k+p_F\|_{\TTT }\le t_0\g^h$. The label $h$ is called the {\sl scale} or {\sl
frequency} label. Note that
\be 1=\sum_{h=h_{L,\b}}^1 \hat f_h(\kk) \ee
and that, since $p_F$ is not uniquely defined at finite volume (we are
interested only to the zero temperature limit), then we can redefine it as
$(2\p/L)(n_F+1/2)$, with $n_F =[Lp_F/(2\p)]$.
Hence,
if $\DD'_L \= \frac{2\p}{L}(\CC+\frac12)$ and $\DD'_{L,\b}=\DD'_L \times
\DD_\b$, we can write:
\be\lb{gdef}
g(\xx-\yy) = g^{(1)}(\xx-\yy) + \sum_{\o=\pm} \sum_{h=h_{L,\b}}^0
e^{-i\o p_F(x-y)} g^{(h)}_\o(\xx-\yy)
\ee
where
\bal
g^{(1)}(\xx-\yy) &= {1\over\b L} \sum_{\kk\in\DD_{L,\b}} e^{-i\kk(\xx-\yy)}
{\hat f_1(\kk)\over -i k_0 + (\cos p_F -\cos k)}\\
g^{(h)}_\o(\xx-\yy) &= {1\over\b L} \sum_{\kk'\in\DD'_{L,\b}}
e^{-i\kk'(\xx-\yy)} {f_h(\kk')\over -i k_0+ E_\o(k')}
\eal
and
\be \lb{dE}
E_\o(k') = \o v_F \sin k' + \cos p_F (1-\cos k')
\ee
Notice we have dropped the important phase factor  $e^{i k_0\d_M}$ from $g$,
for it  plays no explicit role in the following analysis, since the limit
$N\to\io$ is taken before the limits $L\to\io$ and $\b\to\io$. As consequence
of fundamental properties of the Grassmann Gaussian integration, the
decomposition of the covariance \pref{gdef} implies a decomposition of the
field
\be\lb{pdef}
\ps^\e_{\xx,s}= \ps^{\e,(1)}_{\xx,s} + \sum_{\o=\pm} \sum_{h=h_{L,\b}}^0
e^{i\o p_F\e \xx}\ps^{\e,(h)}_{\xx,\o,s}
\ee
where fields with different scale labels or different label $\o$
are independent, and the covariance of
$\ps^{(1)}$ is $g^{(1)}$, while the covariance of $\ps^{(h)}_{\o}$ is
$g^{(h)}_\o$. Basically the label $\o$ refers to either two branches of the
dispersion relation.

Let us now describe the perturbative expansion of the functional $\WW(J,\h)$
defined in \pref{1z}; for simplicity we shall consider only the case $\h=0$. We
can write:
\be\lb{2z}
\bsp
e^{\WW(J,0)} &= \int P(d\psi^{\le 0}) \int P(d\psi^{(1)})\, e^{-\VV(\psi) - \n
\NN(\ps) + \sum_\a \int d\xx J^{(\a)}_\xx
\r^{(\a)}_\xx}=\\
&=e^{-L\b E_0}\; \int P(d\psi^{\le 0}) \, e^{-\VV^{(0)}(\psi^{\le
0})+\BB^{(0)}(\psi^{\le 0},J)}
\esp
\ee
where, if we put $\ux=(\xx_1,\ldots,\xx_{2n})$, $\oo=(\o_1,\ldots,\o_{2n})$ and
$\psi_{\ux,\oo} = \prod_{i=1}^n$ $\psi^+_{\xx_i,\o_i} \prod_{i=n+1}^{2n}
\psi^-_{\xx_i,\o_i}$, the {\it effective potential} $\VV^{(0)}(\ps)$ can be
represented as
\be\lb{3.2aaa} \VV^{(0)}(\psi)= \sum_{n\ge 1} \sum_{\oo} \int d\ux\;
W^{(0)}_{\oo, 2n}(\ux) \psi_{\ux,\oo} \ee
the kernels $W^{(0)}_{\oo,2n}(\ux)$ being analytic functions of $\l$ and $\n$
near the origin; if $|\n|\le C|\l|$ and we put $\uk=(\kk_1, \ldots,
\kk_{2n-1})$, their Fourier transforms satisfy, for any $n\ge 1$, the bounds,
see \S 2.4 of \cite{BM001},
\be |\widehat W^{(0)}_{\oo,2n}(\uk)| \le C^n
|\l|^{\max\{1,n-1\}} \ee
A similar representation can be written for the functional $\BB^{(0)}(\psi^{\le
0},J)$, containing all terms which are at least of order one in the external
fields, including those which are independent on $\psi^{\le 0}$.

The integration of the scales $h\le 0$ is done iteratively in the
following way. Suppose that we have integrated the scale
$0,-1,-2,..,j$, obtaining
\be\lb{61}
e^{\WW(J,0)}=e^{-L\b E_j} \int
P_{Z_j,C_j}(d\psi^{\le j})e^{-\VV^{(j)}(\sqrt{Z_j}\psi^{\le
j})+\BB^{(j)}(\sqrt{Z_j}\psi^{\le j},J)}
\ee
where, if we put $C_j(\kk')^{-1}=\sum_{h=h_{L,\b}}^j f_h(\kk')$,
$P_{Z_j,C_j}$ is the Grassmann integration with propagator
\be\lb{62}
{1\over Z_j}\, g^{(\le j)}_\o(\xx-\yy)= {1\over
Z_j}{1\over\b L} \sum_{\kk\in\DD'_{L,\b}} e^{-i\kk(\xx-\yy)}{C_j^{-1}(\kk)
\over -i k_0+E_\o(k')}
\ee
$\VV^{(j)}(\psi)$ is of the form
\be\lb{3.2aaax} \VV^{(j)}(\psi)= \sum_{n\ge 1} \sum_{\oo} \int
d\ux W^{(j)}_{\oo,2n}(\ux) \psi_{\ux,\oo} \ee
and $\BB^{(j)}(\psi^{\le j},J)$ contains all terms which are at least of order
one in the external fields, including those which are independent on $\psi^{\le
j}$. For $j=0$, $Z_0=1$ and the functional $\VV^{(0)}$ and $\BB^{(0)}$ are
exactly those appearing in \pref{2z}.

First of all, we define a localization operator in the following
way:
\be\lb{rel}
\bsp
\LL \VV^{(j)}(\sqrt{Z_j}\psi) &= \g^j n_j F_\n(\sqrt{Z_j}\psi)+a_j
F_\a (\sqrt{Z_j}\psi)+z_j F_z(\sqrt{Z_j}\psi)\\
&+ l_{1,j} F_1(\psi) + l_{2,j} F_2(\sqrt{Z_j}\psi) + l_{4,j}
F_4(\sqrt{Z_j}\psi)
\esp
\ee
where
\bal
F_\n &= \sum_{\o,s} \int d\xx\, \psi^+_{\xx,\o,s}\psi^-_{\xx,\o,s}\virg &F_1 &=
\frac12 \sum_{\o,s,s'} \int d\xx\, \psi^+_{\xx,\o,s}\psi^-_{\xx,-\o,s}
\psi^+_{\xx,-\o,s'}\psi^-_{\xx,\o,s'}\nn\\
F_\a &= \sum_{\o,s} \int d\xx\, \psi^+_{\xx,\o,s}\DD\psi^-_{\xx,\o,s}\virg &F_2
&= \frac12 \sum_{\o,s,s'} \int d\xx\, \psi^+_{\xx,\o,s} \psi^-_{\xx,\o,s}
\psi^+_{\xx,-\o,s'} \psi^-_{\xx,-\o,s'}\label{ff11}\\
F_z &= \sum_{\o,s}\int d\xx\, \psi^+_{\xx,\o,s}\partial_0
\psi^-_{\xx,\o,s}\virg &F_4 &= \frac12 \sum_{\o,s} \int d\xx\,
\psi^+_{\xx,\o,s} \psi^-_{\xx,\o,s} \psi^+_{\xx,\o,-s} \psi^-_{\xx,\o,-s}\nn
\eal
and $\DD\psi_{\xx,\o,s}=\int d\kk e^{i\kk\xx}\; E_\o(k)
\psi^+_{\kk,\o,s}$ (see definition \pref{dE}). Note
that
\be\lb{init}
l_{4,0}=2\l \hat v(0)+O(\l^2)\quad l_{2,0}=2\l \hat v(0)+O(\l^2)\quad
l_{1,0}=2\l\hat v(2p_F)+O(\l^2)
\ee
and in writing \pref{rel}  the $SU(2)$ spin symmetry has been used. In the
case of local interactions, $\hat v(p)=1$. $F_1$ in \pref{ff11} is called
{\it backward interaction} while $F_2,F_4$ are the {\it forward
interactions}; the {\it umklapp interaction}, defined analogously as
\pref{umm} below, is not present in $\LL \VV^{(j)}$, as well as other terms
quadratic in the fields. The reason is that the condition $\bar p_F\neq 0,
\frac\p2, \p$ says that such terms are vanishing for $j$ smaller than a
suitable constant (depending on $\bar p_F$ and $\l$), because they cannot
satisfy the conservation of the momentum, so there is no need to localize
them (more details are in \cite{M005}).

Moreover, the local marginal operators associated with the densities
\pref{rho} are defined in the following way:
\be\lb{2.20} \LL \BB^{(j)}(\sqrt{Z_j}\psi,J)= \int d\xx \; J^{(\a)}_\xx
\Bigg[\sum_{\a\not=TC_i} Z^{(1,\a)}_j O^{(1,\a)}_\xx(\psi) + \sum_\a
Z^{(2,\a)}_j O^{(2,\a)}_\xx(\psi)\Bigg]
\ee
where $ O^{(1,\a)}$ are the {\it small momentum transfer} correlations, (see
pag. 231 of \cite{So079})
\bal O^{(1,C)}_\xx &= \sum_{\o,s} \psi^+_{\xx,\o,s}
\psi^-_{\xx,\o,s}\nn\\
\lb{op1} O^{(1,S_i)}_\xx &= \sum_{\o,s,s'}
\psi^+_{\xx,\o,s}\s^{(i)}_{s,s'}\psi^-_{\xx,\o,s'}\\
O^{(1,SC)}_\xx &= \sum_{\e,\o,s} s\, e^{2i \e\o p_F x} \psi^\e_{\xx,\o,s}
\psi^\e_{\xx,\o,-s}\nn
\eal
while $O^{(2,\a)}$ are the {\it large momentum transfer} correlations,
(see pag. 221 of \cite{So079})
\bal O^{(2,C)}_\xx &= \sum_{\o,s} e^{2i \o p_F x}
\psi^+_{\xx,\o,s} \psi^-_{\xx,-\o,s}\nn\\
O^{(2,S_i)}_\xx &= \sum_{\o,s,s'}
e^{2i \o p_F x} \psi^+_{\xx,\o,s}\s^{(i)}_{s,s'}\psi^-_{\xx,-\o,s'}\nn\\
\lb{op2} \nn\\[-30pt]
\\
O^{(2,SC)}_\xx &= \sum_{\e,\o,s} s\,
\psi^\e_{\xx,\o,s} \psi^\e_{\xx,-\o,-s}\nn\\
O^{(2,TC_i)}_\xx &= \sum_{\e,\o,s,s'} e^{-i\e \o p_F} \psi^\e_{\xx,\o,s}
\tilde\s^{(i)}_{s,s'} \psi^\e_{\xx,-\o,s'}\nn
\eal
These definitions are such that the difference between $-\VV^{(j)} + \BB^{(j)}$
and $-\LL \VV^{(j)} + \LL \BB^{(j)}$ is made of irrelevant terms.

Note that the factor $e^{-i\e \o p_F}$ in the definition of $O^{(2,TC_i)}_\xx$
comes from the fact that the two $a^\e$ operators in the definition \pref{rho}
of the triplet Cooper density are located in two different lattice
sites
(otherwise the density would vanish). Moreover,
there is no local operator $O^{(1,TC_i)}_\xx$ because
$\sum_{s,s'}\psi^\e_{\xx,\o,s} \tilde\s^{(i)}_{s,s'} \psi^\e_{\xx,\o,s'}\=0$ by
anticommutation of the fermion fields.

We then renormalize the integration measure, by moving to it part of the
quadratic terms in the r.h.s. of \pref{h1}, that is $-z_j (\b L)^{-1}
\sum_{\o,s} \sum_{\kk} [-i k_0 + E_\o(k)] \psi^+_{\kk,\o,s} \psi^-_{\kk,\o,s}$;
equation \pref{61} takes the form:
\be\lb{61a}
e^{\WW(J,0)}=e^{-L\b (E_j+t_j)} \int P_{\tilde Z_{j-1},C_j} (d\psi^{(\le j)})
e^{-\tilde\VV^{(j)}(\sqrt{Z_j} \psi^{\le j}) + \BB^{(j)}(\sqrt{Z_j} \psi^{\le
j},J,\tilde J)}
\ee
where $\tilde\VV^{(j)}$ is the remaining part of the effective interaction,
$P_{\tilde Z_{j-1},C_j}(d\psi^{\le j})$ is the measure whose propagator is
obtained by substituting in \pref{62} $Z_j$ with
\be
\tilde Z_{j-1}(\kk) =Z_j [1+z_j C_j(\kk)^{-1}]
\ee
and $t_j$ is a constant coming from the normalization of the measure. It is
easy to see that we can decompose the fermion field as $\psi^{(\le j)} =
\psi^{(\le j-1)} + \psi^{(j)}$, so that
\be
P_{\tilde Z_{j-1},C_j}(d\psi^{\le j}) = P_{ Z_{j-1}, C_{j-1}}(d\psi^{(\le
j-1)}) P_{ Z_{j-1}, \tilde f_j^{-1}}(d\psi^{(j)})
\ee
where $\tilde f_j(\kk)$ (see eq. (2.90) of \cite{BM001}) has the same support
and scaling properties as $f_j(\kk)$. Hence, if we make the field rescaling
$\psi\to [\sqrt{Z_{j-1}}/ \sqrt{Z_j}]\psi$ and call
$\hat\VV^{(j)}(\sqrt{Z_{j-1}} \psi^{\le j})$ the new effective potential, we
can write the integral in the r.h.s. of \pref{61a} in the form
\be
\int P_{Z_{j-1},C_{j-1}} (d\psi^{(\le j-1)}) \int P_{ Z_{j-1}, \tilde
f_j^{-1}}(d\psi^{(j)}) e^{-\hat\VV^{(j)}(\sqrt{Z_{j-1}} \psi^{(\le j)}) +
\hat\BB^{(j)}(\sqrt{Z_{j-1}} \psi^{(\le j)},J,\tilde J)}\nn
\ee
By performing the integration over $\psi^{(j)}$, we finally get
\pref{61}, with $j-1$ in place of $j$.
In order to analyze the result of this iterative procedure, we
note that $\LL \hat\VV^{(j)}(\psi)$ can be written as
\be \LL \hat\VV^{(j)}(\psi) = \g^j\n_j F_\n(\psi) + \d_j
F_\a(\psi) + g_{1,j} F_1(\psi)+ g_{2,j} F_2(\psi)+ g_{4,j} F_4(\psi)
\ee
where $\n_j=(\sqrt{Z_j}/ \sqrt{Z_{j-1}}) n_j$, $\d_j=(\sqrt{Z_j}/
\sqrt{Z_{j-1}}) (a_j-z_j)$ and $g_{i,j}=(\sqrt{Z_j}/ \sqrt{Z_{j-1}})^2
l_{i,j}$, $i=1,2,4$, are called the {\it running coupling constants} (r.c.c.)
on scale $j$.

In Theorem (3.12) of \cite{BM001} it is proved that the kernels of
$\hat\VV^{(j)}$ and $\hat\BB^{(j)}$ are {\it analytic} as functions of the
r.c.c., provided that they are small enough. One has then to analyze the flow of
the r.c.c.  (the {\it beta function}) as $j\to-\io$. We shall now summarize the
results, following \S 4 and \S 5 of \cite{M005} with some improvement.

\subsection{The flow of the running coupling constants}\lb{sec2.2}

Define  vector notations for the r.c.c.,
\be\lb{vn} \vec v_h\= (v_{1,h}, v_{2,h}, v_{4,h}, v_{\d,h},
v_{\n,h}) =(g_{1,h}, g_{2,h}, g_{4,h},\d_h, \n_h) \=(\vec g_{h},
\d_h, \n_h)\;. \ee
The r.c.c.  satisfy a set of recursive equations, which can be
written in the form
\be\lb{floweq0} v_{\a,j-1}=A_\a v_{\a,j} + \hat \b_\a^{(j)}(\vec
v_j;...,\vec v_0;\l,\n) \ee
with $A_\n=\g$, $A_\a=1$ for $\a\not=\n$. These equations have been already
analyzed in \cite{M005}, where it has been proved that, if $\l$ is real
positive and small enough, then it is possible to choose $\n$ so that, fixed
$\th<1$, $|\n_h|\le C\l\g^{\th h}$, $\forall h\le 0$, and $0< g_{1,h} < \l(1+
\bar a \l |h|)^{-1}$, for some $\bar a>0$, while the other r.c.c. stay bounded
by $C\l$ and converge for $h\to -\io$. In this paper, in order to proof Borel
summability of perturbation theory, we extend the proof to complex values of
$\l$, restricted to the set $D_{\e,\d}$ defined in \pref{dom}; this implies
that we need an analysis a bit more precise of the flow equations
\pref{floweq0}.

To begin with, we put $\n_1\=\n$ and we suppose that the sequence $\{\n_h\}_{h
\le 1}$ are known functions of $\l$, analytic in $D_{\e,\d}$, such that
\be\lb{ne2} |\n_h|\le C|\l|\g^{\th h}\virg h\le 1 \ee
and study the flow equations of the other variables. The idea is that this
restricted flow has properties such that, by a fixed point argument, the
sequence $\{\n_h\}_{h\le 1}$, satisfying the last equation of \pref{floweq0},
can be uniquely determined. Since this point can be treated in the same way as
in spinless case (see \S4.3 of \cite{BM001}), we shall give for granted this
result. Hence, from now on, we shall define ${\bf v}_j = (g_{1,j}, g_{2,j},
g_{4,j},\d_j)$ and we shall consider the restriction of \pref{floweq0} to ${\bf
v}_j$.

The next step is to extract from the functions $\hat \b_\a^{(j)}$ the leading
terms for $j\to-\io$. Observe that the propagator $\tilde g^{(j)}_\o$ of the
single scale measure $P_{ Z_{j-1}, \tilde f_j^{-1}}$, can be decomposed as
\be\lb{ne1}
\tilde g^{(j)}_\o(\xx)= {1\over Z_j}\,
g^{(j)}_{{\rm D},\o}(\xx)+r^{(j)}_\o(\xx)
\ee
where $g^{(j)}_{{\rm D},\o}$ is the  {\it Dirac propagator}
(with cutoff) and  describes the leading asymptotic behavior
\be\lb{gjth}
g^{(j)}_{{\rm D},\o}(\xx):= {1\over\b
L}\sum_{\kk\in\DD_{L,\b}}e^{-i\kk\xx} {\tilde f_j(\kk) \over -i k_0+\o
v_F k}\;,
\ee
while the {\em remainder}
$r^{(j)}_\o$ satisfies, for any $q>0$ and $0<\th<1$, the bound
\be\lb{2.30}
|r^{(j)}_\o(\xx)|\le {\g^{(1+\th)j}\over Z_j} {C_{q,\th}\over 1+(\g^j
|\xx|)^q}\;.
\ee
Let us now call $Z_{{\rm D},j}$ the values of $Z_j$ one would obtain by
substituting $\VV^{(0)}$ with $\LL \VV^{(0)}$ in \pref{2z} and by using for the
single scale integrations the propagator \pref{ne1} with $r^{(i)}_\o(\xx)\=0$
for any $i\ge j$. It can be proved by an inductive argument, see \S4 of
\cite{BM001}, that, if all the r.c.c.  stay of order $\l$,
\be\lb{ne} \left| {Z_{j}\over Z_{j-1}}-
{Z_{{\rm D}, j}\over Z_{{\rm D}, j-1}}\right|\le C \e_j^2\g^{\th j} \ee
where
$$\e_j=\max\{ |\l|,
\max_{0\ge h\ge j}|\vec g_h|, \max_{0\ge h\ge j}|\d_h| \}\;.
$$
It is then convenient to decompose the functions $\hat \b_\a^{(j)}$ as
\be\lb{bal} \hat \b_\a^{(j)}(\vec v_j;...,\vec v_0;\l,\n)
= \b_\a^{(j)}(\vv_j,...,\vv_0) +\bar \b_\a^{(j)}(\vec v_j;...,\vec v_0;\l)
\ee
where $\b_\a^{(j)}(\vv_j,...,\vv_0)$ is given by the sum of all trees
containing only endpoints with r.c.c. $\d_h, \vec g_h$, $0\ge h\ge j$, modified
so that the propagators $g^{(h)}_\o$ and the wave function renormalizations
$Z_h$, $0\ge h\ge j$, are replaced by $g^{(h)}_{{\rm D},\o}$ and $Z_{{\rm D},
h}$; $\bar\b_\a^{(j)}$ contains the correction terms together with the
remainder of the expansion.
\begin{lemma}
\be\lb{2.34}
|\bar \b_\a^{(j)}(\vec v_j;...,\vec v_0;\l)|\le
\begin{cases}
C \e_j^2 \g^{\th j} & \text{if $\a\not=\d$}\\
(\bar c\,\e_0 + C\e_j^2) \g^{\th j} & \text{if $\a=\d$}
\end{cases}
\ee
\end{lemma}
As showed in \cite{M005}, this lemma is basically a consequence of
\pref{ne} and \pref{ne2}. Therefore the leading term in \pref{bal}
is $\b_\a^{(j)}$, that we further decompose as
\be\lb{2.43a}
\b_\a^{(j)}(\vv_j,...,\vv_0) = \tilde\b_\a^{(j)}(\vv_j) +
r_{\a,j}(\vv_j,...,\vv_0)
\ee
where $\tilde \b_\a^{(j)}(\vv) = \b_\a^{(j)}(\vv,...,\vv)$. We can write:
\be\lb{2.43b}
\tilde\b_\a^{(j)}(\vv_j) = \sum_{i=0,1} b_{\a,i}^{(j)}(\vv_j) + b_{\a,\ge
2}^{(j)}(\vv_j)
\ee
where $b_{\a,i}^{(j)}(\vv_j)$ is the contribution of order $i$ in $g_{1,j}$,
wile $b_{\a,\ge 2}^{(j)}(\vv_j)$ is the contributions of all trees with at
least two endpoints of type $g_1$. The crucial property is the following lemma.
\begin{lemma}[partial vanishing of  the beta function]
\be\lb{beta23}
|b_{\a,i}^{(j)}(\vv_j)| \le C \e_j^2 \g^{\th j} \virg i=0,1
\ee
\end{lemma}
The above property was proven in \S 5.3 of \cite{M005}, extending the proof for
the spinless case in \cite{BM004,BM005}, and it will be reviewed in App.
\ref{appC}. Now, let us extract from $\tilde \b_\a^{(j)}(\vv_j)$ the second
order contributions, which all belong to $b_{\a,\ge 2}^{(j)}(\vv_j)$; we get:
\be\lb{2.46}
\tilde \b_\a^{(j)}(\vv_j) = -a_\a g_{1,j}^2+ \sum_{i=0,1} b_{\a,i}^{(j)}(\vv_j)
+ \tilde r_{\a,j}(\vv_j)\ee
with $a_1=a>0$, $a_2=a/2$, $a_4=a_\d=0$, and, for some $b_1>0$,
\be\lb{2.47a} |\tilde r_{\a,j}(\vv_j)| \le b_1 \e_j |g_{1,j}|^2\;.
\ee
In the limit $L,\b=\io$, if $g_{{\rm D},\o}^{(\ge h)} \= \sum_{j=h}^0 g_{{\rm
D},\o}^{(j)}$,
\be\lb{defa} a = 2 \lim_{h\to-\io} \frac{1}{|h|}
\int \frac{dk}{(2\p)^2} \hg_{{\rm D},+}^{(\ge h)}(\kk)
\hg_{{\rm D},-}^{(\ge h)}(\kk) =
\frac{\log\g}{\p v_F}
\ee
Let us now analyze in more detail the functions $r_\a^{(j)}(\vv_j,...,\vv_0)$,
which appear in \pref{2.43a}. If we define, for $j'\ge j+1$,
\be
D_\a^{(j,j')}(\vv_j,...,\vv_0) =
\b_\a^{(j)}(\vv_j,...,\vv_j,\vv_{j'},...,\vv_0)-
\b_\a^{(j)}(\vv_j,...,\vv_j,\vv_{j},...,\vv_0)
\ee
we can decompose $r_\a^{(j)}(\vv_j,...,\vv_0)$ in the following way:
\be\lb{2.50}
r_{\a,j}(\vv_j,...,\vv_0) = \sum_{j'=j+1}^0 D_\a^{(j,j')}(\vv_j,...,\vv_0)
\ee
Note that $D_\a^{(j,j')}(\vv_j,...,\vv_0)$ is obtained from
$\b_\a^{(j)}(\vv_j,...,\vv_0)$, by changing the values of the r.c.c. in the
following way: the r.c.c. associated to endpoints of scales lower than $j'$ are
put equal to the corresponding r.c.c. of scale $j$; those of scale greater than
$j'$ are left unchanged; at least one of the r.c.c. $v_{r,j'}$ is substituted
with $v_{r,j'} - v_{r,j}$. By using the {\em short memory property} (see \eg
(4.31) of \cite{BM001}), we can show that , if $\e_j$ is small enough,
\be\lb{3.22}
|D_\a^{(j,j')}(\vv_j,...,\vv_0)| \le b_3 \e_j \g^{-(j'-j)\th} |\vv_{j'} -\vv_j|
\ee
for some $b_3>0$. If we insert in the flow equation \pref{floweq0} the
equations \pref{bal}, \pref{2.43a}, \pref{2.46}, \pref{2.50} and use the bounds
\pref{2.34}, \pref{beta23}, \pref{2.47a} and \pref{3.22}, we get, if $\e_j$ is
small enough,
\be\lb{vdiff0}
|\vv_{j-1} -\vv_j| \le (a+ b_1\e_j) |g_{1,j}|^2 + (\bar c\,\e_0 + b_2 \e_j^2)
\g^{\th j} + b_3\e_j \sum_{j'=j+1}^0 \g^{-\th(j'-j)} |\vv_{j'} -\vv_j|
\ee
for some $b_2>0$. The form of this bound implies that, in order to control
the flow, it is sufficient to prove that $g_{1,j}$ goes to $0$ as $j\to -\io$
so fast that $|g_{1,j}|^2$ is summable on $j$. Hence, we have to look more
carefully to the flow equation of $g_{i,j}$. By proceeding as before, we can
write
\be\lb{2.53a}
g_{1,j-1} = g_{1,j} - a g_{1,j}^2 + \tilde r_{1,j} + r_{1,j} + \bar r_{1,j}
\ee
\be |\bar r_{1,j}| \le b_2\e_j^2 \g^{\th j} \virg |\tilde r_{1,j}| \le b_1 \e_j
|g_{1,j}|^2 \virg |r_{1,j}| \le  b_3\e_j \sum_{j'=j+1}^0 \g^{-\th(j'-j)}
|\vv_{j'} -\vv_j|
\ee
It is easy to show that, if $\e_0$ is small enough, there is a constant $c_4$,
such that, if $g_{1,0}\in D_{\e_0,\d}$ and $c_4 |j_0| |g_{1,0}|^2\le
|g_{1,0}|^{2-\h}$, $\h<1$, then, for $j\ge j_0$,
\be\lb{2.55} g_{1,j}\in D_{2\e_0,\d/2} \virg |g_{1,0}|/2 \le |g_{1,j}|\le
2|g_{1,0}| \virg \e_j\le 2\e_0 \ee
Hence we put $j_0 = -(c_4|g_{1,0}|^{1/2})^{-1}$ and suppose $\e_0$ so small
that
\be\lb{2.57c} \e_{j_0} \g^{\frac{\th}2 j_0} \le 2 c_5 |g_{1,j_0}|
\g^{\frac{\th}2 j_0} \le |g_{1,j_0}|^3\ee
where we also used the fact that, since $\hat v(2p_F)>0$, $\e_0\le c_5
|g_{1,0}|$, for some constant $c_5$.

\begin{lemma}\lb{lm2.4}
If $g_{1,0}\in D_{\e_0,\d}$  and $j\ge j_0$, then, if $\e_0$ is small enough,
\be\lb{vdiff1}
|\vv_{j-1} -\vv_j| \le 2a |g_{1,j}|^2 + 2\bar c \e_0 \g^{\frac{\th}{2} j}
\ee
\end{lemma}

\0{\bf Proof} - We shall proceed by induction. By \pref{2.55}, if $\e_0$ is
small enough, $\bar c\,\e_0 + b_2 \e_j^2 \le (3/2)\bar c\,\e_0$ and
$a+b_1\e_j \le 3a/2$; hence, \pref{vdiff1} is true for $j=0$. Let us suppose
that \pref{vdiff1} is verified for $j>h> 0$. By \pref{2.55}, if $j\ge h\ge
j_0$, $|g_{1,j}|/|g_{1,h}|\le 4$; hence, by using \pref{vdiff0} and
\pref{vdiff1}, we get:
\be\nn
\bsp
&|\vv_{h-1} -\vv_h| \le (3/2) a |g_{1,h}|^2 + (3/2) \bar c\,\e_0 \g^{\th h}  +\\
&b_3\e_h \sum_{j=h+1}^0 \g^{-\th(j-h)} (j-h) \max_{h< j' \le j} \left[2a
|g_{1,j'}|^2 + 2\bar c\e_0 \g^{\frac{\th}{2} j'} \right]\\
&\le |g_{1,h}|^2 \left[(3/2) a + 64 a b_3\e_0 \sum_{n=0}^\io n\, \g^{-\th n}
\right] + \g^{\frac{\th}{2} h} \e_0 \left[(3/2) \bar c + 4\bar c b_3\e_0
\sum_{n=0}^\io n \g^{-\frac{\th}{2} n}\right] \esp
\ee
Hence, \pref{vdiff1} is verified also for $j=h$, if $\e_0$ is small
enough.\Halmos

The previous analysis implies that the flow is essentially trivial up to
values of $j$ of order $|g_{1,0}|^{-1/2}$ (or even $|g_{1,0}|^{-\h}$,
$0<\h<1$). If $j\le j_0$, we write \pref{2.53a} in the form
\be\lb{2.57b}
g_{1,j-1} = g_{1,j} - a_j\, g_{1,j}^2 \virg a_j\= a - \frac{\tilde r_{1,j} +
r_{1,j}+ \bar r_{1,j}}{g_{1,j}^2}
\ee
and we define $A_{j_0}=0$ and, for $j< j_0$,
\be A_j= \frac1{j_0-j} \sum_{j'=j+1}^{j_0} a_{j'}
\qquad
\lb{gtilde}
\tilde g_{1,j} = \frac{g_{1,j_0}}{1+A_j g_{1,j_0}(j_0-j)}
\ee

\begin{lemma}\lb{lm2.5}
There are constants $c_1, c_2, c_3$ such that, if $g_{1,0}\in D_{\e_0,\d}$ and
it $\e_0$ is small enough, then the following bounds are satisfied, for all
$j<j_0$.
\be\lb{bej} \e_j\le c_3\e_0\ee
\be\lb{vdiff}
|\vv_j -\vv_{j+1}| \le c_1 |g_{1,j+1}|^2
\ee
\be\lb{gerr}
|g_{1,j} - \tilde g_{1,j}| \le |\tilde g_{1,j}|^{3/2}
\ee
\be\lb{bAj} |a_j - a| \le c_2 |g_{1,j_0}|\ee
\end{lemma}

\0{\bf Proof} - We shall proceed by induction. By using \pref{2.57c},
\pref{vdiff1} and \pref{2.55}, we see that the bounds \pref{bej} and
\pref{vdiff} are satisfied for $j=j_0$, if $c_3\ge 2$, $c_1\ge 3a$ and $4\bar c
\e_0 \le a$. Moreover, $g_{1,j_0} = \tilde g_{1,j_0}$ and, by proceeding as in
the proof of Lemma \ref{lm2.4} and using \pref{2.57c}, it is easy to prove that
there is a constant $\bar c_2$, such that
$$|a_{j_0}-a| \le \bar c_2 |g_{1,j_0}|$$
Hence, all the bounds are verified (for $\e_0$ small enough) for $j=j_0$, if
$c_1\ge 3a$, $c_2\ge \bar c_2$ and $c_3\ge 2$. Suppose that they are verified
for $j_0 \ge j\ge h$.

The validity of \pref{gerr} for $j=h-1$ follows from Prop. \ref{propA2},
which only rests on the bound \pref{bAj} for $j\ge h$. On the other hand,
\pref{gerr} implies that, if $\e_0$ is small enough, $2^{-1} |\tilde g_{1,j}|
\le |g_{1,j}| \le 2 |\tilde g_{1,j}|$; hence, using \pref{gtilde}, we get,
for $j>h$
\be\lb{Rjh} \left| \frac{g_{1,j}}{g_{1,h}} \right| \le 4 \frac{|1+A_h
g_{1,j_0}(j_0-h)|}{|1+A_j g_{1,j_0}(j_0-j)|} \ee
Let us now define, as in App. \ref{appA}, $A_j=\a_j+i\b_j$, $\a_j = \Re A_j$, and suppose that
\be\lb{cond0} 2 c_2 \e_0 \le a/2 \ee
so that, by \pref{2.55}, $\a_j\ge a/2$, $|\b_j|\le 2 c_2 \e_0$, $|A_j|\le 3a/2$,
for $j>h$. By proceeding as in the proof of the bound \pref{AA4} in App.
\ref{appA}, we get, if $j>h$ and $|\text{Arg}\; g_{1,0}|\le \p-\d$, $\d>0$ (so
that $|\text{Arg}\; g_{1,j_0}|\le \p-\d/2$, see \pref{2.55}),
$$|1+g_{1,j_0} \a_j (j_0-j)| \ge \frac{1}{3} \sin (\d/2) [1+|g_{1,j_0}| \a_j (j_0-j)]$$
and, if we put $1+A_j g_{1,j_0}(j_0-j) = 1+\a_j g_{1,j_0}(j_0-j) + w_j$, we choose $\e_0$ so that
\be\lb{cond01}\frac{|w_j|}{|1+g_{1,j_0} \a_j (j_0-j)|} \le \frac{6 c_2\e_0
|g_{1,j_0}|(j_0-j)}{\sin (\d/2) |g_{1,j_0}| (a/2) (j_0-j)} = \frac{12
c_2\e_0}{a \sin (\d/2)} \le \frac12\ee
Then, by using \pref{Rjh}, we get
\be\lb{2.56a} \left| \frac{g_{1,j}}{g_{1,h}} \right| \le
\frac{24}{\sin(\d/2)} \frac{1+ (3a/2) |g_{1,j_0}|(j_0-h)|}{1+(a/2)
|g_{1,j_0}|(j_0-j)|} \le C_\d (j-h) \ee
for some constant $C_\d$, only depending on $\d$ and $a$.
Moreover, since $\e_h \le c_3\e_0$, then $\bar c\,\e_0 + b_2 \e_h^2 \le 2\bar
c\,\e_0$ and $a+b_1\e_j \le 2a$, if
\be\lb{cond1} b_2 c_3^2 \e_0 \le \bar c\virg \hbox{and\ } b_1 c_3 \e_0 \le a\ee
Hence, by using the bounds \pref{vdiff0}, \pref{vdiff}, \pref{2.57c} and
\pref{2.56a}, we get
\be\nn
\bsp
|\vv_{h-1} -\vv_h| &\le 2a |g_{1,h}|^2 + 2\bar c\,\e_0 \g^{-\th(j_0- h)}
|g_{1,j_0}|^2 + c_1 b_3\e_h \sum_{j=h+1}^0 \g^{-\th(j-h)} (j-h)
\max_{h< j' \le j} |g_{1,j'}|^2\\
&\le |g_{1,h}|^2 \left[ 2a + 2\bar c\,\e_0 C_\d^2 \max_{n\ge 0} \g^{-n\th} n^2
+ c_1 \e_h b_3 C_\d^2 \sum_{n=0}^\io \g^{-\th n} n^3 \right]\esp
\ee
It follows that \pref{vdiff} is satisfied also for $j=h$, if
\be\lb{cond2}
2a + 2\bar c\,\e_0 C_\d^2 \max_{n\ge 0} \g^{-n\th} n^2 + 2 c_1 c_3 \e_0 b_3
C_\d^2 \sum_{n=0}^\io \g^{-\th n} n^3 \le c_1
\ee
Moreover, by using \pref{vdiff} and $|g_{1,j}| \le 2 |\tilde g_{1,j}|$, we get,
for some $b_4>0$, only depending on $a$, under the condition \pref{cond0}:
$$\e_{h-1} \le \e_0 + \sum_{j=h}^0 |\vv_{j-1} -\vv_j| \le \e_0 + b_4 c_1 \e_0$$
so that $\e_{h-1} \le c_3\e_0$, if
\be\lb{cond3} 1+b_4 c_1 \le c_3\ee
The bound for $a_{h-1}-a$ can be done in the same way; it is easy to see that
\be |a_{h-1}-a| \le \left[ b_1 c_3 + b_2 c_3^2 \e_0 C_\d^2 \max_{n\ge 0}
\g^{-n\th} n^2 + 2 c_1 c_3 b_3
C_\d^2 \sum_{n=0}^\io \g^{-\th n} n^3\rgt] \e_0
\ee
Hence, \pref{bAj} is verified for $j=h-1$, if
\be\lb{cond4}
\tilde c_2 \= 2\bar c\, C_\d^2 \max_{n\ge 0} \g^{-n\th} n^2 + 2 c_1 c_3 b_3
C_\d^2 \sum_{n=0}^\io \g^{-\th n} n^3 \le c_2
\ee
The conditions \pref{cond0}, \pref{cond01}, \pref{cond1}, \pref{cond2},
\pref{cond3} and \pref{cond4} can be all satisfied, by taking, for example,
$c_1=4a$, $c_3=1+4a b_4$ and $c_2= \max\{\bar c_2, \tilde c_2\}$, if $\e_0$
is small enough.\Halmos

Lemma \ref{lm2.5} implies that, if $g_{1,0}\in D_{\e,\d}$, with $\e$ small
enough (how small depending on $\d$), $g_{1,j}$ goes to $0$, as $j\to -\io$,
and $\sum_{j=h}^0 |g_{1,j}|^2 \le C\d^{-1}|\l|$, uniformly in $h$. This is an
easy consequence of the condition \pref{gerr} and the condition $\hat
v(2p_F)>0$; note that the power $3/2$ in the r.h.s. of \pref{gerr} could be
replaced $2-\h$, $\h>0$, but $2$ is not allowed.
The form of the flow \pref{floweq0} then implies also that $g_{2,j}$, $g_{4,j}$
and $\d_j$ converge, as $j\to -\io$, to some limits $g_{2,-\io}$, $g_{4,-\io}$
and $\d_{-\io}$ of order $\l$, such that
\bal
\lb{2.42} g_{2,-\io} &= g_{2,0}-{1\over 2}g_{1,0}+O(|\l|^{3/2})=
\left[2\hat v(0)-\hat v(2p_F) \right]\l+ O(|\l|^{3/2})\\
\lb{2.43} g_{4,-\io} &= g_{4,0} +O(\l^2) = 2\l\hat v(0)+O(\l^2) \cr \d_{-\io}
&= O(\l) \eal
Let us now suppose that $\l$ is a (small) positive number; the previous bounds imply
that $g_{1,j}>0$, for any $j\le 0$. The following Lemma will allow to control the
logarithmic corrections to the power law fall-off of the correlations.

\begin{lemma}\lb{lm2.6}
There are four sequences $w_{i,h}$, $\d_{i,h}$, $i=1,2$, $h\le j_0$, such
that
\be\lb{g1sum} \sum_{j=h}^{j_0} g_{1,j}= (1+ w_{1,h}) \frac1{a} \log [1+a
g_{1,j_0} (j_0-h)] + \d_{1,h} \ee
\be\lb{g2diff} \sum_{j=h}^{j_0} \big[ g_{2,j} - g_{2,-\io} \big] = (1+
w_{2,h}) \frac1{2a} \log [1+a g_{1,j_0} (j_0-h)] + \d_{2,h} \ee
with
\be\lb{wb} |w_{i,h}| \le C\l \virg |\d_{i,h}| \le C\l^{1/2}\ee
\be\lb{wb1} |w_{i,h-1} - w_{i,h}| \le \frac{C\l}{[1+a g_{1,j_0} (j_0-h)] \log
[1+a g_{1,j_0} (j_0-h)]}\ee
\end{lemma}

\0{\bf Proof} - Let us put $g_0=g_{1,j_0}$, and $a(s)$ the function of $s\ge
0$, such that $a(s)=a_{j_0-n}$, if $n\le s <n+1$. Then, by using
\pref{gtilde}, \pref{gerr} and \pref{bAj}, it is easy to see that
\be\lb{wb2} \lft| \sum_{j=h}^{j_0} g_{1,j} - I_{j_0-h} \rgt| \le C\l^{1/2}
\virg I_n= \int_0^n ds \frac{g_0}{1+ g_0 \int_0^s dt\, a(t)} \ee
On the other hand, \pref{bAj} also implies that $a(s)=a +\l r(s)$, with
$|r(s)|\le C$; hence
\be\nn I_n= \int_0^n ds \frac{g_0}{1+ g_0 a s} -\l \int_0^n ds \frac{g_0^2\,
\int_0^s dt\, r(t)}{[1+ g_0 \int_0^s dt\, a(t)] [1+ g_0 a s ]}\ee
implying that
\be\nn \lft| I_n - \frac1a \log(1+ a g_0 n)\rgt| \le \frac{4C\l}{a^2}
\int_0^{ag_0 n} dx \frac{x}{(1+x)^2} < \frac{4C\l}{a^2} \log(1+ a g_0 n) \ee
Hence there is a constant $\tilde w_n$ such that $I_n = (1/a + \tilde
w_n)\log(1+ a g_0 n)$, with $|\tilde w_n| \le C\l$; this bound, together with
the bound in \pref{wb2}, proves \pref{wb} for $i=1$. To prove \pref{wb1},
note that
\be\nn |I_{n+1} - I_n| \le \int_n^{n+1} ds \frac{g_0}{1+g_0 \frac{a}2 s} =
\frac2{a} \log \lft(1+ \frac{a g_0}{2+a g_0 n} \rgt)\ee
\be\nn I_{n+1} - I_n = (1/a + \tilde w_{n+1})\log \lft(1+ \frac{a g_0}{1+a
g_0 n} \rgt) + (\tilde w_{n+1} - \tilde w_n)\log(1+ a g_0 n)\ee
so that, if $\l$ is small enough,
$$|\tilde w_{n+1} - \tilde w_n| \log(1+ a g_0 n) \le \lft(\frac3{a} +
C\l\rgt) \log \lft(1+ \frac{a g_0}{1+a g_0 n} \rgt) \le \frac{4 g_0}{1+
a g_0 n}$$
To prove \pref{wb} and \pref{wb1} for $i=2$, note that, by \pref{2.46} and
Lemma \ref{lm2.5}, if $j\le j_0$,
\be\nn\bsp g_{2,j} - g_{2,-\io} &= \sum_{h=-\io}^{j} \lft[\frac{a}2 +
O(\l)\rgt] g_{1,h}^2 = \lft[\frac{a}2 + O(\l)\rgt] \int_{|j|}^\io ds
\frac{g_0^2}{(1+a g_0 s)^2} + O(\tilde g_{1,j}^{3/2})\\
&=\lft[\frac12 + O(\l)\rgt]  \tilde g_{1,j} + O(\tilde g_{1,j}^{3/2})\esp\ee
Hence, the proof of \pref{wb} is almost equal to the previous one, while the
proof of \pref{wb1} needs a slightly different algebra; we omit the
details.\Halmos

\subsection{The flow of renormalization constants}\lb{sec2.4}
The renormalization constant of the free measure satisfies
\bal
\lb{ffgz} {Z_{j-1}\over Z_j} &= 1+ \b_z^{(j)}(\vec
g_j,\d_j,...,\vec g_0,\d_0)+ \bar\b_z^{(j)}(\vec v_j;..,\vec v_0;\l)\;;
\eal
while the renormalization constants of the densities, for $\a=C,S_i,SC,TC_i$ and
$i=1,2$, satisfy the equations
\bal
\lb{ffg} {Z^{(i,\a)}_{j-1}\over Z^{(i,\a)}_j} &= 1+ \b_{(i,\a)}^{(j)}(\vec
g_j,\d_j,...,\vec g_0,\d_0)+ \bar\b_{(i,\a)}^{(j)}(\vec v_j;..,\vec v_0;\l)\;.
\eal
In these two formulas,
by definition, the $\b_t^{(j)}$ functions, with $t=z$ or $(i,\a)$, are
given by a sum of multiscale graphs, containing only vertices with r.c.c.  $\vec
g_h,\d_h$, $0\ge h \ge j$, modified so that the propagators $g^{(h)}_{\o}$ and the
renormalization constants $Z_h$, $Z^{(i,\a)}_h$, $0\ge h\ge j$, are replaced by
$g^{(h)}_{{\rm D},\o}$, $Z_h^{(D)}$, $Z^{(D,i,\a)}_h$ (the definition of
$Z^{(D,i, \a)}_h$ is analogue to the one of $Z_h^{(D)}$); the $\bar\b_t^{(j)}$
functions contain the correction terms together the remainder of the expansion.
Note that, by definition, the constants $Z_j^{(D)}$ are exactly those generated
by \pref{ffgz} and \pref{bal} with $\bar\b_z^{(j)}=\bar\b_\a^{(j)}=0$. Note
also that $|\bar \b_z^{(j)}|\le C \bar v_j^2 \g^{\th j}$, while
$|\bar\b_{(i,\a)}^{(j)}|\le C \bar v_j \g^{\th j}$.

By using \pref{ffgz} and \pref{ffg}, we can write
\be
{Z_{j-1}^{(1,\a)} \over Z_{j-1}}= {Z_{j}^{(1,\a)}\over Z_{j}} \lft[1+
\b_{z,(1,\a)}^{(j)}(\vec g_j,\d_j) + \hat\b_{z,(1,\a)}^{(j)}(\vec v_j;..,\vec
v_0;\l)\rgt]
\ee
with $|\hat\b_{z,(1,\a)}|\le C \bar v_j \g^{\th j}$. If we define $\tilde
\b_{z,(1,\a)}^{(j)}(\vec g,\d)$ the value of $\b_{z,(1,\a)}^{(j)}(\vec
g_j,\d_j;...;\vec g_0,\d_0)$ at $(\vec g_i,\d_i)=(\vec g,\d)$, $j\le i\le 0$
and $\tilde \b_{z,(1,\a)}^{(j,\le 1)}(\vec g,\d)$ the sum of its terms of order
$0$ and $1$ in $g_{1,h}$, it turns out that
\be
\lb{beta44}|\tilde \b_{1,\a}^{(1)(j)}(\vec g_j,\d_j)|\le C [\max \{|g_{1}|
|g_{2}|, |g_{4}|, |\d|\}]^2 \g^{\th h}\virg \mbox{if $\a=C$}
\ee
This bound, as crucial as the analogous bound \pref{beta23}, has been proved in
\cite{M007_3}; the proof will be sketched in App. \ref{appC}.
The bound \pref{beta44}, together with $\sum_{k=j}^0 |g_{1,k}|^2 \le C|\l|$ and
the fact that $Z^{(1,S_i)}_h=Z^{(1,C)}_h$ by the SU(2) spin symmetry, imply
that
\be\lb{2.48}
\left| \frac{Z_{j}^{(1,\a)}}{Z_j}-1 \right|\le C |\bar\e_j^2| \virg \a=C,S_i
\ee
Regarding the flow of the other renormalization constants, we can write
\be\lb{2.53} Z^{(t)}_j= \g^{-\h_t j} \hat Z^{(t)}_j \ee
where $Z^{(z)}_j=Z_j$ and, by definition,
\be\lb{fff}
\h_t\=\lim_{j\to -\io}\h_{t,j} :=
\log_\g \Big[ 1+ \b_t^{(0,j)}(g_{2,-\io},
g_{4,-\io},\d_{-\io};...;g_{2,-\io},g_{4,-\io} ,\d_{-\io}) \Big]
\ee
Note that the exponents $\h_t$ are functions of $\vec v_{-\io}$ only, an
observation which will play a crucial role in the following. Moreover, by an
explicit first order calculation, we see that
\be\lb{expt} \h_t=
\begin{cases}
(2\p v_F)^{-1} g_{2,-\io} + O(\l^2)& t=(2,C), (2,S_i)\\
-(2\p v_F)^{-1}  g_{2,-\io} + O(\l^2)& t=(2,SC), (2,TC_i)\\
\qquad O(\l^2) & \text{otherwise}
\end{cases}
\ee
while
\bal {\hat Z^{(t)}_{h-1}\over \hat Z^{(t)}_h} &= 1 + O(\tilde g_{1,h}\l) +
r^{(t)}_h \virg t=z,(1,\a), \a\not=TC_i\nn\\
{\hat Z^{(2,C)}_{h-1}\over \hat Z^{(2,C)}_h} &=1- a g_{1,h} + \frac{a}2
(g_{2,h}-g_{2,-\io})+O(\tilde g_{1,h}\l) + r^{(2,C)}_h\nn\\
\lb{2.56} {\hat Z^{(2,S_i)}_{h-1}\over \hat Z^{(2,S_i)}_h} &=1 +
\frac{a}2 (g_{2,h}-g_{2,-\io}) + O(\tilde g_{1,h}\l) + r^{(2,S_i)}_h\\
{\hat Z^{(2,SC)}_{h-1}\over \hat Z^{(2,SC)}_h} &=1 - \frac{a}2 g_{1,h} -
\frac{a}2 (g_{2,h}-g_{2,-\io}) + O(\tilde g_{1,h}\l) + r^{(2,SC)}_h \nn\\
{\hat Z^{(2,TC_i)}_{h-1}\over \hat Z^{(2,TC_i)}_h} &=1 +  \frac{a}2 g_{1,h} -
\frac{a}2 (g_{2,h}-g_{2,-\io}) +O(\tilde g_{1,h}\l) + r^{(2,TC_i)}_h\nn
\eal
where $a$ and $\tilde g_{1,h}$ are defined as in \pref{defa} and
\pref{gtilde}, respectively, and $\sum_{h=-\io}^0 |r^{(t)}_h| \le C|\l|^2$.
Let us define:
\be\lb{2.88a} q^{(h)}_t = {\log \hat Z^{(t)}_h\over \log (1 + a g_{1,0}|h|)}
\ee
Hence, by using \pref{g1sum}, \pref{g2diff} and \pref{wb}, we get
\be\lb{2.57} \bsp |q^{(h)}_t| \le C\l\virg
& t=z,(1,\a), \a\not=TC_i\\
|q^{(h)}_t -\frac12 \bar\z_\a| \le C\l\virg & t=(2,\a) \esp \ee
where the constants $\bar\z_\a$ are those of \pref{zalfa}.

\subsection{Proof of Theorem \ref{th1.1}}\lb{sec2.5}

In order to prove the representation \pref{asymp} of the density correlations
$\O_\a(\xx)$, we can proceed exactly as in \S 5 of \cite{BM001}, where a
similar problem was treated in all details; hence we shall only describe the
result. We can write for $\O_\a(\xx-\yy)$ a convergent tree expansion, whose trees
have two endpoints associated to density operators
({\em special endpoints}) and
an arbitrary number of interaction endpoints ({\em normal endpoints}). As one
could expect, $\O_\a(\xx-\yy)$ behaves, as $|\xx-\yy| \to \io$, as the function
$\tilde\O_\a(\xx-\yy)$ calculated by taking only the trees with two special
endpoints of scale $h\le 0$ and no normal endpoints. This function can be
obtained by the following procedure. Let us consider the expression
$$
\la O^{(1,\a)}_\xx;  O^{(1,\a)}_\yy\ra^T_{{\rm D}}
+ \la O^{(2,\a)}_\xx ; O^{(2,\a)}_\yy\ra ^T_{{\rm D}}
$$
where the operators $O^{(i,\a)}_\xx$, $i=1,2$, are those defined in \pref{op1}
and \pref{op2} and $\la \cdot\ra^T_{{\rm D}}$
is the truncated expectation evaluated with
covariance $\sum_h Z_h^{-1} g^{(h)}_{{\rm D},\o}$. Hence, this
expression can be written as a sum of terms, each one proportional to $Z_h^{-1}
g^{(h)}_{{\rm D},\o}(\xx-\yy)$ $Z_{h'}^{-1} g^{(h')}_{{\rm D},\o'}(\xx-\yy)$, for some
values of $h$, $h'$, $\o$ and $\o'$. $\tilde\O_\a(\xx-\yy)$ is obtained by
multiplying each one of these terms by $[Z^{(1,\a)}_{h\vee h'}]^2$ ($h\vee h'
=\max\{h,h'\}$), if it appears in the calculation of
$\la O^{(1,\a)}_\xx O^{1,\a}_\yy\ra^T_{{\rm D}}$,
otherwise by $[Z^{(2,\a)}_{h\vee h'}]^2$. Let us
consider first the case $\a=C$; we have
\bal
\tilde\O_C(\xx) &= \O^{(1,C)}(\xx) + \cos(2p_F x) \O^{(2,C)}(\xx)\nn\\
\O^{(1,C)}(\xx) &= 2\sum_\o \sum_{h,h'} \frac{[Z^{(1,C)}_{h\vee h'}]^2}{Z_h
Z_{h'}}
g^{(h)}_{{\rm D},\o}(\xx) g^{(h')}_{{\rm D},\o}(\xx)\\
\O^{(2,C)}(\xx) &= 4 \sum_{h,h'} \frac{[Z^{(2,C)}_{h\vee h'}]^2}{Z_h Z_{h'}}
g^{(h)}_{{\rm D},+}(\xx) g^{(h')}_{{\rm D},-}(\xx)\nn \eal
Let us now observe that, for any $N>0$, $|g^{(h)}_{{\rm D},\o}(\xx)| \le C_N \g^h
[1+ (\g^h |\xx|)^N]^{-1}$. Hence, if $|\xx|\ge 1$, in the previous sums the
main contribution is given by the terms with $|h|$ and $|h'|$ of the same size
as $\log_\g |\xx|$, so that one expects that the asymptotic behavior of
$\O^{(i,C)}(\xx)$, $i=1,2$, is the same of the the function
$\bar\O^{(i,C)}(\xx)$, obtained by the substitutions of $\g^{-h}$ and
$\g^{-h'}$ with $|\xx|$ in the asymptotic expressions of the renormalization
constants, given by \pref{2.53} and \pref{2.56}, that is
\be\lb{substZ} \frac{[Z^{(i,C)}_{h\vee h'}]^2}{Z_h Z_{h'}} \rightarrow
|\xx|^{2(\h_{i,C}-\h_z)} \Big[ 1+f(\l) \log |\xx| \Big]^{2\lft(
q^{(h_\xx)}_{i,C}-q^{(h_\xx)}_z\rgt)} \ee
where the coefficients $q^{(h)}_t$ are defined as in \pref{2.88a}, $h_\xx=
\inf \{h:\g^h |\xx|\ge 1\}$, and, by \pref{init}, \pref{defa}, \pref{gtilde}
and Lemma \ref{lm2.6},
\be\lb{2.57a} f(\l)=\frac{a g_{1,j_0}}{\log\g} = \frac{2\l \hat v(2p_F)}{\p
v_F} + O(\l^{3/2}) \ee
In order to justify the substitution \pref{substZ}, let us put
$\h_i=2(\h_{i,C}-\h_z)$ and $q_i(\xx)$ any continuous interpolation between
$2[ q^{(h_\xx)}_{i,C}-q^{(h_\xx)}_z]$ and $2[ q^{(h_\xx-1)}_{i,C}
-q^{(h_\xx-1)}_z]$. Note that, thanks to the bounds \pref{wb} and \pref{wb1},
$q_i(\xx)$ is a bounded function of order $\l$, defined up to fluctuations
bounded, for $|\xx|\ge 1$, by $C\l [L(\xx) \log L(\xx)]^{-1}$, with $L(\xx)=
1+f(\l) \log |\xx|$; hence, its precise definition modifies the following
expressions only for a factor $1+O(\l)$. Let us now note that
\be\lb{2.61} \bsp &|\O^{(i,C)}(\xx) - \bar\O^{(i,C)}(\xx)| \le C_N
|\xx|^{\h_i-2} [1+f(\l) \log |\xx|]^{q_i(\xx)} \sum_{h,h'} \frac{\g^h |\xx|}{
1+ (\g^h |\xx|)^N} \frac{\g^{h'}
|\xx|}{ 1+ (\g^{h'} |\xx|)^N}\;\cdot\\
&\cdot\; \left| \frac{(\g^h |\xx|)^{\h_z} (\g^{h'} |\xx|)^{\h_z}} {(\g^{h\vee
h'}|\xx|)^{2\h_{i,C}}} \left[ \frac{L(|\xx|)}{L(\g^{|h|})}
\right]^{q^{(h)}_z} \left[ \frac{L(|\xx|)}{L(\g^{|h'|})} \right]^{q^{(h')}_z}
\left[ \frac{L(|\xx|)}{L(\g^{|h\vee h'|})} \right]^{-2q^{(h\vee h')}_{i,C}}
\frac{c_h c_{h'}}{\tilde c_{h\vee h'}^2} -1\right| \esp \ee
where
$$L(t)=1+f(\l)\log t \virg c_h = L(\g^{|h|})^{q^{(h)}_z}/\hat Z_h^{(z)} \virg \tilde
c_h = L(\g^{|h|})^{-q^{(h)}_{i,C}}/\hat Z_h^{(i,C)}$$
By \pref{2.56}, \pref{g1sum} and \pref{g2diff}, $c_h = 1 + O(\l^{1/2})$ and
$\tilde c_h =1 + O(\l^{1/2})$. On the other hand, if $r>0$ and $t\not=0$,
$$|r^t -1| \le |t \log r| (r^t + r^{-t})$$
and, if $q\not=0$,
$$\left| \left[ \frac{L(|\xx|)}{L(\g^{|h|})} \right]^{q} -1\right|
\le C_q \left[ |f(\l)\log(\g^h|\xx|)| + |f(\l)\log(\g^h|\xx|)|^{|q|+1}\right]$$
These two bounds, together with the bound
$$\sum_{h=-\io}^0 \frac{(\g^h r)^\a |\log(\g^h r)|^\b}{1+(\g^h r)^N} \le
C_{N,\a,q}$$
valid for any $\b$, $r>0$, $a>0$ and $N>\a$, imply that
\be |\O^{(i,C)}(\xx) - \bar\O^{(i,C)}(\xx)| \le C_N \l^{1/2} |\xx|^{\h_i-2}
[1+f(\l) \log |\xx|]^{q_i(\xx)} \ee
By the remark after \pref{gtilde}, the factor $\l^{1/2}$ can be improved up to
$\l^{1-\th}$, $\th<1$.

By proceeding as in \S5 of \cite{BM001}, it is possible to prove that a bound
of this type is satisfied also from the sum over all the other trees. Hence,
in order to complete the proof of \pref{asymp} in the case $\a=C$, we have
only to calculate $\bar\O^{(1,C)}(\xx)$ and $\bar\O^{(2,C)}(\xx)$. By using
\pref{2.48}, we see that $\h_{1,C}=\h_z$ and $q^{(h)}_{1,C}=q^{(h)}_z$, so
that, if we define $X_C=1-\h_{2,C}-\h_z$ and $\z_C(\xx)= 2[q_{2,C}(\xx) -
q_z(\xx)]$, we get
\be\bsp
\bar\O^{(1,C)}(\xx) &= 2\sum_\o g_{{\rm D},\o}(\xx) g_{{\rm D},\o}(\xx)\\
\bar\O^{(2,C)}(\xx) &= 4 |\xx|^{2(1-X_C)} [1+f(\l) \log |\xx|]^{\z_C(\xx)}
g_{{\rm D},+}(\xx) g_{{\rm D},-}(\xx) \esp \ee
where $g_{{\rm D},\o}(\xx) = \sum_{h=-\io}^0 g^{(h)}_{{\rm D},\o}(\xx)$. On
the other hand, it is easy to see that, for any $N\ge 2$,
$$g_{{\rm D},\o}(\xx) = \frac{1}{2\p} \frac1{v_F x_0 + i\o x} + O(|\xx|^{-N})$$
It follows that, up to terms that we put in the ``remainder'' $\hat R_C(\xx)$,
\be \bar\O^{(1,C)}(\xx) = \frac1{\p^2 \tilde\xx^2} \bar\O_0(\xx) \virg
\bar\O^{(2,C)}(\xx) = \frac{L(\xx)^{\z_C(\xx)}}{\p^2 |\tilde\xx|^{2X_C}} \ee
where the functions $\bar\O_0(\xx)$ and $L(\xx)$ are defined as in Theorem
\ref{th1.1}. Hence, by using \pref{expt} and \pref{2.57a}, we get
\pref{asymp} for $\a=C$, together with the fact that $\z_C(\xx)=-3/2 +
O(\l)$, in agreement with \pref{zalfa}, and $2X_C=2-b\l +O(\l^2)$, in
agreement with \pref{1.8}. Note also that, in Theorem \ref{th1.1}, we have
modified the function $f(\l)$ by erasing the terms of order greater than $1$
in $\l$; the only effect of this modification is a change of the function $R_C(\xx)$,
which does not change its bound.

The proof of \pref{asymp} in the other cases is done in the same way. In
particular, in the case $\a=S_i$ we have to use again the bound \pref{2.48},
while the fact that there is no oscillating contribution to the leading term of
$\O_{TC_i}$ is due to the fact there is no local marginal term which can
produce it, by the remark after \S\ref{op2}.

Finally, the proof of the scaling relations \pref{1.8} follows from the
important fact, proved in \S\ref{ss2.5d}, that they are the same as those of
the effective model. Hence they follow from the explicit calculations of \S
\ref{ss4.4}.


%
%
%
%


\subsection{Proof of Theorem \ref{th1.4}}\lb{sec2.6}
First consider  the free energy \pref{fe},
%
%
We can decompose it as $E(\l)= \sum_{h=-\io}^0 E_h(\l)$, where $E_h(\l)$ is the
contribution of the trees whose root has scale $h$, hence depends only on the
running couplings $\vec v_j$ with scale $j> h$. The tree expansion implies that
there exists $\e_0$, such that, if
\be\lb{cond00} \bar\l_h = \max_{j\ge h}|\vec v_j|\le \e_0 \ee
then $|E_h| \le c_2 \g^{2h} \e_0$, with $c_2$ independent of $h$. The
analysis of \S\ref{sec2.2} implies that, given $\d\in(0,\p/2)$, there exists
$\e$ such that, if $\l\in D_{\e,\d}$, the condition \pref{cond00} is verified
uniformly in $h$; then it is easy to see that $E(\l)$ is analytic in $
D_{\e,\d}$ and continuous in its closure. The domain of analyticity of
$E_h(\l)$ is in fact larger; the form of the beta function immediately
implies that there exist two constants $c_3$ and $\bar c$ such that $\bar\l_0
\le c_3|\l|$ and, if $\bar\l_h\le \e_0$, then $\bar\l_{h-1} \le \bar\l_h +
\bar c \bar\l_h^2$; hence, if $c_3|\l| \le \min \{\e_0/2, 1/[4 \bar c
(|h|+1)]\}$, then, if $j>h$ and $\bar\l_j \le 2\bar\l_0$,
$$\bar\l_{j-1} \le \bar\l_0 + |j| \bar c \bar\l_j^2 \le \bar\l_0(1+ 4|j|\bar c \bar\l_0)
\le 2\bar\l_0$$
It follows that $E_h(\l)$ is analytic in the set \pref{domh}, with
$c_0=c_3^{-1} \min\{\e_0/2, 1/(4\bar c)\}$, and that $|E_h(\l)| \le c_1 \g^{2
h}$, with $c_1=c_2 \e_0$; hence $E(\l)$ satisfies \pref{Lesn} with
$\k=2 \log \g$.

Let us now consider the 2-point Schwinger function $S_2(\xx)$. By using the
tree expansion (similar to that written in \cite{BGPS994} for the infrared part
of the spinless continuous Fermi gas), we can write $S_2(\xx) = \sum_{h=-\io}^0
S_{2,h}(\xx)$, where $S_{2,h}(\xx)$ is the contribution of the trees whose root
has scale $h$. By proceeding as in \S6 of \cite{BGPS994}, we can prove that, if
\pref{cond00} is verified (possibly with a smaller $\e_0$), then, for any $N>0$,
\be
|S_{2,h}(\xx)| \le c_N\sum_{\bar h=h+1}^0 \g^{-\frac{\bar h-h}{2}}
\frac{\g^{\bar h}}{Z_{\bar h}} \frac1{1+(\g^{\bar h} |\xx|)^N}\virg
\lft|\frac1{Z_h} \rgt| \le \g^{|h|\over 4}
\ee
with $c_N$ independent of $h$. Hence, if we define $h_{\xx}$ so
that $\g^{h_\xx}|\xx|\in [1,\g)$, then, if $h_\xx>h$
\be |S_{2,h}(\xx)| \le c_2 \lft[ \sum_{\bar h=h+1}^{h_\xx} \g^{-{\bar
h\over 2}}\g^{{3\over 4}\bar h}\g^{h\over 2}+\sum_{\bar
h=h_{\xx}}^{0} \g^{-{\bar h\over 2}}\g^{{3\over 4}\bar
h}\g^{h\over 2}\g^{2(h_\xx-\bar h)} \rgt]\le \tilde c_2
\, \g^{\frac{h}2}\g^{\frac{h_\xx}4} \ee
and a similar bound holds for $h_\xx<h$ so that
\be |S_{2,h}(\xx)| \le c_s \g^{\frac{h}2} (1+|\xx|)^{-1/4} \ee
and we can proceed as in free energy case, so proving \pref{Lesn} for
$S_2(\xx)$, with $c_1=c_s (1+|\xx|)^{-1/4}$ and
$\frac{\k}{\log \g}=\frac12$ (this value could be
improved up to any value smaller than $1$, at the price of lowering $\e_0$ down
to $0$).

The previous argument can be extended to the generic $2n$-point Schwinger
function, by using the same strategy used in \S2.3 of \cite{BFM007} to analyze
the corresponding tree expansion in the case of the Thirring model. The proof
of the Theorem in the case of the density correlations is very similar to that
of the 2-point function case, if one uses the description of the tree expansion
given in \S5 of \cite{BM001}. We shall not give any further detail; the idea at
the base of the proof is always that, in the tree expansion of the correlations
at fixed space-time points, the contribution of the trees with the root at
scale $h$ must decrease exponentially with the distance from some fixed scale
depending on the space-time points.

\section{RG Analysis of the  Effective Model} \lb{ss2.5b}

\subsection{Introduction} \lb{ss2.5c}

We introduce an {\it effective model} (related to the g-ology model introduced
in \cite{So079} but differing because the interaction is
time-dependent),
describing fermions with linear dispersion relation and a non-local
interaction; such model will be used to prove the crucial bounds \pref{beta23}
and \pref{beta44} (on which the previous analysis is based) and for the proof
of the Luttinger liquid relations \pref{1.8}, together with Theorems
\ref{th1.2} and \ref{th1.3}.
The model is expressed in terms of the following Grassmann integral:
\be\lb{vv1}
\bsp
e^{\WW_{[l,N]}(\phi,J)} &= \int\! P_Z(d\psi^{[l,N]}) \exp \left\{ -\tilde
V(\sqrt{Z}\psi^{[l,N]}) + \sum_{\o,s} \int d\xx J_{\xx,\o,s}
\psi^{[l,N]+}_{\xx,\o,s} \psi^{[l,N]-}_{\xx,\o,s}\right.\\
& + \left. \sum_{\o,s} \int d\xx [\psi^{[l,N]+}_{\xx,\o,s} \phi^-_{\xx,\o,s} +
\phi^+_{\xx,\o,s} \psi^{[l,N]-}_{\xx,\o,s}]\right\}\;, \esp
\ee
where $\xx\in\tilde\L$ and $\tilde\L$ is a square subset of $\RRR^2$ of size
$\g^{-l}$, say $\g^{-l}/2 \le |\tilde\L|\le \g^{-l}$, $P_Z(d\psi^{l, N})$ is
the fermionic measure with propagator
\be\lb{gth} g^{[l,N]}_{{\rm D},\o}(\xx-\yy)={1\over Z}{1\over
L^2}\sum_{\kk}e^{i\kk(\xx-\yy)}{\chi^{[l,N]}(|\tilde \kk|)\over -ik_0+\o c k}
\virg \tilde\kk = (ck, k_0)\virg c=v_F(1+\d)
\ee
where $Z>0$ and $\d$ are two parameters, to be fixed later, $v_F$ is defined as
in Theorem \ref{th1.1} and $\chi_{l,N}(t)$ is a cut-off function depending on a
small positive parameter $\e$, nonvanishing for all $\kk$ and reducing, as
$\e\to 0$, to a compact support function equal to $1$ for $\g^{l}\le |\kk|\le
\g^{N+1}$ and vanishing for $|\kk|\le \g^{l-1}$ or $|\kk|\ge \g^{N+1}$ (its
precise definition can be found in (21) \cite{BM002}); $\g^l$ is the {\em
infrared cut-off} and $\g^N$ is the {\em ultraviolet cut-off}. The limit
$N\to\io$, followed from the limit $l\to-\io$, will be called the {\it limit of
removed cut-offs}. The interaction is
\be\lb{fs1} \tilde V(\ps)=g_{1,\perp}
V_{1,\perp}(\ps)+ g_\arp V_\arp(\ps) + g_\erp V_\erp(\ps) + g_4
V_4(\ps) \ee
with
\be\lb{five1}
\bsp
V_{1,\perp}(\ps) &= \frac12 \sum_{\o,s}\int d\xx d\yy h(\xx-\yy)
\ps^+_{\xx,\o,s}\ps^-_{\xx,\o,-s} \ps^-_{\yy,-\o,s} \ps^+_{\yy,-\o,-s}\\
V_\arp(\ps) &= \frac12 \sum_{\o,s}\int d\xx d\yy h(\xx-\yy)
\ps^+_{\xx,\o,s}\ps^-_{\xx,\o,s}\ps^+_{\yy,-\o,s}\ps^-_{\yy,-\o,s}\\
V_\erp(\ps) &= \frac12 \sum_{\o,s}\int d\xx d\yy h(\xx-\yy)
\ps^+_{\xx,\o,s}\ps^-_{\xx,\o,s}\ps^+_{\yy,-\o,-s}\ps^-_{\yy,-\o,-s}\\
%
%
V_4(\ps) &= \frac12 \sum_{\o,s}\int d\xx d\yy h(\xx-\yy)
\ps^+_{\xx,\o,s}\ps^-_{\xx,\o,s}\ps^+_{\yy,\o,-s}\ps^-_{\yy,\o,-s}
\esp
\ee
where $h(\xx-\yy)$ is a rotational invariant potential, of the form
\be h(\xx-\yy)={1\over L^2}\sum_{\pp} \hat h(\pp)
e^{i\pp(\xx-\yy)}\;, \ee
with $|\hat h(\pp)|\le C e^{-\m |\pp|}$, for some constants $C$,
$\m$, and $\hat h(0)=1$.

The model with $g_{1,\perp}=0$ is invariant under the global phase
transformation $\psi^\pm_{\xx,\o,s} \to e^{\pm i\a_{\o,s}}
\psi^\pm_{\xx,\o,s}$, with the constant phase $\a_{\o,s}$
depending both on $\o$ and $s$. However, if $g_{1,\perp}\not=0$,
the model is only invariant under the transformation
$\psi^\pm_{\xx,\o,s} \to e^{\pm i\a_{\o}} \psi^\pm_{\xx,\o,s}$
with the phase independent of $s$.

The removal of the ultraviolet cut-off is controlled by an easy extension of
the analysis given in \S 2 of \cite{M007_2} or in \S 3 of \cite{BFM009} for a
spinless model with interaction $\l V_{\arp}(\psi)$; the presence of the other
terms (including the $g_3$-interaction considered in App. \ref{appB}) produces
more lengthy expressions but introduces no extra difficulties. The crucial idea
is to use an improvement respect to the power counting bounds due to the
non-locality of the interaction, and to use that the "fermionic boubble" (see
(2.39) of \cite{M007_2} or (3.17) of \cite{BFM010}) is exactly vanishing.

Regarding the removal of the infrared cut-off, we perform a multiscale analysis
very similar to the one given in \S\ref{sec2}, that we shall sketch in App.
\ref{appB} and App. \ref{appC} (we will refer to \S 4 of \cite{M005} for more
details). It turns out that the infrared cut-off cannot be removed by this
technique for all values of the couplings, but we are able to consider only two
situations leading to a bounded flow:
\begin{enumerate}
\item the case $g_{1,\perp}=0$
\item the case $g_\arp=g_{\erp}-g_{1,\perp}$ and $g_{1,\perp}>0$
\end{enumerate}
In the first case, when the ultraviolet and infrared cut-offs are
removed, the model becomes exactly solvable, a property related to
the invariance under the {\em local} phase transformation
$\psi^\pm_{\xx,\o,s} \to e^{\pm i\a_{\xx,\o,s}}
\psi^\pm_{\xx,\o,s}$. Indeed, as we will see in
\S\ref{sec4.1}--\S\ref{ss4.4}, the functional integrals generating
the correlations can be exactly computed, up to corrections which
are proved to be vanishing in the removed cutoffs limit. This will
allow us to prove \pref{beta23} and that the exponents of this
model and the Hubbard model (also analyzed via functional
integrals) are the same, if the parameters of the effective model
are suitable chosen (see \S\ref{ss2.5d} below), so that the
universal relations \pref{1.8} follow. In the second case the
model is not solvable, but still some correlations can be exactly
computed, see \S\ref{ss2.5e}, and this, again via a fine tuning of
the effective model parameters, allows us to prove the relation
\pref{hh1}.

\subsection{Ward Identities in the $g_{1,\perp}=0$ case}\lb{sec4.1}

In the $g_{1,\perp}=0$ case we can derive a set of Ward Identities
(WI). If we make in the generating functional \pref{vv1} the gauge
transformation $\psi^\e_{\xx,\o,s} \to e^{i\a_{\xx,\o,s}}
\psi^\e_{\xx,\o,s}$, we obtain, in the limit of removed cutoffs,
the {\em functional Ward identity} (WI):
\be\lb{grez}
\bsp
D_{\m}(\pp){\partial \WW(J,\h)\over \partial \hJ_{\pp,\m,s}}
&-D_{-\m}(\pp)\sum_{\s,r}\n^{\m\s}_{sr}(\pp) {\partial \WW(J,\h)\over\partial
\hJ_{\pp,-\s,r}}= \\
&-D_{-\m}(\pp){\hJ_{-\pp,\m,s}\over 4\p Z^2 c} + B_{\pp,\m,s}(J,\h) \esp
\ee
where $D_\m(\pp) = -ip_0 +c\o p$,
$$
B_{\pp,\m,s}(J,\h)= \frac1Z \int\!{d\kk\over (2\p)^2}
 \lft[\hh^+_{\kk+\pp,\m,s}
 {\partial \WW\over \partial \hh^+_{\kk,\m,s}}-
 {\partial \WW\over \partial \hh^-_{\kk+\pp,\m,s}}
 \hh^-_{\kk,\m,s}\rgt]
$$
and
\be\lb{4.2}
\n^{\o}_{s}(\pp) = \lft[\d_{\o,1}\lft(\d_{s,-1}g_\erp + \d_{s,1}g_\arp\rgt)+
\d_{\o,-1}\d_{s,-1}g_4\rgt] \frac{\hat h(\pp)}{4\p c}
\ee
Summing \pref{grez} over $s$ we obtain the {\it charge Ward identity}:
\be\lb{char}
\bsp
\Big[D_{\m}(\pp) &-\n_4(\pp) D_{-\m}(\pp)\Big] \sum_s{\partial \WW\over
\partial\hJ_{\pp,\m,s}} -2\n_\r(\pp) D_{-\m}(\pp) \sum_s{\partial \WW\over
\partial\hJ_{\pp,-\m,s}}\\
&=\sum_s\lft[-D_{-\m}(\pp){\hJ_{\pp,\m,s}\over 4\p Z^2 c}
+B_{\pp,\m,s}(J,\h)\rgt] \esp
\ee
with
$$\n_4(\pp)= g_4 \hat h(\pp)/(4\p c), \quad 2\n_\r(\pp)=
(g_\arp + g_\erp) \hat h(\pp)/(4\p c)$$
Multiplying \pref{grez} by $s$ and summing over $s$ we obtain the {\it spin
Ward identity}:
\be\lb{spin}
\bsp
\Big[D_{\m}(\pp) &+\n_4(\pp) D_{-\m}(\pp)\Big] \sum_ss{\partial \WW\over
\partial\hJ_{\pp,\m,s}} -2\n_\s(\pp) D_{-\m}(\pp) \sum_ss{\partial \WW\over
\partial\hJ_{\pp,-\m,s}}\\
&=\sum_s s\lft[-D_{-\m}(\pp){\hJ_{\pp,\m,s}\over 4\p Z^2 c}
+B_{\pp,\m,s}(J,\h)\rgt] \esp
\ee
with
\be 2\n_\s(\pp)= (g_\arp - g_\erp) \hat h(\pp)/(4\p c)\ee

The proof of \pref{char} and \pref{spin} with $J=0$ is essentially identical
the one in \S 2 of \cite{M007_2} or in \S 3 of \cite{BFM009}, while the
presence of the term linear in $J$ in the r.h.s. is explained in
Sec. IV B of \cite{Fa10} (see also App. A of
\cite{BM010}); hence, we will omit here the details.

Note the presence, in the WI \pref{char} and \pref{spin}, of the $\n_\s$,
$\n_\r$, $\n_4$ terms, which are called {\it anomalies}; they appear as a
consequence of the breaking of local symmetries in the functional integral
\pref{vv1}, due the momentum cut-off $\chi_{l,N}(\kk)$ in the fermionic
integration. This symmetry breaking produces extra terms in the WI which do not
vanish when the cut-offs are removed. Note also that such anomalies are linear
in the coupling. Such a property is called {\it anomaly non renormalization},
and is crucially related to the non locality of the interaction \cite{M007_2};
in presence of local interactions, like in the massless Thirring model, it can
be violated \cite{BFM007}. Another important point to be stressed is that
\pref{char} is true also when $g_{1,\perp}>0$, while \pref{spin} is not; this
remark will be used in the proof of Theorem \ref{th1.2}.

By some other simple algebra we obtain from \pref{MM1},\pref{MM2}
the identity
\be\lb{wis} {\dpr \WW\over \dpr\hJ_{\pp,\m',s'}} =\sum_{\m,s}
{M^\r_{\m',\m}(\pp)+s's M^\s_{\m',\m}(\pp)\over 2}
\lft[-D_{-\m}(\pp){\hJ_{-\pp,\m,s}\over 4\p Z^2 c} +B_{\pp,\m,s}(J,\h)\rgt]
\ee
where, if $\g=\r,\s$, and setting $\n_{4,\r}=-\n_{4,\s}=\n_4$,
\be\lb{MM}
\bsp M^\g_{\m',\m}(\pp)
&={\Big[D_{-\m}(\pp)-\n_{4,\g}(\pp)D_\m(\pp)\Big]\d_{\m',\m}+
\Big[2\n_{\g}(\pp) D_{\m}(\pp)\Big]\d_{\m',-\m}\over
\Big[D_+(\pp)-\n_{4,\g}D_-(\pp)\Big]\Big[D_-(\pp)-\n_{4,\g}(\pp)D_+(\pp)\Big]
-4\n_{\g}^2(\pp) D_+(\pp)D_-(\pp)}\\
&={u_{\g,+}(\pp)\d_{\m',\m}+w_{\g,+}(\pp)\d_{\m',-\m} \over
-iv_{\g,+}(\pp)\big(p_0+i\m v_{\g}(\pp)c p_1\big)} +
{u_{\g,-}(\pp)\d_{\m',\m}+w_{\g,-}(\pp)\d_{\m',-\m} \over
-iv_{\g,+}(\pp)\big(p_0-i\m v_{\g}(\pp)c p_1\big)} \esp
\ee
for
\be\lb{MM1}
\bsp
u_{\g,\m}(\pp) &={1\over 2}\lft[{1-\n_{4,\g}(\pp)\over v_{\g,+}(\pp)}
+\m{1+\n_{4,\g}(\pp)\over v_{\g,-}(\pp)}\rgt]\\
w_{\g,\m}(\pp) &=\n_\g(\pp)\lft[{1\over v_{\g,+}(\pp)} -\m{1\over
v_{\g,-}(\pp)}\rgt] \esp
\ee
\be\lb{MM2} v^2_{\g,\m}(\pp)=\Big(1-\m\n_{4,\g}(\pp)\Big)^2
-4\n_\g^2(\pp)\virg v_\g(\pp) =v_{\g,-}(\pp)/v_{\g,+}(\pp)
\ee

By doing suitable functional derivatives of the functional WI \pref{char}, we
can get many WI between the correlation functions. For example, if
we make two derivatives w.r.t $\hh^+_{\pp+\kk,\o,s}$ and
$\hh^-_{\kk,\o,s}$ in both sides of \pref{char}, we sum over $\m$ and we put
$\h=J=0$, we get the following {\em charge vertex Ward
identity}:
\be\lb{chara}
\bsp
\hspace{-.3cm} -ip_0 \Big[1-2\n_\r(\pp)-\n_4(\pp)\Big]
&G_{\r;\o,s}(\pp;\pp+\kk)+ cp_1
\Big[1-2\n_\r(\pp)+\n_4(\pp)\Big] G_{j;\o,s}(\pp;\pp+\kk)\\
&=\frac1Z \lft[G_{2;\o,s}(\kk)-G_{2;\o,s}(\pp+\kk)\rgt] \esp
\ee
where
\bal
G_{\r;\o,s}(\pp;\pp+\kk) &= \sum_{\m,t}
{\dpr^3 \WW\over \dpr\hJ_{\pp,\m,t}
\dpr\hh^+_{\pp+\kk,\o,s}\dpr\hh^-_{\kk,\o,s}}\lb{vertr}\\
G_{j;\o,s}(\pp;\pp+\kk) &= \sum_{\m,t} \m\,
{\dpr^3 \WW\over \dpr\hJ_{\pp,\m,t}
\dpr\hh^+_{\pp+\kk,\o,s}\dpr\hh^-_{\kk,\o,s}}\lb{vertj}\\
G_{2;\o,s} & ={\dpr^2
\WW\over\dpr\hh^+_{\kk,\o,s}\dpr\hh^-_{\kk,\o,s}}
\eal
In a similar way, the functional WI \pref{wis} can be used to obtain
a closed expression for the correlations of the density operator $\r_{\xx,\o,s} =
\psi^+_{\xx,\o,s} \psi^-_{\xx,\o,s}$. In fact, if we take
\pref{wis} with $\h=0$ and we perform a derivative w.r.t.
$\hJ_{-\pp,\m,s}$, we obtain:
\be\lb{wis1} \la\hat\r_{\pp,\o',s'} \hat\r_{-\pp,\o,s}\ra_T  =
-D_{-\o}(\pp) {\hat h(\pp)\over 4\p Z^2 c} {M^\r_{\o',\o}(\pp)+
s's M^\s_{\o',\o}(\pp)\over 2} \ee
which implies that
\be\lb{wis2} \la\r_{\xx,\o',s'} \r_{\yy,\o,s}\ra_T = \frac12 \lft[
G^\r_{\o',\o}(\xx-\yy) + s' s\, G^\s_{\o',\o}(\xx-\yy) \rgt] \ee
where
$$G^\g_{\o',\o}(\xx-\yy) = {1\over 4\p Z^2 c} \int{d\pp\over (2\p)^2}\;
e^{i\pp(\xx-\yy)} {p_0^2 + c^2 p^2 \over D_\o(\pp)} M^\g_{\o',\o}(\pp)
$$

\subsection{Schwinger-Dyson and Closed equations.}

By substituting the Ward Identities found in the previous section in the
Schwinger-Dyson equations, one get a set of closed equations, up to corrections
which are vanishing in the limit of removed cut-offs; this is due to the non
locality of the interaction and the proof is essentially identical to the one
in \cite{M007_2} or in \S 4 of \cite{BFM009} for the spinless case. The
presence of the spin makes however the resulting closed equations much more
complex, and new properties emerge from their solution, like the spin-charge
separation phenomenon.

Given any $F(\psi)$ which is a power series in the field, the Wick Theorem says
that, if $<\cdot>$ denotes the expectation w.r.t. the free measure,
\be \la \hp^-_{\kk,\o,s}F(\ps)\ra_0 =\hg_{{\rm D},\o}(\kk) \la {\dpr
F(\ps)\over \dpr \hp^+_{\kk,\o,s}}\ra_0. \ee
It follows that
\be {\partial e^{\WW} \over\partial \hh^+_{\kk,\o,s}}(0,\h) = \la
\hp^-_{\kk,\o,s}e^{\VV(\psi,0,\h)}\ra_0 =\hg_{{\rm D},\o}(\kk) \la {\dpr\over \dpr
\hp^+_{\kk,\o,s}}e^{\VV(\psi,\h)}\ra_0 \ee
Hence, by using \pref{fs1} and \pref{4.2}, we find, in the removed cutoffs
limit:
\be\lb{SDE}
\bsp
D_\o(\kk) &{\partial e^{\WW} \over\partial \hh^+_{\kk,\o,s}}(0,\h) =
\frac1Z \hh^-_{\kk,\o,s} e^{\WW(0,\h)}\\
&-Z \sum_{\o',s'}\int\!{d\pp\over(2\p)^2}\ 4\p c \n^{\o\o'}_{ss'}(\pp)
{\partial^2 e^{\WW}\over
\partial\hJ_{\pp,-\o',s'} \partial\hh^+_{\kk+\pp,\o,s}}(0,\h)
\esp
\ee
that we shall call the {\em Schwinger-Dyson equation}.  Let us now take the WI
in the form \pref{wis}, with $J=0$ and $\m'=-\o'$, derive it w.r.t.
$\hh^+_{\pp+\kk.\o,s}$ and insert the resulting expression in \pref{SDE}; we
obtain the following closed equation
\be\bsp\lb{ce}
D_\o(\kk) {\partial e^{\WW} \over\partial \hh^+_{\kk,\o,s}} = \frac1Z
\hh^-_{\kk,\o,s} e^{\WW} -\sum_{\m,t} \int\! {d \pp d\qq\over (2\p)^4}\
\hF^{-\o\m}_{-\o,st}(\pp)\cdot\\
\cdot \left[\hh^+_{\qq+\pp,\m,t} {\partial^2 e^{\WW}\over \partial
\hh^+_{\qq,\m,t} \partial\hh^+_{\kk+\pp,\o,s}}- {\partial^2 e^{\WW}\over
\partial\hh^+_{\kk+\pp,\o,s} \partial \hh^-_{\qq+\pp,\m,t}}
\hh^-_{\qq,\m,t}\rgt]  \;, \esp
\ee
where, if $M^\g_{\o,\o'}$ is defined as in \pref{MM},
$$
\hF^{\m}_{\o,s}(\pp)=
4\p c \sum_{\o', s'}\n^{\o\o'}_{ss'}(\pp)M^{s'}_{\o',\o\m}(\pp) \virg
M^s_{\m,\m'} = \frac12 (M^\r_{\m,\m'} + s M^\s_{\m,\m'})
$$
%
%

\subsection{The two point function}\lb{ss4.2}

By using \pref{ce}, we easily get:
$$
\la \psi^-_{\xx,\o,s}\psi^+_{\yy,\o',s'}\ra = \d_{\o,\o'}\d_{s,s'}
S_{\o}(\xx-\yy)
$$
where $S_{\o}(\xx)$ is the solution of the equation:
\be\lb{eqx} \lft(\dpr_\o S_{\o}\rgt)(\xx) -F^{-}_{-\o,+}(\xx)
S_{\o}(\xx) = \frac1Z \d(\xx)\;, \ee
with $\dpr_\o=\dpr_{x_0}+i\o c\dpr_{x_1}$. The solution of \pref{eqx} is:
\be\lb{Som+} S_{\o}(\xx)= e^{\D^{-}_{+}(\xx|0)}g_\o(\xx) \virg
g_\o(\xx) = \frac1{2\p Z} \frac1{cx_0+i\o x}\;, \ee
having defined $\D^\e_{s}$ such that $\dpr_{\o}^\xx \D^\e_{s}(\xx|\zz)=
F^\e_{-\o,s}(\xx)$:
\be\lb{DD} \D^\e_{s}(\xx|\zz) = \int\!{d\kk\over (2\p)^2} \
{e^{-i\kk\xx} - e^{-i\kk\zz} \over D_\o(\kk)}
\hF^\e_{-\o,s}(-\kk)=\D^{\e}_\r(\xx|\zz)+s\D^{\e}_\s(\xx|\zz)\;. \ee
for
$$
\hat \D_\r^\e(\pp)=g_\r \hat h(-\pp) {M^\r_{-\o,-\o\e}(-\pp)\over
D_\o(\pp)}+ {g_4\over 2} \hat h(-\pp) {M^\r_{\o,-\o\e}(-\pp)\over
D_\o(\pp)}
$$
$$
\hat \D_\s^\e(\pp)=g_\s \hat h(-\pp) {M^\s_{-\o,-\o\e}(-\pp)\over
D_\o(\pp)}- {g_4\over 2} \hat h(-\pp) {M^\s_{\o,-\o\e}(-\pp)\over
D_\o(\pp)}
$$
In order to evaluate the asymptotic behavior of $\D^\e_{s}(\xx|0)$, we need to
study functions of the type
\be\lb{II} I_{\o,\e}(\xx) = \int{d^2\pp\over (2\p)^2}\ a(\pp)
{e^{-i\pp\cdot\xx}-1\over (p_0+i\o c p_1)[v_+(\pp) p_0-i\e\o v_-(\pp) c p_1]}
\ee
where $a(\pp)$ and $v_s(\pp)>0$ are even smooth functions of fast decrease. It
is easy to show that
\be\lb{II1} I_{\o,\e}(\xx) = {a(0)\over v_+(0)} \tilde
I_{\o,\e}(\xx) + A + O(1/|\xx|) \ee
where $A$ is a {\it real} constant and, if $v=v_-(0)/v_+(0)$,
$$
\tilde I_{\o,\e}(\xx) = \int_{-1}^{+1} {d p_1\over (2\p c)}
\int_{-\io}^{+\io} {d p_0\over (2\p)} {e^{-i(p_o x_0 + p_1 x_1/c)}-1\over
(p_0+i\o p_1)(p_0- i\e\o v p_1)}
$$
One can see that, if $v>0$, $v\not=1$ and $\xx\not=0$,
$$
\tilde I_{\o,\e}(\xx) = {1\over 2\p c(1+\e v)} \left[ F(x_0,\o x_1/c) +
\e F(v x_0, -\e\o x_1/c) \right]
$$
where
$$
F(x_0,x_1) = \int_0^1 {dp_1\over p_1} \left[ e^{-p_1(|x_0|+i {\rm
sgn}(x_0) x_1)} -1\right]= \ln|z| +i{\rm Arg}\big({\rm
sgn}(x_0)z\big)+B+{\rm O}(1/z)
$$
where $z=x_0+ix_1$, $B$ is a real constant and $|{\rm Arg}(z)|\le \p$. Since
$$
{\rm Arg}\big({\rm sgn}(x_0)z\big)= {\rm Arg}(z)-\th(x_0){\rm
sgn}(x_1)\p
$$
the function $F(\xx)$ (considered only for $|\xx|>1$) is discontinuous at
$x_0=0$, while $\tilde I_{\o,\e}(\xx)$ is continuous. We can then write
\be\lb{II2} \tilde I_{\o,\e}(\xx) = -{1\over 2\p c(1+\e v)} \left[
\log (x_0+i\o x_1/c) + \e \log (v x_0 -i\e \o x_1/c) \right] + C +O(1/|\xx|)
\ee
where $C$ is again a real constant. By using \pref{MM}, \pref{II}, \pref{II1}
and \pref{II2}, one can easily check that
\be\lb{delta}
\bsp
\D^\e_\g(\xx|0)= &-{H^\e_{\g,\e}\over 4\p c} \ln\lft(v^2_\g x^2_0+
(x_1/c)^2\rgt) -{H^\e_{\g,-}+H^\e_{\g,+}\over 4\p c} \ln{x_0+i\o x_1/c\over
v_\g x_0+i\o x_1/c}\\
&+ C^\e_\g +O(1/|\xx|)
\esp
\ee
for
$$
H^+_{\g,\e}={2g_\g u_{\g,\e}+g_{4,\g} w_{\g,\e}\over v_{\g,+}+\e
v_{\g,-}}={\e g_\g\over v_{\g,+}v_{\g,-}}
$$
$$
H^-_{\g,\e}={2g_\g w_{\g,\e}+ g_{4,\g} u_{\g,\e}\over v_{\g,+}-\e
v_{\g,-}}=-{4\p\e\over 2\n_{\g,+}\n_{\g,-}} \lft[1-{(v_{\g,-}-\e
v_{\g,+})^2\over 4}\rgt]
$$
where $C^\pm_\g$ are real constants and $v_\g=v_{\g,+}(0)/v_{\g,-}(0)$ (and
$g_{4,\r}=g_4$ while $g_{4,\s}=-g_4$).

By using \pref{MM1} and \pref{MM2},
\be {H^+_{\g,+}\over 4\p c}= {\n_{\g}\over
v_{\g,+}v_{\g,-}}={\z_\g\over 2} \qquad {H^+_{\g,-}+H^+_{\g,+}\over 4\p c}=0
\ee
\be {H^-_{\g,-}\over 4\p c}= {1-\frac14 (v_{\g,+} + v_{\g,-})^2\over 2
v_{\g,+}v_{\g,-}}={\h_\g\over 2} \qquad {H^-_{\g,-}+H^-_{\g,+}\over 4\p
c}=-{1\over 2} \ee
Note that this expression is continuous in $v_\g=1$, as one expects, and that,
at least at small coupling, $\h_\g \ge 0$.

By using \pref{Som+} and \pref{DD}, we finally get
\be\lb{4.23}
S_\o(\xx) = \frac1{2\p Z} { (c^2 v_\r^2 x_0^2 + x_1^2)^{-\h_\r/2} (c^2v_\s^2
x_0^2 +x_1^2)^{-\h_\s/2} \over (c v_\r x_0+i\o x_1)^{1/2} (c v_\s x_0+i\o
x_1)^{1/2}} e^{C +O(1/|\xx|)}
\ee
where $C$ is a real constant $O(g)$ and $z^{1/2} = |z|^{1/2} e^{i Arg(z)/2}$.
Note that the leading term is well defined and continuous at any $\xx\not=0$.

Note also that, if $g_4=0$, $v_\r=v_\s=1$ and $\h_\r=\h_\s\=\h/2$, so that
\be
S_\o(\xx) = \frac1{2\p Z} { (c^2 x_0^2 + x_1^2)^{-\h/2} \over c x_0+i\o x_1}
e^{C +O(1/|\xx|)}
\ee
If we also put $g_\erp=0$ and $g_\arp=\l$, we get for $\h$ the value found for
the regularized Thirring model, that is $\h=2\t^2/(1-\t^2)$, with $\t=\l/(4\p
c)$; see eq. (4.21) of \cite{BFM009}.

\subsection{The four point functions and the densities correlations}\lb{ss4.4}

We want to calculate the truncated correlations $\la O^{(t)}_\xx O^{(t)}_\yy\ra^T$
of the local quadratic operators $O^{(t)}_\xx$, $t=(1,\a)$ or $(2,\a)$, defined
as the analogous operators of the Hubbard model, see \pref{op1} and \pref{op2};
note that $p_F$ has no special meaning in the effective model, but it is left there
since we want to compare the correlations in the two models.

Our UV regularization implies that $\la O^{(t)}_\xx\ra=0$ for any $t$; hence we can
make the calculation very simply, by using the explicit expressions of the four
points functions which follow from the closed equation \pref{ce} and then
evaluating them so that the two coordinates corresponding to each $O^{(t)}$
operator coincide, if this is meaningful. This works for all values of $t$,
except $(1,C)$ and $(1,S_3)$, where there is a singularity, related to the fact
that the operators $\r_{\xx,\o,s} = \psi^+_{\xx,\o,s} \psi^-_{\xx,\o,s}$ are
not well defined in the limit $N\to\io$, because of the singularity of the free
propagator at $\xx=0$. However, in these cases we can use directly the WI
\pref{wis1} for the density correlations, which allows us to calculate
correctly, in the limit $N\to \io$, the correlations of $O_\xx^{(1,C)}$ and
$O_\xx^{(1,S_3)}$, by using \pref{wis2}, \pref{MM} and the equations \pref{II},
\pref{II1}, \pref{II2}. We get, for $|\xx|>1$,
\be\nn
\bsp
G^\g_{\o,\o}(\xx) &\simeq {1-v_\g^2\over 8\p^2 c^2 Z^2} \lft[{u_{\g,+}\over
v_{\g,+}-v_{\g,-}} { 1\over (v_\g x_0+i\o x_1/c)^2}- {u_{\g,-}\over
v_{\g,+}+v_{\g,-}} { 1\over (v_\g x_0-i\o x_1/c)^2}\rgt]\\
G^\g_{-\o,\o}(\xx) &\simeq {1-v_\g^2\over 8\p^2 c^2 Z^2}
\lft[{w_{\g,+}\over v_{\g,+}-v_{\g,-}} { 1\over (v_\g x_0+i\o x_1/c)^2}-
{w_{\g,-}\over v_{\g,+}+v_{\g,-}} { 1\over (v_\g x_0-i\o x_1/c)^2}\rgt]
\esp
\ee
the corrections being of order $1/|\xx|^3$. This implies that, for $|\xx|>1$,
\be
\la O^{(1,C)}_{\bf 0} O^{(1,C)}_\xx \ra^T = \frac{v_\r^2 (1-\n_4+2\n_\r) +
(1+\n_4-2\n_\r)}{2\p Z^2 c^2 v_{\r,+} v_{\r,-}} {v^2_\r x^2_0-x^2/c^2\over
(v^2_\r x^2_0+x^2/c^2)^2} +O(1/|\xx|^3)
\ee
while $\la O^{(1,S_3)}_{\bf 0} O^{(1,S_3)}_\xx \ra^T$ is obtained from this
expression, by replacing $\n_4$ with $-\n_4$ and $\n_\r$ with $\n_\s$ (hence
also $v_\r$, $v_{\r,+}$ and $v_{\r,-}$ with $v_\s$, $v_{\s,+}$ and $v_{\s,-}$).

One can see, see Lemma 4.1 of \cite{BFM009_1}, that the same
result could be obtained starting from the four point function, if
we take the limit $\e\to 0$ of the expression obtained by the
substitution of the density operator $\r_{\xx,\o,s}$ with the
regularization
$$\r^\e_{\xx,\o,s} = \int d\uu \d_\e(\uu-\xx) \psi^+_{\xx,\o,s}
\psi^-_{\uu,\o,s}$$
where $\d_\e(\xx)$ is a smooth approximation of the delta function, rotational
invariant (in agreement with our UV regularization), whose support does not
contain the point $\xx=0$.

In order to calculate the other correlations, we first note that the only four
points functions different from zero are those defined by the equation
$$
G^{\o_1,\o_2}_{s_1,s_2}(\xx,\yy,\uu,\vv)
= \la \psi^-_{\xx,\o_1,s_1} \psi^-_{\yy,\o_2,s_2}
\psi^+_{\uu,\o_2,s_2} \psi^+_{\vv,\o_1,s_1}\ra
$$
By \pref{ce}, $G^{\o_1,\o_2}_{s_1,s_2}(\xx,\yy,\uu,\vv)$ is the solution of the
equation:
\be\bsp
\lft(\dpr_{\o_1}^\xx G^{\o_1,\o_2}_{s_1,s_2}\rgt)(\xx,\yy,\uu,\vv) =
\d(\xx-\vv) S_{\o_2}(\yy-\uu)-\d_{\o_1,\o_2}\d_{s_1,s_2} \d(\xx-\uu)
S_{\o_1}(\yy-\vv)+ \\
\Big[ -F^{-\o_1\o_2}_{-\o_1,s_1s_2}(\xx-\yy)
+F^{-\o_1\o_2}_{-\o_1,s_1s_2}(\xx-\uu) +F^{-}_{-\o_1,+}(\xx-\vv)\Big]
G^{\o_1,\o_2}_{s_1,s_2}(\xx,\yy,\uu,\vv) \esp
\ee
For the two-points correlation of $O^{(2,\a)}_\xx$ we are interested in the
case $\o_1=-\o_2=\o$. For $G^\o_s(\xx,\yy,\uu,\vv)
=G^{\o,-\o}_{s',ss'}(\xx,\yy,\uu,\vv)$ we find
\be\lb{eqG} G^\o_s(\xx,\yy,\uu,\vv) = e^{-\Big[\D^+_s(\xx-\yy|\vv-\yy) -
\D^+_s(\xx-\uu,\vv-\uu)\Big]} S_{\o}(\xx-\vv) S_{-\o}(\yy-\uu)\;. \ee

Therefore, for $\a=C,S_3$ we set $\xx=\uu$, $\yy=\vv$ and $s=+$, while for
$\a=S_1.S_2$ we set $s=-$; for $TC_1,TC_3$ we set $\uu=\vv$, $\xx=\yy$ and
$s=+$; while for $TC_2,SC$ we set $s=-$.

For the two-points correlation of $O^{(1,\a)}_\xx$, $\a\not=C,S_3$, we are
interested in the case $\o_1=\o_2=\o$. If $\bar G^\o_s(\xx,\yy,\uu,\vv)
=G^{\o,\o}_{s',ss'}(\xx,\yy,\uu,\vv)$ we find
\be\lb{eqGbar}
\bsp \bar G^\o_s(\xx,&\yy,\uu,\vv) = e^{-\Big[
\D^-_s(\xx-\yy|\vv-\yy) - \D^-_s(\xx-\uu,\vv-\uu)\Big]} S_{\o}(\xx-\vv)
S_{\o}(\yy-\uu)\\ &-\d_{s,+} e^{-\Big[ \D^-_+(\xx-\yy|\uu-\yy) -
\D^-_+(\xx-\vv,\uu-\vv)\Big]} S_{\o}(\xx-\uu) S_{\o}(\yy-\vv)\;. \esp
\ee
For $\a=SC$ we set $\xx=\yy$, $\uu=\vv$ and $s=-$; for $\a=S_1,S_2$ we set
$\xx=\uu$, $\yy=\vv$ and $s=-$; for $\a=S_3,C$ we set $\xx=\uu$, $\yy=\vv$ and
$s=+$.

Therefore, it is easy to see, by using \pref{DD} and \pref{4.23}, that, for
$|\xx|>1$,
\bal
&\la O^{(2,\a)}_{\bf 0} O^{(2,\a)}_\xx \ra^T = \frac{1}{\p^2 Z^2 c^2}
{\cos(2p_F x)^{m_\a}\over (v^2_\r x^2_0+x^2/c^2)^{x_{\r,t}}} {1\over (v^2_\s
x^2_0+x^2/c^2)^{x_{\s,t}}} +O(1/|\xx|^3)\virg \forall\a\nn\\
&\la O^{(1,SC)}_{\bf 0} O^{(1,SC)}_\xx \ra^T = -\frac{1}{\p^2 Z^2 c^2}
{\cos(2p_F x)\over (v^2_\r x^2_0+x^2/c^2)^{2\h_\r}} {v^2_\r x^2_0-x^2/c^2\over
(v^2_\r x^2_0+x^2/c^2)^2} +O(1/|\xx|^3) \lb{4.28}\\
&\la O^{(1,\a)}_{\bf 0} O^{(1,\a)}_\xx \ra^T = \frac{1}{\p^2 Z^2 c^2} {1\over
(v^2_\s x^2_0+x^2/c^2)^{2\h_\s}} {v^2_\s x^2_0-x^2/c^2\over (v^2_\s
x^2_0+x^2/c^2)^2}+O(1/|\xx|^3),\; \a=S_1,S_2\nn
\eal
where $m_\a=1$, if $\a=C,S_i$, while $m_\a=0$, if $\a=SC, TC_i$, and
\be x_{\g,t} =
\begin{cases}
\h_\g-\z_\g+1/2 & t=(2,C), (2,S_3)\\
\h_\g-s(\g)\z_\g+1/2 & t=(2,S_1), (2,S_2)\\
\h_\g+\z_\g+1/2 & t=(2,TC_1), (2,TC_3)\\
\h_\g+s(\g)\z_\g+1/2 & t=(2,SC), (2,TC_2)
\end{cases}
\ee

Let us now consider the special case $g_\s=0$ (i.e. $\h_\s=\z_\s=0$), which we
use as a effective model for the Hubbard model. In this case, the equations
\pref{4.28} imply that $\la O^{(t)}_{\bf 0} O^{(t)}_\xx \ra$ decays, for
$|\xx|\to\io$, as $|\xx|^{-2 X_t}$, with
\be\lb{4.30} 2X_t =
\begin{cases}
2+ 2\h_\r - 2\z_\r & t=(2,C), (2,S_i)\\
2+ 2\h_\r + 2\z_\r & t=(2,SC), (2,TC_i)\\
2+ 4\h_\r & t=(1,SC)\\
2 & t=(1,C), (1,S_i)
\end{cases}
\ee

Note that
\be \h_\r=-{1\over 2}+ {4-v_{\r+}^2- v_{\r-}^2\over 4 v_{\r+} v_{\r-}}\virg
\x_\r={2 \n_\r\over v_{\r+} v_{\r-}}
\ee
Let us now define $K=2 X_{2,C}-1$ and $\tilde K=2 X_{2,SC}-1$. By using
\pref{MM2}, we see that
\be\lb{KK} K=\frac{(1-2\n_\r)^2 - \n_4^2}{v_{\r+} v_{\r-}} = \sqrt{(1-\n_4)-2\n_\r\over
(1-\n_4)+2\n_\r} \sqrt{(1+\n_4)-2\n_\r\over (1+\n_4)+2\n_\r}\ee
$$\tilde K=\frac{(1+2\n_\r)^2 - \n_4^2}{v_{\r+} v_{\r-}}=K^{-1}$$
$$4\h_\r = K+\tilde K -2$$
These equations imply that all the critical indices $X_t$ and the parameter
$\h_\r$ can be expressed in terms of the single parameter $K$, only depending
on $g_2/c$ and $g_4/c$. In the following section we will show that the coupling
of the model \pref{vv1} can be chosen so that its exponents coincide with the
Hubbard ones; then, by some simple algebra, one can check the validity of the
scaling relations \pref{1.8}.

\subsection{Fine tuning of the parameters of the effective model}
\lb{ss2.5d}

Let us call $\tilde v_h=(\tilde g_{2,h},\tilde g_{4,h},\tilde \d_h)$, $h\le 0$,
the running coupling constants in the effective model with ultraviolet cutoff
$\g^N$ and parameters
\be\lb{5.1} g_{1,\perp}=0\virg g_{\arp}=g_{\erp}=\tilde g_{2,N}
\virg g_4=\tilde g_{4,N} \virg \d =\tilde \d_N \ee
so that, in particular, $c=v_F(1 +\tilde \d_N)$, and put $\tilde v_N=(\tilde
g_{2,N}, \tilde g_{4,N}, \tilde \d_N)$. We call $v_h=(g_{2,h}, g_{4,h}, \d_h)$,
$h\le 0$, the analogous constants in the Hubbard model, while $\vec v_h$ will
be defined as in \S\ref{sec2.2}, that is $\vec v_h=(v_h,g_{1,h},\n_h)$. The
analysis of the RG flow given in \S\ref{sec2} and App. \ref{appB} implies that,
for $h\le 0$,
\bal
\lb{asasa} \tilde v_{h-1} &=\tilde v_h + \beta^{(0,h)}(\tilde v_h,..,\tilde
v_0) + \tilde r^{(h)}(\tilde v_h,..,\tilde v_0,\tilde v_N)\\
\lb{asasa2} v_{h-1} &= v_h + \beta^{(0,h)}(v_h,..,v_0) + r^{(h)}(\vec
v_h,..,\vec v_0,\l)
\eal
where $\beta^{(0,h)}(\tilde v_h,..,\tilde v_0)$ is the beta function of the
effective model with parameters \pref{5.1}, modified so that the endpoints have
scale $\le 0$. Note that $\beta^{(0,h)}(v_h,..,v_0)$ is the function
$\beta^{(h)}(v_h,..,v_0)$ defined in \pref{bal}, modified so that, in its tree
expansion, no trees containing endpoints of type $g_1$ appear and the space
integrals are done in terms of continuous variables, instead of lattice
variables (the difference is given by exponentially vanishing terms). The
crucial bound \pref{van} and the short memory property imply that $|\tilde
r^{(h)}(\tilde v_h,..,\tilde v_N)|\le C [\max_{k\ge h} |\tilde v_k|]^2\;
\g^{\th h}$, while the analysis of \S\ref{sec2.2} implies that $r^{(h)}(\vec
v_h,..,\vec v_0,\l)$ satisfies a bound similar to \pref{3.22}.

\begin{lemma}\lb{lm5.1}
Given the Hubbard model with coupling $\l$ such that $g_{1,0}\in D_{\e,\d}$, it
is possible to choose $\tilde v_N$ as analytic function of $\l$, so that
\be\lb{5.4} \tilde g_{2}=2\l \left[\hat v(0)-{1\over 2}\hat v(2p_F)
\right] + O(\l^{3/2})\virg \tilde g_{4}=2\l\hat v(0)+O(\l^2)\virg
\tilde\d=O(\l) \ee
and, if $\tilde v_h$ are the r.c.c.  of the effective model with parameters
satisfying \pref{5.1}, while $v_h$ are the r.c.c.  of the Hubbard model, then,
$\forall h\le 0$,
\be\lb{144} |v_h-\tilde v_h|\le C {|g_{1,0}|\over 1+(a/2) |g_{1,0}| \,|h|}
\ee
Moreover, the r.c.c.  $\tilde v_h$ have a well definite limit as $N\to +\io$ and
this limit still satisfies \pref{144}.
\end{lemma}

\0{\bf Proof} - We have seen in the previous sections that the flows
\pref{asasa} and \pref{asasa2} have well defined limits $\tilde v_{-\io}$ and
$v_{-\io}$, as $h\to -\io$, if the initial values are small enough and
$g_{1,0}\in D_{\e,\d}$. Moreover, the proof of this property for the flow
\pref{asasa} implies that $\tilde v_{-\io}$ is a smooth invertible function of
$\tilde v_N$, such that $\tilde v_{-\io}=\tilde v_N+O(\tilde v_N^2)$; let us
call $\tilde v_N(\tilde v_{-\io} )= \tilde v_{-\io} + O(\tilde v_{-\io}^2)$ its
inverse. It is also clear that $\tilde v_N(\tilde v_{-\io} )$ has a well
defined limit as $N\to\io$, that we shall call $\tilde v(\l)$, and that this is
true also for the r.c.c.  $\tilde v_h$, $h\le 0$.

The previous remarks, together with \pref{2.42} and \pref{2.43}, imply that it
is possible to choose $\tilde v_N$, satisfying \pref{5.4}, so that
\be\lb{147}\tilde v_{-\io}-v_{-\io}=0\ee
In order to prove \pref{144}, we note that, because of the bound \pref{van} and
the short memory property, in the effective model with couplings satisfying
\pref{5.4},
\be |\tilde v_h - \tilde v_{-\io}| \le C \l^2 \g^{\th h}\ee
On the other hand, from Lemma 2.4 and 2.5
\be |v_h - v_{-\io}| \le C \sum_{j=-\io}^h
[|g_{1,j}|^2+\l\g^{{\th\over 2}j}] \le C_1 {|g_{1,0}|\over 1+(a/2) |g_{1,0}| \,|h|} \ee
These two bounds immediately imply \pref{144}.\Halmos

Let us now note that the critical indices of the effective model can be
calculated in terms of $\tilde v_{-\io}$ by the same procedure used for the
Hubbard model in \S\ref{sec2.4} and that we get an equation like \pref{fff},
{\it with the same function $\b_t^{(0,j)}$}. Hence, the above lemma allows us
to conclude that the critical indices in the Hubbard model and in the effective
model coincide, provided that the value of $\tilde v= \lim_{N\to\io} \tilde
v_N$ is chosen properly. It follows that all the indices are given by the
equations \pref{4.30}, with
\be\lb{5.9}
\bsp
\n_\r &= {\tilde g_2(\l)\over 4\pi c}= \l {\hat v(0)-\hat v(2p_F)/2\over 2\pi
\sin \bar p_F} + O(\l^{3/2})\\
\n_4 &= {\tilde g_4(\l)\over 4\pi c} = \l {\hat v(0)\over 2\pi \sin \bar p_F}
+O(\l^2)\esp
\ee
where \pref{5.4} has been used, together with $c=\sin \bar p_F +
O(\l)$. Moreover, \pref{5.9} and \pref{KK} imply that $K= 1- 2\l
[\hat v(0)-\hat v(2p_F)/2]/(\pi \sin \bar p_F)+ O(\l^{3/2})$, in agreement with \pref{g1}.


\section{
Spin-Charge Separation}
\lb{ss2.5dhy}

If $\kk\not=0$, the Fourier transform $\hat S_2(\kk+\o \pp_F)$ of the two-point
Schwinger function $S_2(\xx)$ in the Hubbard model can be written as a tree
expansion, in a way similar to eq. (2.64) of \cite{BFM007}, whom we shall refer
to for the notation:
\be \hat S_2(\kk+\pp^\o_F) = \sum_{n=0}^\io \sum_{j_0=-\io}^{0}
\sum_{\t\in\TT_{j_0,n,2,0}} \sum_{\bP\in \PP \atop |P_{v_0}|=2}
\hG^{2}_{\t,\o}(\kk)\ee
where $\pp_F^\o=(\o p_F,0)$, $\o=\pm$. Here $\hG^{2}_{\t,\o}(\kk)$
represents the contribution of a single tree $\t$ with $n$
endpoints and root of scale $j_0$; if $|\kk|\in [\g^{h_\kk},
\g^{h_\kk+1})$, it obeys the bound:
\be |\hG_{\t,\o}^{2}(\kk)| \le C \g^{-(h_\kk-j_0)} {\g^{-h_\kk}\over
Z_{h_\kk}} \sum_{n=0}^\io \sum_{j_0=-\io}^{0} \sum_{\t\in\TT_{j_0,n,\kk}}
\sum_{\bP\in \PP \atop |P_{v_0}|=2} (C|\l|)^n \prod_{\rm v\ not\ e.p.}
\g^{-d_v}{Z_{h_v}\over Z_{h_v-1}} \;, \ee
where $\TT_{j_0,n,\kk}$ denotes the family of trees whose special vertices
(those associated with the external lines) have scale $h_\kk$ or $h_{\kk+1}$.
Moreover, $d_v>0$, except for the vertices belonging to the path connecting the
root with $v^*$, the higher vertex (of scale $h^*$) preceding both the two
special endpoints, where $d_v$ can be equal to $0$. These vertices can be
regularized by using a factor $\g^{-(h^*-j_0)}$, extracted from the factor
$\g^{-(h_\kk-j_0)}$, so that we can safely perform the sum over all the trees
with a fixed value of $h^*$ and we get
\be\lb{mmm} |\hat S_2(\kk+\pp^\o_F)| \le C {\g^{-h_\kk}\over
Z_{h_\kk}} \sum_{h^*=-\io}^{h_\kk} \g^{-(h_\kk - h^*)} \le C
{\g^{-h_\kk}\over Z_{h_\kk}} \ee
A similar bounds can be obtained for the effective model with $g_{1\erp}=0$ and
couplings chosen as in Lemma \ref{lm5.1}. We shall call $\hat S_\o^M(\kk)$ and
$\tilde Z_h$ the two-point function Fourier transform and the renormalization
constants, respectively, in this model.

Let us put
\be \hat S_2(\kk+\pp^\o_F)={1\over Z_{h_\kk}} \bar G^2_\o(\kk)
\virg \hat S_\o^M(\kk) = {1\over \tilde Z_{h_\kk}} \bar
G^{2,M}_\o(\kk) \ee
We can write
\be \hat S_2(\kk+\pp^\o_F) = {\tilde Z_{h_\kk}\over Z_{h_\kk}}
{\bar G^2_\o(\kk)\over \tilde Z_{h_\kk}} = {\tilde Z_{h_\kk}\over
Z_{h_\kk}} \hat S_\o^M(\kk)+ {1\over Z_{h_\kk}} [\bar
G^2_\o(\kk)-\bar G^{2,M}_\o(\kk)] \ee
Note now that $\bar G^2_\o(\kk)$ differs from $\bar G^{2,M}_\o(\kk)$ for three
reasons:

\begin{itemize}

\item[1)] the propagators are different, which produces a difference
    exponentially small thanks to \pref{2.30}, \pref{ne} and the short
    memory property;

\item[2)] the r.c.c.  $v_h$ and $\tilde v_h$ are different, which produces a
    difference of order $\tilde g_{1,h_\kk}$, thanks to \pref{144} and the
    short memory property;

\item[3)] in the tree expansion of $\bar G^2_\o(\kk)$ and of the ratios
    $Z_j/Z_{j-1}$ there are trees with endpoints of type $g_1$, not present
    in the tree expansion of $\bar G^{2,M}_\o(\kk)$ and $\tilde Z_j/\tilde
    Z_{j-1}$; this fact produces again a difference of order $\tilde
    g_{1,h_\kk}$.

\end{itemize}

\0 These remarks, together with the fact that there is no tree with only one
endpoint in the tree expansion, implies that
\be\lb{bs1}
\left|{1\over Z_{h_\kk}} [\bar G^2_\o(\kk)-\bar G^{2,M}_\o(\kk)] \right| \le
C|\l \tilde g_{1,h_\kk}| {\g^{-h_\kk}\over Z_{h_\kk}}
\ee

For similar reason, we have
\be\lb{bs2}
{\tilde Z_{h_\kk}\over Z_{h_\kk}} = \prod_{j=h_\kk}^0 {\tilde Z_{h_j}\over
Z_{h_{j-1}}} {\tilde Z_{h_0}\over Z_{h_0}} = [1+O(\l^2)] e^{O(\l)
\sum_{j=h_\kk}^0 \tilde g_{1,j}} = [1+O(\l)] L(|\kk|^{-1})^{O(\l)}
\ee
where $L(t)$, $t\ge 1$, is the same function defined in Theorem \ref{th1.1}.

Theorem 1.3 easily follows from \pref{bs1}, \pref{bs2} and the explicit
expression \pref{4.23} of $S_\o(\xx)$ in the effective model, applied to the
case $g_\s=0$, $c=v_F (1+\tilde\d)$.


\section{Susceptibility and Drude weight}
\lb{ss2.5e}

The effective model is not invariant under a gauge transformation with the
phase depending both on $\o$ and $s$, if $g_{1,\perp}>0$; however, it is still
invariant under a gauge transformation with the phase only depending on $\o$.
This is true, in particular, if the interaction is spin symmetric, that is if
$g_\arp=g_{\erp}-g_{1,\perp}$, see item d in App. \ref{appB}. Since also the
Hubbard model is spin symmetric, it is natural to see if one can use this
``restricted " gauge invariance to get some useful information on the
asymptotic behavior of the Hubbard model, by comparing it with the effective
model with $g_{1,\perp}>0$.

Let us put $g_{\erp}\equiv \bar g_2$, $g_{1\perp}\equiv \bar g_1$ and
$g_\arp=\bar g_2- \bar g_1$. We want to show that we can choose the parameters
of the effective model $\bar g_1$, $\bar g_2$, $\bar g_4$, $\bar\d$, so that
the running coupling constants are asymptotically close to those of the Hubbard
model. This result is stronger of the similar one contained in Lemma
\ref{lm5.1}, since now all the running couplings are involved, and this implies
also that the values of $\bar g_2$, $\bar g_4$ and $\bar\d$ are {\it different}
with respect to the analogous constants defined in Lemma \ref{lm5.1}. The main
consequence of these considerations is that we can use the restricted WI of
this new effective model to get non trivial information on some Hubbard model
correlation functions, not plagued by logarithmic corrections.

Let $\vec l_h=(\bar g_{1,h}, \bar g_{2,h}, \bar g_{4,h}, \bar\d_h)$, $h\le 0$,
be the running coupling constants appearing in the integration of the infrared
part of the effective model. The smoothness properties of the integration
procedure imply that, in the UV limit, $\vec l_0$ is a smooth invertible
function of the interaction parameters $\vec l= (\bar g_{1}, \bar g_{2}, \bar
g_{4}, \bar\d)$, whose inverse we shall call $\vec l(\vec l_0)$; hence we can
fix the effective model by giving the value of $\vec l_0$ and by putting $\vec
l= \vec l(\vec l_0)$. In a similar way we call $\vec g_h=(g_{1,h}, g_{2,h},
g_{4,h}, \d_h)$, $h \le 0$, the running couplings of the Hubbard model with
coupling $\l$.

We now define $\vec x_h=\vec l_h-\vec g_h$, $h\le 0$. By using the
decomposition \pref{bal} for $\vec g_h$ and the similar one for $\vec l_h$, we
can write
\be \vec x_{h-1}=\vec x_h+ [\vec\b^{(1)}_h(\vec g_h,..,\vec
g_0)-\vec\b^{(1)}_h(\vec l_h,..,\vec l_0)] +\vec\b^{(2)}_h(\vec
g_h,\n_h...\vec g_0,\n_0,\l)+ \vec\b^{(3)}_h(\vec l_h,..,\vec
l_0,\vec l) \ee
where $\vec\b^{(1)}_h$ coincides with the function $\beta^{(h)}$ defined in
\pref{bal}. In the usual way, one can see that
\be |\vec\b^{(1)}_h(\vec g_h,..,\vec g_0)-\vec\b^{(1)}_h(\vec
l_h,..,\vec l_0)|\le C \left[ |\l|+\sup_{k\ge h}|\vec l_k| \right] \sum_{k=h}^0
\g^{-\th (k-h)}|\vec x_k| \ee
and that $|\vec\b^{(2)}_h|\le C|\l|\g^{\th h}$, $|\vec\b^{(3)}_{h}|\le C
[\sup_{k\ge h}|\bar l_k|]^2$. Note that the different power in the coupling of
these two bounds is due to the terms linear in $\l$ in the beta function for
$\d_h$, which are present in the Hubbard model, while similar terms are absent
in the effective model, see remark after \pref{bf} in App. \ref{appB}.

We want to show that, given $\l$ positive and small enough, it is possible to
choose  $\vec l_0$, hence $\vec x_0$, so that $\vec x_{-\io}=0$; we shall do
that by a simple fixed point argument. Note that $\vec x_{-\io}=0$ if and only
if
\be\lb{5.3} \vec x_{\bar h}=-\sum_{h=-\io}^{\bar h} \{
[\vec\b^{(1)}_h(\vec g_h,..,\vec g_0)-\vec\b^{(1)}_h(\vec l_h,..,\vec l_0)]
+\vec\b^{(2)}_h+ \vec\b^{(3)}_h\}\ee
We consider the Banach space $\MM_\th$, $\th<1$, of sequences $\vec x=\{\vec
x_h\}_{h\le 0}$ with norm $\|\vec x\|=\sup_{k\le 0}|\vec x_k|\g^{-(\th/2)k}$
and the operator ${\bf T}:\MM_\th\to \MM_\th$, defined as the r.h.s. of
\pref{5.3}. Given $\x>0$, let $\BB_\x = \{\vec x\in \MM_\th: \|\vec x\| \le \x
\l\}$; if $\l$ is small enough, say $\l\le \e_0$ and $\x\l \le \e_0$, and $\vec
l_h=\vec g_h + \vec x_h$, the functions $\vec\b^{(1)}_h(\vec l_h,..,\vec l_0)$
and $\vec\b^{(3)}_h(\vec l_h,..,\vec l_0,\vec l)$ are well defined and satisfy
the bounds above, even if $\vec x$ is not the flow of the effective model
corresponding to $\vec l_0$. Hence, we have:
\be \g^{-(\th/2)h} |{\bf T}(\vec x)_h|\le c_0 \l(\x
\l+1) \sum_{k=-\io}^{h-1}\g^{{\th\over2} k}\le c_1 \l (1+\x \l)\ee
so that $\BB_\x$ is invariant if $\x= 2 c_1$ and $\l\le \e_1=\min \{\e_0,
\e_0/(2c_1), 1/(2c_1)\}$. Moreover
\be\nn\bsp {\bf T}(\vec x)_h-{\bf T}&(\vec x')_h = \sum_{h=-\io}^{\bar h}
\big\{ [\vec\b^{(1)}_h( \{\vec g_k + \vec x_k\}_{k\ge h}) -
\vec\b^{(1)}_h(\{\vec g_k + \vec x'_k\}_{k\ge h})]\\
&+ [\vec\b^{(3)}_h(\{\vec g_k + \vec x'_k\}_{k\ge h})  - \vec\b^{(3)}_h(\{\vec
g_k + \vec x_k\}_{k\ge h})] \big\}\esp\ee
and $|{\bf T}(\vec x)_h-{\bf T}(\vec x')_h|\le c_2\l \|x-x'\|$, thanks to the
fact that all the terms in the r.h.s. of this equation are of the second order
in the running couplings. It follows that, if $c_2\l<1$, ${\bf T}$ is a
contraction in $\BB_\x$, so that \pref{5.3} has a unique solution $\vec
x^{(0)}$ in this set; moreover, if we put $\vec l_h = \vec g_h + \vec
x^{(0)}_h$, $\{\vec l_h\}_{h\le 0}$ is the flow of the effective model
corresponding to a value of $\vec l$ such that
\be\lb{5.5} |\vec g_h-\vec l_h|\le C|\l|\g^{{\th \over 2}h}\ee
Finally, this solution is such that $\vec l$ is equal to $\vec g_0$ at the
first order; hence, by using \pref{init}, we get
\be\lb{7.8}
\bar g_1= 2\l\hat v(2p_F)+O(\l^2)\virg \bar g_2=2\l \hat v(0) + O(\l^2)\virg
\bar g_4=2\l \hat v(0)+O(\l^2)\virg \bar\d=O(\l)
\ee

Thanks to the bound \pref{5.5}, this choice of $\vec l$, allows us to extend to
the Hubbard model Lemma 1 of \cite{BM010}, proved for the spinless fermion
model. Hence, we can say that there are constants $Z=1+O(\l^2)$, $Z_3=1+O(\l)$
and $\tilde Z_3=v_F+O(\l)$ such that, if $\k \le 1$ and $|\pp|\le \k$,
\be\lb{fff2}\bsp
\hat\O_C(\pp) &=Z_3^2 \la\hat\r_\pp
\hat\r_{-\pp}\ra^{(g)} +A_c + o(\pp)\\
\hat D(\pp) &= -\tilde Z_3^2\la\hat j_\pp \hat j_{-\pp}\ra^{(g)} +A_j +o(\pp)
\esp\ee
where $\la\cdot\ra^{(g)}$ denotes the expectation in the effective model satisfying
\pref{5.5}, $A_c$ and $A_j$ are suitable $O(1)$ constants and
\be \r_\xx=\sum_{\o,s}\psi^+_{\xx,\o\,s}\psi_{\xx,\o\,s}\quad\quad
j_\xx=\sum_{\o,s}\o\psi^+_{\xx,\o\,s}\psi_{\xx,\o\,s}\;.\ee
Moreover, if we put $\pp_F^\o=(\o p_F,0)$ and we suppose that $0<\k\le
|\pp|,|\kk'|,|\kk'-\pp|\le 2\k$, $0<\th<1$, then
\bal
\hat G^{2,1}_\r(\kk'+ \pp_F^\o, \kk'+\pp+\pp_F^\o) &=Z_3 \la\hat\r_\pp
\psi_{\kk',\o}\psi^+_{\kk'+\pp,\o}\ra^{(g)}[1+O(\k^\th)]\nn\\
\hat G^{2,1}_j(\kk'+ \pp_F^\o, \kk'+\pp+\pp_F^\o) &=\tilde Z_3 \la\hat
j_\pp \psi_{\kk',\o}\psi^+_{\kk'+\pp,\o}\ra^{(g)}[1+O(\k^\th)]\lb{h10a} \\
\hat S_2(\kk'+\pp_F^\o) &= \la\psi^-_{\kk',\o,\s}\psi^+_{\kk,\o,\s}
\ra^{(g)}[1+O(\k^\th)]\;.\nn
\eal
where $G^{2,1}_\r(\xx)$ and $G^{2,1}_j(\xx)$ are defined after \pref{eqm},
while the functions $\la\hat\r_\pp \psi_{\kk',\o}\psi^+_{\kk'+\pp,\o}\ra^{(g)}$ and
$\la\hat j_\pp \psi_{\kk',\o}\psi^+_{\kk'+\pp,\o}\ra^{(g)}$ coincide with the
functions \pref{vertr} and \pref{vertj}, respectively, with $c=v_F(1+\bar\d)$.
As already mentioned, if $g_1>0$, the effective model is still invariant under a
spin-independent phase transformation; hence the WI \pref{grez} is satisfied,
if we sum both sides over $s$ and we substitute $\n_s^\m(\pp)$ with
\be\lb{4.2a}
\bar\n^{\m}_{s}(\pp) = \lft\{\d_{\o,1} [\d_{s,-1}\bar g_2 + \d_{s,1} (\bar g_2-
\bar g_1)]+ \d_{\o,-1}\d_{s,-1}\bar g_4\rgt\} \frac{\hat h(\pp)}{4\p \bar
c}\virg \bar c=v_F(1+\bar\d)
\ee
Therefore we get the a WI similar to \pref{chara}, that is
\bal\label{ref4}
&-ip_0 [1-\bar\n_4(\pp)-2\bar \n_\r(\pp)] \la\hat\r_\pp
\psi_{\kk',\o}\psi^+_{\kk'+\pp,\o}\ra^{(g)}+ \bar cp [1+\bar\n_4(\pp)-
2\bar \n_\r(\pp)]  \la\hat j_\pp
\psi_{\kk',\o}\psi^+_{\kk'+\pp,\o}\ra^{(g)}
\notag\\
&=\frac1Z \lft[\la\psi^-_{\kk,\o,\s}\psi^-_{\kk,\o,\s}\ra^{(g)}
-\la\psi^-_{\kk+\pp,\o,\s}\psi^-_{\kk+\pp,\o,\s}\ra^{(g)}\rgt]
\eal
where
\be\lb{7.13} \bar\n_4(\pp) =\bar g_4{\hat h(\pp)\over 4\p
\bar c}\virg \bar \n_\r(\pp)= {\bar g_2-\bar g_1/2\over 4\p \bar c}\hat h(\pp)
\ee
By replacing \pref{h10a} in \pref{ref4}, and comparing with
\pref{ref1} we get, if $\bar\n_4(0)\equiv \bar\n_4$, $\bar
\n_\r(0)=\bar \n_\r$
\be\lb{5.13} {Z_3\over Z}=(1-\bar\n_4-2\bar \n_\r)\virg {\tilde
Z_3\over Z}=\bar c(1+\bar\n_4-2\bar \n_\r) \ee
Moreover, by proceeding as in derivation of \pref{wis1}, we get:
\be
\bsp
D_\o(\pp) \la\r^{(c)}_{\pp,\o} \r^{(c)}_{-\pp,\o'}\ra^{(g)} -\bar\n_4(\pp)
D_{-\o}(\pp) \la\r^{(c)}_{\pp,\o} \r^{(c)}_{-\pp,\o'}\ra^{(g)}\\
-2\bar\n_\r(\pp) D_{-\o}(\pp) \la\r^{(c)}_{\pp,-\o}
\r^{(c)}_{-\pp,\o'}\ra^{(g)} = -\d_{\o,\o'} {D_{-\o}(\pp)\over
2\pi Z^2 \bar c} \esp \ee
Hence, by some simple algebra, we get:
\bal
\la\r^{(c)}_{\pp,\o} \r^{(c)}_{-\pp,\o}\ra^{(g)} &= {1\over 2\pi Z^2 \bar c\bar
v^2_{\r,+}} { D_{-\o}(\pp) [D_{-\o}(\pp) - \bar\n_4 D_\o(\pp)]\over p_0^2 +
\bar c^2 \bar v_\r^2 p^2} + O(\pp)\\
\la\r^{(c)}_{\pp,\o} \r^{(c)}_{-\pp,-\o}\ra^{(g)} &= {1\over 2\pi Z^2 \bar c\bar
v^2_{\r,+}} { 2\bar\n_\r D_{\o}(\pp) D_{-\o}(\pp) \over p_0^2 + \bar c^2 \bar
v_\r^2 p^2} + O(\pp) \eal
where
\be
\bar v_\r= \bar v_{\r,-}/ v_{\r,+} \virg \bar v_{\r,\m}^2 = (1-\m\bar\n_4)^2-
4\bar\n_\r^2
\ee
Therefore the charge and current density correlations are given
by:
\be\label{xc}
\bsp \la\r^{(c)}_{\pp} \r^{(c)}_{-\pp}\ra^{(g)} &= {1\over \pi Z^2 \bar c\bar
v^2_{\r,+}} { -p_0^2 (1- \bar\n_4 + 2\bar\n_\r) + \bar c^2 p^2 (1 +\bar\n_4
-2\bar\n_\r) \over p_0^2 + \bar c^2 \bar v_\r^2 p^2} +
O(\pp)\\
\la j^{(c)}_{\pp} j^{(c)}_{-\pp}\ra^{(g)} &= {1\over \pi Z^2 \bar c\bar v^2_{\r,+}}
{ -p_0^2 (1- \bar\n_4 - 2\bar\n_\r) + \bar c^2 p^2 (1 +\bar\n_4 +2\bar\n_\r)
\over p_0^2 + \bar c^2 \bar v_\r^2 p^2} + O(\pp) \esp
\ee
From the WI \pref{ref1} we see that \be \hat\O_C(0,p_0)=0\virg \hat D(p,0)=0
\ee
and this fixes the values of the constants $A_c$ and $A_j$ in \pref{fff2}, so
that
\be
\bsp \hat \O_C(\pp) &= {Z_3^2\over \pi Z^2 \bar c\bar v^2_{\r,+}} [ (1+\bar\n_4-
2\bar\n_\r)+\bar v_\r^2(1-\bar\n_4+2\bar\n_\r)] {\bar c^2 p^2\over
p_0^2+\bar v_\r^2 \bar c^2 p^2}+o(\pp)\\
\hat D(\pp) &= {\tilde Z_3^2\over \pi Z^2 \bar c\bar v^2_{\r,+} v_\r^2} [
(1+\bar\n_4+ 2\bar\n_\r)+\bar v_\r^2(1-\bar\n_4-2\bar\n_\r)] {p_0^2\over
p_0^2+\bar v_\r^2 \bar c^2 p^2}+o(\pp)\esp
\ee
If we insert \pref{5.13} in the previous equations, we get, for the
susceptibility \pref{kk} and the Drude weight \pref{cc}, the values
\be
\bsp \k &={(1-\bar\n_4-2\bar \n_\r)^2 \over \pi  \bar c\bar v^2_{\r,+}\bar v_\r^2}[
(1+\bar\n_4- 2\bar\n_\r)+\bar v_\r^2(1-\bar\n_4+2\bar\n_\r)] =
\frac{\bar K}{\p \bar c\bar v_\r}\\
D &= {\bar c(1+\bar\n_4-2\bar \n_\r)^2\over \pi\bar v^2_{\r,+} \bar v_\r^2} [
(1+\bar\n_4+ 2\bar\n_\r)+\bar v_\r^2(1-\bar\n_4-2\bar\n_\r)]= \frac{\bar K \bar
c\bar v_\r}{\p}\esp
\ee
where
\be\lb{bKK}
\bar K=\frac{(1-2\bar\n_\r)^2 - \bar\n_4^2}{\bar v_{\r+} \bar v_{\r-}} =
\sqrt{(1-\bar\n_4)- 2 \bar\n_\r\over (1- \bar\n_4) + 2\bar\n_\r} \sqrt{(1+
\bar\n_4)-2\n_\r\over (1+ \bar\n_4)+2\bar\n_\r}
\ee
so that
\be {\k\over D}= {1\over \bar c^2\bar v_\r^2}
\ee
and this completes the proof of Theorem \ref{th1.2}.
\vskip.3cm

{\bf Remark} We are unable to see if $\bar c\bar v_\r=v_F(1+\bar\d)\bar v_\r$
coincides with the velocity $c v_\r = v_F(1+\tilde\d) v_\r$ appearing in the
two-point function asymptotic behavior \pref{4.23}, with $v_\r$ given (see
\pref{MM2}) by
\be v_\r= {((1+\n_4)^2-4\n_\r^2)\over ((1-\n_4)^2-4\n_\r^2)}\ee
with $\n_\r$ and $\n_4$ defined as in \pref{5.9}. In fact, it is easy to see
that $\tilde\d$ is equal to $\bar\d$ at the first order and this is true also
for $\bar v_\r$ and $v_\r$ by \pref{7.8}, \pref{7.13} and \pref{5.9}; however,
our arguments are not able to exclude that the two velocities are different.
Moreover, \pref{bKK} and \pref{KK} imply that $\bar K=K$ at first order, but
they also could be different. Note that the equality of $\bar K$ and $K$, would
imply that $\k=K/v$, with $v=\bar c\bar v_\r$ being the charge velocity, a
relation proposed in \cite{Ha080} which, together with \pref{1.8} and
\pref{hh1}, would allow the exact determination of the exponents in terms of
the susceptibility and the Drude weight.
%
%
%
%


%
%
%


\appendix

\section{The $g_1$ map}
\lb{appA}

Let us consider the following map on the complex plane:
\be\lb{A1} g_{n+1}=g_n - a_n g_n^2 \ee
where $a_n$ is a sequence depending on $g_0$, such that, if $|g_0|$ is small
enough,
\be
\lb{A1a} a_n = a+\s_n\virg |\s_n|\le c_0 |g_0|\;,
\ee
for some positive constants $a$ and $c_0$. We want to study the trajectory of
the map \pref{A1}, under the condition that
\be g_0\in D_{\e,\d}=\{z\in\CCC: |z|<\e, |\text{Arg } (z)|\le \p-\d
\} \virg \d\in (0,\p/2) \ee
We shall first study the properties of a sequence $\tilde g_n$, which turns out
to be a good approximation of $g_n$.
Let us define:
\be\lb{A1d} A_n= \frac1n \sum_{k=0}^{n-1} a_k \ee

\begin{lemma}\lb{lmA1}

Given $\d \in (0,\p/2)$, there exists $\e_0(\d)$ such that, if $\e\le\e_0(\d)$
and $g_0\in D_{\e,\d}$, the sequence
\be\lb{Agn} \tilde g_n = \frac{g_0}{1+g_0 n A_n} \ee
at any step $n\ge 0$ is well defined  and does not exit the  larger domain
$D_{\e_1,\d_1}$, for $\e_1=2\e/(sin\, \d)$ and $\d_1=\d/2$.
\end{lemma}

\0{\bf Proof} - First of all, we choose $\e$ so that
\be\lb{A4} c_0\e\le a/2\quad \Rightarrow\quad a/2\le \Re\, a_n \le 3a/2 \virg
|\Im\, a_n| \le c_0|g_0|
\ee
where $c_0$ is the constant defined in \pref{A1a}; we can write
\be A_n=\a_n + i \b_n \virg \a_n\ge a/2 \virg |\b_n| \le c_0|g_0|\;.
\ee
Define
$\tilde z_n := 1+g_0 n A_n := 1+g_0 n \a_n + \tilde w_n$; then, if $g_0\in
D_{\e,\d}$,
\be\lb{AA4} |1+g_0 n \a_n| \ge  \max \left\{\sin\d,\ \frac{\sin\ \d}{3}(1+|g_0| n\a_n)
\right\}\ee
In fact, it is trivial to show that $|1+g_0 n \a_n|\ge \sin\d$; on the other
hand, if $|g_0| n\a_n \ge 2$,

$$|1+g_0 n \a_n|\ge |g_0| n \a_n - 1 = (|g_0| n \a_n  + 2|g_0| n \a_n - 3)/3
\ge (|g_0| n \a_n  + 1)/3$$
By using \pref{AA4}, we get
\be\lb{A5} \frac{|\tilde w_n|}{|1+g_0 n \a_n|} \le \frac{6c_0}{a\sin\d}
|g_0|\;. \ee
It follows that, if $\e$ is small enough,
\be\lb{A3} |\tilde z_n| \ge  \frac12 \sin\d \ee
so that, in particular, the definition \pref{Agn} is meaningful.

Now we want to prove that $\tilde g_n\in D_{\e_1,\d_1}$, with $\e_1=2\e/(
\sin\,\d)$ and $\d_1 = \d/2$, if $\e$ is small enough. Let $g_0=\r_0 e^{i
\theta_0}$; by using \pref{AA4} and \pref{A5}, we see that, if $\e$ is small
enough,
\be\lb{A3b} |\tilde g_n| \le \frac{2\,|g_0|}{|1+\a_n g_0 n|} \le
\frac{2\e}{\sin \d}\;; \ee
besides it is easy to see that
$$\lft| \text{Arg } \left( \frac{g_0}{1+ \a_n g_0 n}
\right)\right| = \lft| \text{Arg } \left( \frac{\r_0}{e^{-i\theta_0}+ \a_n \r_0 n}
\right)\right|\le |\theta_0| \le \p-\d\;.
$$
Then, since $\tilde g_n = \frac{\r_0}{e^{-i\theta_0}+ \a_n \r_0 n} (1 +
w_n)$, with $w_n$ of order $g_0$, for $\e$ small enough,
\be\lb{A6} |\text{Arg } (\tilde g_n)|\le \p-\d/2
\ee
\Halmos

\begin{proposition}\lb{propA2}

Given $\d \in (0,\p/2)$, there exists $\e_0(\d)$, such that, if $\e\le\e_0(\d)$
and $g_0\in D_{\e,\d}$, then
\be\lb{A2c} g_n\in D_{\e_2,\d_2} \virg \e_2=\frac{3 \e}{\sin\, \d}
\virg \d_2=\frac{\d}4 \ee
Moreover, if $\tilde g_n$ is defined as in \pref{Agn},
\be\lb{A2b} |g_n - \tilde g_n| \le |\tilde g_n|^{3/2} \ee
\end{proposition}

\0{\bf Proof} - We shall proceed by induction on the condition \pref{A2b},
which is true for $n=0$. Suppose that it is true for $n\le N$; then, by using
\pref{A3b} and \pref{A6}, we see that, if $\e$ is small enough and $n\le N$,
\be\lb{A8}|g_n|\le 3|\tilde g_n|/2 \le 3\e/\sin \d\virg |\text{Arg }
(g_n)|\le \p-\d/4
\ee
which proves \pref{A2c}. Moreover, by \pref{A1}, if $\e$ is small enough,
\be\lb{A7} |g_{N+1}| \le 2|g_N| \le 3|\tilde g_N|\ee

Note now that
\be\lb{A11}
\frac{1}{g_{n+1}} - \frac{1}{g_n} = \frac{a_n}{1-a_n g_n} = a_n + a_n^2 g_n +
\D_n =\frac{1}{\tilde g_{n+1}} - \frac{1}{\tilde g_n} + a_n^2 g_n + \D_n
\ee
where $\D_n$ is a quantity which can be bounded by $c_1 |g_n|^2$, for some
constant $c_1$. We can rewrite \pref{A11} in the form
\be\lb{A10}
\frac{1}{g_{n+1}} - \frac{1}{\tilde g_{n+1}} = \frac{1}{g_n} - \frac{1}{\tilde
g_n} + a_n^2 g_n + \D_n
\ee
By using \pref{A4}, \pref{AA4}, \pref{A5}, \pref{A8}, \pref{A7} and \pref{A10},
we get, if $\e$ is small enough,
\be
\bsp &|g_{N+1} - \tilde g_{N+1} | = |g_{N+1}|\,|\tilde g_{N+1}|
\left| \frac1{g_{N+1}} - \frac1{\tilde g_{N+1}} \right| \\
\le 3|\tilde g_N|\,|\tilde g_{N+1}| &\sum_{n=0}^N [6 a^2 |\tilde g_n| + \frac94
c_1|\tilde g_n|^2] \le c_2|\tilde g_N|^{3/2} \frac{|g_0|^{1/2}}{(1+\frac{a}2
|g_0| N)^{1/2}} \sum_{n=0}^N \frac{|g_0|}{1+\frac{a}2 |g_0| n}\\
&\le |\tilde g_N|^{3/2} \frac{c_3|g_0|^{1/2}}{(1+\frac{a}2 |g_0| N)^{1/2}} \log
\lft(1+\frac{a}2 |g_0| N\rgt) \le |\tilde g_N|^{3/2}\esp
\ee
where $c_2$ and $c_3$ are two suitable constants.\Halmos

\section{Symmetries of the Effective Model and RG Flow}
\lb{appB}

The RG analysis of the effective model will be done by exploiting some symmetry
properties of a more general model, obtained by adding to the interaction
\pref{fs1} the term $g_3 V_3(\ps)$, with
\be
V_3(\ps) = \frac12 \sum_{\o,s}\int d\xx d\yy h(\xx-\yy)
\ps^+_{\xx,\o,s}\ps^+_{\xx,\o,-s}
\ps^-_{\yy,-\o,s}\ps^-_{\yy,-\o,-s}\label{umm} \ee

The integration of the positive (ultraviolet) scales $N,N-1,..,0$ is
essentially identical to the one described in \S 2 of \cite{M007} or \S 3 of
\cite{BFM009}, for the spinless case; it does not need any localization
procedure.

The integration of the infrared scales is done in a way similar to the one in
the Hubbard model described in \S\ref{sec2}. However, before starting the
multiscale IR integration, we have to perform some technical operations, which
will make possible to compare the flow of the running couplings with that of
the Hubbard model.

After the integration of the UV scales up to $j=1$, the free measure propagator
is given by $g^{[l,1]}_{{\rm D},\o}(\xx)$, defined as in \pref{gth} with $N=1$. In
this expression, the velocity $c$ has the role of the Fermi velocity $v_F$ of
the Hubbard model. In order to match the asymptotic behavior of the two models,
we can not choose $c=v_F$; for this reason we introduced the parameter $\d$.
However, it is not possible to compare the RG flows of the two models if the
two velocities are different; hence, we have to move from the free measure to
the interaction the term proportional to $\d$. Moreover, since also the cutoff
function $\chi^{[l,1]}(|(k_0, c k)|)$ depends on $\d$, we have to ``modify'' it
in $\chi_{[l,1]}(|(k_0, v_F k)|)$.

The simplest way of performing these operations without introducing spurious
singularities is the following one. We start with a free measure of the form
\be\lb{B2}
P(d\psi^{[l,1]}) = \NN^{-1} \exp\lft\{ - {Z\over L^2} \sum_{\kk,\o,s} C_l(\kk)
[-ik_0+\o v_F (1+\d)k] \hat\psi_{\kk,\o,s}^{+[l,1]}
\psi_{\kk,\o,s}^{-[l,1]}\rgt\}
\ee
where $C_l(\kk)=\chi^{[l,1]}(|(k_0, c k)|)^{-1}$. We can move to the scale 1
effective interaction the term
$$- \d {Z\over L^2} \sum_{\kk,\o,s}
\o v_F k \hat\psi_{\kk,\o,s}^{+[l,1]} \psi_{\kk,\o,s}^{-[l,1]} =
-\d V_\d(\psi)\virg \text{with}$$
\be
V_\d(\psi) = \sum_{\o,s}\int d\xx \ps^+_{\xx,\o,s} (i\o v_F \partial_x)
\ps^-_{\xx,\o,s}
\ee
The new free measure differs from \pref{B2} because $\d$ is multiplied by
$u_l(\kk) = 1 -\chi^{[l,1]}(|(k_0, c k)|)$. On the other hand, since $\d$ will
be chosen of order $\l$, $\chi^{[l,1]}(|(k_0, c k)|)$ and $\chi^{[l,1]}(|(k_0,
v_F k)|)$ differ only for values of $\kk$ of size $\g$ or $\g^l$ and
\be\lb{ul}
u_l(\kk)=0\virg \text{if\ } \chi^{[l+1,0]}(|(k_0, v_F k)|)>0 \ee
Hence, we can write
\be
\chi^{[l,1]}(|(k_0, c k)|) = \bar \chi^{(1)}(\kk) + \chi^{[l+1,0]}(|(k_0, v_F
k)|) + \bar \chi^{(l)}(\kk)
\ee
with $\bar \chi^{(1)}(\kk)$ and $\bar \chi^{(l)}(\kk)$ smooth functions, whose
support is on values of $\kk$ of size $\g$ or $\g^l$, respectively; moreover,
if we define
\be
\tilde C_l(\kk) = \lft[\chi^{[l+1,0]}(|(k_0, v_F k)| + \bar
\chi^{(l)}(\kk))\rgt]^{-1}
\ee
then $\tilde C_l(\kk)=1$, if $1\ge |(k_0, v_F k)|\ge \g^{l+1}$ and $\tilde
C_l(\kk)^{-1} \bar \chi^{(l)}(\kk) \le 1$.
It follows that the free measure $P(d\psi^{[l,1]})$ can be written as
$P(d\bar\psi^{(1)}) P(d\tilde\psi^{[l,0]})$, where $\bar\psi^{(1)}$ is a field
whose covariances has the same scale properties of $\psi^{(1)}$, while
\be
P(d\tilde\psi^{[l,0]}) = \NN^{-1} \exp\lft\{ - {Z\over L^2} \sum_{\kk,\o,s}
\tilde C_l(\kk) [-ik_0+\o v_F (1+ u_l(\kk) \d)k] \tilde\psi_{\kk,\o,s}^{+[l,0]}
\tilde\psi_{\kk,\o,s}^{-[l,0]} \rgt\}
\ee
The integration of the single scale field $\bar\psi^{(1)}$ can be done without
any problem. At this point, we start the multiscale integration of the field
$\tilde\psi^{[l,0]}$, by performing the effective potential localization and
the free measure renormalization as in the Hubbard model. Thanks to the support
properties of $u_l(\kk)$, the steps from $j=0$ to $j=l+1$ will give the same
result we should get if the propagator of $\tilde\psi^{[l,0]}$ were equal to
$${1\over Z}{1\over L^2} \sum_{\kk} e^{i\kk\xx} {\chi_{l,0}(|(k_0, v_F k)|)\over
-ik_0+\o v_F k}$$
This means that the renormalized single scale propagator will have the form
\pref{gjth}, corresponding to the leading behavior of the single scale
propagator in the Hubbard model. This property is not true only in the last
step, $j=l$, but this is not a problem, since we have to study the RG flow at
fixed $j$ and $l\to-\io$ and, moreover, the contribution of the IR scale
fluctuations to the Schwinger functions at fixed space-time coordinates
vanishes as $l\to-\io$.

Let us now analyze in more detail the RG flow of the effective model for $h\le
0$. The main difference with respect to the Hubbard model is that \pref{rel}
has to be replaced by
\be\lb{fs2a}
\bsp \tilde V^{(j)}(\sqrt{Z_j}\ps) &= g_{1,\perp,j} F_{1,\perp}(\sqrt{Z_j}\ps)+
g_{\arp,j} F_\arp(\sqrt{Z_j}\ps) +\\
+ g_{\erp,j} F_\erp(\sqrt{Z_j}\ps) &+g_{3,j} F_3(\sqrt{Z_j}\ps)+g_{4,j}
F_4(\sqrt{Z_j}\ps) + \d_j V_\d(\sqrt{Z_j}\ps)\esp
\ee
where the functions $F_\a(\psi)$ are defined as the functions $V_\a(\psi)$ of
\pref{five1} with $h(\xx-\yy)=\d(\xx-\yy)$; the absence of local terms
proportional to $\psi^+\psi^-$ is a consequence of the oddness in $\kk$ of the
free propagator. The running couplings verify equations of the form
\be\lb{bf}
\bsp g_{\a,h-1}-g_{\a,h} &= B_{\a}^{(h)}\lft(\vec g_h, \d_h,..\vec g_0,
\d_0, \vec g, \d\rgt)\\
\d_{h-1}- \d_h &= B_{\d}^{(h)}\lft(\vec g_h, \d_h,..\vec g_0, \d_0, \vec g,
\d\rgt)\esp
\ee
where $\a=1\perp,\arp,\erp,3,4$ and $\vec g_j=(g_{1,\perp,j}, g_{3,j},
g_{\arp,j}, g_{\erp,j}, g_{4,j})$. Note that the functions $B_{\a}^{(h)}$ and
$B_{\d}^{(h)}$ are of the second order in their arguments; in the case of
$B_{\d}^{(h)}$, this follows from the structure of $\tilde V(\psi)$ (see
\pref{five1}), which does not allow to build Feynman graphs of the first order
in $\vec g$. For the same reason
\be \d_0=\d + O(\e_0^2)\virg \e_0=\max\{|\vec g|,|\d|\}\ee
and this relations can be inverted, if $\e_0$ is small enough.

\vspace{.3cm}

There are some symmetries which is important to exploit. For notational
simplicity, we will write $(G_{1,\perp},G_3,G_\arp,G_\erp, G_4, \D)$ or $(\vec
G,\D)$ in place of $\lft(\vec g_h, \d_h,..\vec g_0, \d_0, \vec g, \d\rgt)$ .

\bd \item{a.} {\it Spin U(1).} Free measure and interactions are invariant
under the transformation
$$
\ps^\e_{\xx,\o,s}\to e^{i\e\a_s}\ps^\e_{\xx,\o,s}
$$
where $\a_s$ is a spin-dependent angle. This means that the only possible local
effective interactions must have as many $\ps^+_s$ as $\ps^-_s$, for each given
$s$. Therefore, all the allowed local quartic interactions are the ones listed
in \pref{five1}; it is also clear from the symmetries $\o\to -\o$ and $s\to -s$
that they must occur in the same linear combinations.
\item{b.} {\it Particle-hole symmetry.} The free measure is invariant if the
    particle hole switching involves the spin $s'$ only, i.e.
    $\hp^\e_{\kk,\o,s}\to \hp^{-ss'\e}_{-ss'\kk,\o,s}$; interactions are not
    invariant and we find:

\be \lb{s13} B_{1,\perp}^{(h)}(G_{1,\perp},G_3,G_\arp,G_\erp, G_4, \D)
=-B_3^{(h)}(-G_3,-G_{1,\perp},G_\arp,- G_\erp, -G_4,\D) \ee
\be \lb{salfa}
\left. \bsp B_\arp^{(h)}(G_{1,\perp},G_3,G_\arp,G_\erp, G_4,\D)
&= B^{(h)}_\arp(-G_3,-G_{1,\perp},G_\arp,-G_\erp, -G_4,\D)\\
B_\erp^{(h)}(G_{1,\perp},G_3,G_\arp,G_\erp, G_4,\D)
&=-B_\erp^{(h)}(-G_3,-G_{1,\perp},G_\arp,-G_\erp,-G_4,\D)\\
B_4^{(h)}(G_{1,\perp},G_3,G_\arp,G_\erp, G_4,\D) &=-B_4^{(h)}
(-G_3,-G_{1,\perp},G_\arp,- G_\erp, -G_4,\D) \esp\right\}
\ee
\be \lb{sdel} B_\D^{(h)}(G_{1,\perp},G_3,G_\arp,G_\erp, G_4, \D)
=B_\D^{(h)}(-G_3,-G_{1,\perp},G_\arp,-G_\erp, -G_4, \D) \ee

\item{c.} {\it Chiral U(1).} The free measure is invariant under the
    transformation
\be\lb{u1} \hp^\e_{\kk,\o,s}\to e^{i\e\a_\o}\hp^\e_{\kk,\o,s} \ee
for $\a_\o$ a chirality-dependent angle. All the interactions are $U(1)$
invariant, but for $V_3$; if $g_3=0$, then an interaction $V_3$ won't be
generated by the flow.  This fact can be seen also graph by graph. Indeed, in
the graphs for $B_\arp$, $B_\erp$, $B_4$ and $B_\d$, the number of the
half-lines $\ps^+_\o$ has to equal the number of the half-lines $\ps^-_\o$
(regardless the spin label); this can only happen when there is an even number
of interactions $V_3$. Therefore these three beta functions are even in $G_3$
and then, by \pref{salfa} and \pref{sdel}, also even in $G_{1,\perp}$; hence,
if $\a=\arp,\erp,4,\d$,
\be\lb{ralfa} B_\a^{(h)}(G_{1,\perp},G_3,G_\arp,G_\erp, G_4, \D) =
\bar B_\a^{(h)}(G^2_{1,\perp},G^2_3,G_\arp,G_\erp, G_4,\D) \ee
where $G_\a^2$ denotes the tensor $\{g_{\a,j} g_{\a,j'}\}_{j.j'\ge h}$. By a
similar argument, $B_3^{(h)}$ has to be odd in $G_3$ and $B_1^{(h)}$ even;
then, by \pref{s13}, $B_1^{(h)}$ is also odd in $G_1$ and $B_3^{(h)}$ even, so
that
\be\lb{r13}
\bsp B_{1,\perp}^{(h)}(G_{1,\perp},G_3,G_\arp,G_\erp, G_4, \D) &=
G_{1,\erp} \bar B^{(h)}(G^2_{1,\perp},G^2_3,G_\arp,G_\erp, G_4, \D)\\
B_3^{(h)}(G_{1,\perp},G_3,G_\arp,G_\erp, G_4, \D) &= G_3 \bar
B{(h)}(G^2_3,G^2_{1,\perp},G_\arp,-G_\erp,- G_4, \D)\esp
\ee
where $G_\a \bar B_\a^{(h)}$ is a shorthand for $\sum_{j\ge h} g_{\a,j}
B_\a^{(h,j)}$.

In this way we have found two invariant surfaces in the space of the
interaction parameters $(\vec g,\d)$:
$$
\CC_1=\{\vec g,\d :g_{1,\perp}=0\}\qquad \CC_3=\{\vec g, \d:g_3=0\}
$$

\item{d.} {\it Spin SU(2).} It is convenient to rewrite the interaction as
\be\lb{fs2}
\bsp \tilde V(\ps) &= g_{1,\perp} \lft(V_{1,\perp}(\ps)-V_\arp(\ps)\rgt) +
(g_\arp+g_{1,\perp}-g_\erp) V_\arp(\ps) +\\
&+g_\erp \lft(V_\erp(\ps)+ V_\arp(\ps)\rgt) + g_3 V_3(\ps) + g_4 V_4(\ps) +
\d_h V_\d(\ps)\esp
\ee
It is evident that $V_{1,\perp}-V_\arp$, $ V_\erp+ V_\arp$, $V_3$ $V_4$ and $V_\d$, as
well as the free measure, are invariant under the transformation of the fields
$$
\hp^-_{\kk,\o,s}\to \sum_{s'}U_{s,s'}\hp^-_{\kk,\o,s'} ,\quad
\hp^+_{\kk,\o,s}\to \sum_{s'}\hp^+_{\kk,\o,s'} U^\dag_{s',s}
$$
for $U\in SU(2)$. While $V_\arp$ isn't: if $g_\arp+g_1-g_\erp=0$ it will remain
zero. Thus we find four other invariant surfaces:
$$
C_{1,+}=\{\vec g,\d:g_{1,\perp}=g_{\erp}-g_{\arp}\} \virg
C_{3,+}=\{\vec g,\d:g_{3}=g_{\erp}+g_{\arp}\}
$$
$$
C_{1,-}=\{\vec g ,\d:-g_{1,\perp}=g_{\erp}-g_{\arp}\} \virg
C_{3,-}=\{\vec g,\d:-g_{3}=g_{\erp}+g_{\arp}\}
$$
(in fact $\CC_{1,-}$, $\CC_{3,+}$ and $\CC_{3,-}$ are obtained form $\CC_{1,+}$
through \pref{s13}, \pref{salfa}, \pref{ralfa} and \pref{r13}). $C_{3,+}$ is
also called Fowler invariant (see \cite{So079}, page 220).

\item{e.} {\it Vector-Axial Symmetry.} All the terms in $\tilde V(\ps)$, but
    $V_1(\psi)$ and $V_3(\psi)$, are invariant under the transformation
$$
\ps^\e_{\xx,\o,s}\to e^{i\e\th_{\o,s}}\ps^\e_{\xx,\o,s}
$$
therefore the surface $g_{1,\perp}=g_3=0$ is invariant. \ed

\vspace{.3cm}

Finally we consider the flow of $Z_h$ and the renormalization constant
$Z^{(1)}_h$ associated with the density operator
$\r_{\xx,\o,s}=\psi^+_{\xx,\o,s} \psi^-_{\xx,\o,s}$ in the generating
functional \pref{vv1}; $Z^{(1)}_h$ is defined as $Z^{(1,C)}_h$ in \pref{2.20}.
It is easy to see, by using the symmetry properties of the model as before,
that $Z_{h-1}/Z_h = 1+ B_z^{(h)}(\vec G, \D)$ and $Z^{(1)}_{h-1}/Z^{(1)}_h =
1+B_\r^{(h)}(\vec G, \D)$, with $B_\a^{(h)}(\vec G, \D) = \bar
B_\a^{(h)}(G^2_1,G^2_3,G_\arp,G_\erp, G_4,\D)$ for $\a=z,\r$. Hence
\be \frac{Z^{(1)}_{h-1}}{Z_{h-1}} = \frac{Z^{(1)}_h}{Z_h} [1 +
\tilde B^{(h)}(\vec G,\D)] \ee
with \be\lb{ralpha1} \tilde B^{(h)}(\vec G,\D)=\bar
B_2^{(h)}(G^2_{1,\perp},G^2_3,G_\arp,G_\erp, G_4,\D) \ee
%

%

%
%

\section{Vanishing of the Beta Function}
\lb{appC}

We recall the main ideas of the proof of \pref{beta23} and
\pref{beta44} (see
\cite{BM002},\cite{BM005},\cite{M005},\cite{M007_3} for more
details). We consider the model \pref{vv1} with $g_{1,\perp}=0$;
$\d$, $g_\arp$ and $g_\erp$ are small but arbitrary parameters. We
take the limit $N\to\io$ at fixed $l$; if $|\kk|=\g^l$ (so that,
in particular, $\hat f_l(\kk)=1$)
\be
\hat S_\o(\kk) \= \la\hat\psi_{\kk,\o,s}^- \hat\psi^+_{\kk,\o,s}\ra_l={1\over Z_l
D_{l,\o}(\kk)}[1+W_2^{(l)}(\kk)] \ee
where $\la\cdot\ra_l$ denotes the expectation with propagator \pref{gth},
$$D_l(\kk)=-i k_0+\o v_F(1+\d_l)k$$
and \be \lb{kajkj}|W_2^{(l)}(\kk)| \le C (\e_l^2 + \bar g_l
\g^{\th l})\ee with $\bar g_l=\max_{h\ge l}
\max\{|g_{2,h}|,|g_{4,h}|\}$ and $\e_l=\max_{j\ge l} \max\{\bar
g_j, |\d_j|\}$, Note that we have included in $D_l(\kk)$ a
correction of the free Fermi velocity due to the local term $\d_l
V_\d$, so that $W_2^{(l)}(\kk)$ does not contain terms of order
$\e_l$, except those associated to irrelevant terms.
Moreover, by the analogue of (112) of \cite{BM002}
\be
\la\hat\r_{\pp,\o,s} \hat\psi^-_{\kk,\o,s} \hat\psi^+_{\kk-\pp,\o,s}\ra_l =
Z^{(1)}_l \hat S_\o(\kk) \hat S_\o(\kk-\pp) [1+W_{2,1}^{(l)}(\kk)]
\ee
where $|W_{2,1}^{(l)}(\kk)| \le C(\e_l^2 + \bar g_l \g^{\th l})$.
In \S\ref{sec4.1} we have written the WI directly in the $l\to-\io$ limit, but
they are true also when $l$ is finite, up to corrections studied in detail in
\S 4 of \cite{BM002} and \S 4 of \cite{BM005}; using the bounds is such papers
for the corrections we derive from the finite $l$ analogue of \pref{grez}
\be
D_N(\pp) \la\hat\r_{\pp,\o,s} \hat\psi^-_{\kk,\o,s} \hat\psi^+_{\kk-\pp,\o,s}\ra_l=
\hat S_\o(\kk-\pp) - \hat S_\o(\kk) + D_N(\pp) R_l(\kk,\pp)
\ee
where $\kk$ and $\kk-\pp$ are of size $\g^l$,
$$D_N(\kk)=-i k_0+\o v_F(1+\d)k$$
and
$$|R_l(\kk,\pp)| \le C \g^{-2l} \frac{Z^{(1)}_l}{Z_l^2} (\e_l^2 +
\bar g_l \g^{\th l})$$
Hence, if we put $\kk=\bar\kk$ and $\pp=2\bar\kk$, with $|\bar\kk|=\g^l$, we
get $D_N(\bar\kk) [Z^{(1)}_l/Z_l] = D_l(\bar\kk) [1+ \D(\bar\kk)]$, with
$|\D(\kk) \le C(\e_l^2 + \bar g_l \g^{\th l})$, which implies that $\d_l = \d+
O(\e_l^2)$ and $Z^{(1)}_l/Z_l= 1 + O(\e_l^2)$. On the other hand, the value of
$\d_h$ is independent of the infrared cutoff, if $h\ge l+1$, and $|\d_{l+1} -
\d_l| \le C\e_l^2$. It follows that, for any $j\in [l,N]$ and uniformly in the
cutoffs,
\bal
\lb{h1} \d_j &= \d+ O(\e_j^2)\\
\lb{h1Z} \frac{Z^{(1)}_j}{Z_j} &= 1 + O(\e_j^2)
\eal


In the same way, we can also use the the Schwinger-Dyson equation
combined with the Ward Identities with a finite infrared cut-off;
we get \pref{ce} with a correction term, which can be bounded as
in \S4 of \cite{BM005}. If we restrict the resulting expression to
the four point function with momenta at the infrared scale, we
get, see \cite{BM005},\cite{M005}
\be\lb{h3} g_{\arp,j}=g_{\arp}+ O(\e_j^2) \virg
g_{\erp,j}=g_{\erp}+ O(\e_j^2) \virg g_{4,j}=g_4 + O(\e_j^2) \ee
Let us call $b_{\a}^{(j)}(g_\arp, g_\erp, g_4, \d)$ the function which is
obtained by $B_{\a}^{(j)}\lft(\vec g_h, \d_h,..\vec g_0, \d_0, \vec g, \d\rgt)$
(defined in \pref{bf}), by subtracting the contribution of the trees containing
endpoints of scale grater than $0$ and by putting $(\vec g_j, \d_j) = (0, 0,
g_\arp, g_\erp, g_4, \d)$, $\forall j=h, \ldots. 0$ (the value of the r.c.c.  is
independent of the scale). Then, by Lemma 3.4 of \cite{BM004}, we get the
bound:
\be\lb{van} \lft|b_{\a}^{(j)}(g_\arp, g_\erp, g_4, \d) \rgt| \le C
[\max \{ |g_\arp|, |g_\erp|, |g_4|, |\d| \}]^2 \g^{\th j}\virg
\a=\arp, \erp,4,\d \ee
In we define in a similar way the function $\tilde b^{(j)}(g_\arp, g_\erp, g_4,
\d)$ in terms of $\tilde B^{(j)}\lft(\vec g_j, \d_j,..\vec g_0, \d_0, \vec g,
\d\rgt)$ (defined in \pref{ralpha1}), \pref{h1Z} implies the bound
\be\lb{vanZ} \lft|\tilde b^{(j)}(g_\arp, g_\erp, g_4, \d) \rgt|
\le C [\max \{ |g_\arp|, |g_\erp|, |g_4|, |\d| \}]^2 \g^{\th j}\ee
The bounds \pref{van} and \pref{vanZ} allow to show that, in the model
\pref{vv1}, the infared cut-off can be removed by an analysis very similar to
the one in \S\ref{sec2.2}-\ref{sec2.4}.

Let us now consider the Hubbard model. In \pref{bal} we have written its Beta
function as sum of two terms, the second of which is asymptotically negligible,
by \pref{2.34}; the first term, denoted in \pref{bal} by $\b^{(j)}_\a(\vec
g_j,\d_j;...;\vec g_0,\d_0)\equiv \b^{(j)}_\a(G_{1},G_{2},G_{4},\D)$ (using a
notation similar to that used after \pref{bf}) coincides with the Beta function
of the effective model on the invariant surface $\CC_{1,+}\cap \CC_3$, if we
subtract from it the contribution of the trees containing endpoints of scale
grater than $0$. Hence, by using \pref{ralfa} and \pref{r13}, we get
\be
\bsp \b_1^{(j)}(G_1,G_2,G_4,\D) &= G_1 \bar
\b_{1\perp}^{(j)}(G_1^2,0,G_2-G_1,G_2,G_4,\D)\\
\b_2^{(j)}(G_1,G_2,G_4,\D) &= \bar
\b_{\perp}^{(j)}(G_1^2,0,G_2-G_1,G_2,G_4,\D)\\
\b_4^{(j)}(G_1,G_2,G_4,\D) &= \bar
\b_4^{(j)}(G_1^2,0,G_2-G_1,G_2,G_4,\D)\\
\b_\d^{(j)}(G_1,G_2,G_4,\D) &= \bar \b_\d^{(j)}(G_1^2,0,G_2-G_1,G_2,G_4,\D)
\esp
\ee
where $\bar \b_\a^{(j)}(G_1^2,0,G_2-G_1,G_2,G_4,\D)$ denotes the value of $\bar
B_\a^{(j)}(G_1^2,0,G_2-G_1,G_2,G_4,\D)$, after the subtraction of the trees
containing endpoints of scale grater than $0$. Therefore, if $\a\not=1$, the
contributions of order $0$ and $1$ in $G_1$ of $\b_\a^{(j)}(\vec G,\D)$ are the
same as the contributions of the same order of $\bar \b_\a^{(j)}(0,0, G_2-G_1,
G_2, G_4, \D)$. By using \pref{van}, one sees immediately that these
contributions are of order $\g^{\th j}$ as functions of $j$, so that
\pref{beta23} is proved. In the same way, using \pref{vanZ} and \pref{ralpha1}
we can prove \pref{beta44}.

This completes the proof of the boundedness of the RG flow in the Hubbard
model, as well as in the effective model with $g_{1,\perp}>0$ and
$g_\arp=g_{\erp}-g_{1,\perp}$. Note that, in order to prove the boundedness of
the flow of the spin-symmetric Hubbard model, we need information from a non
spin symmetric model; in fact, we have derived \pref{beta23} from the model
\pref{vv1} with $g_{\erp}\not=g_{\arp}$ and $g_{1,\perp}=0$.


\vskip.3cm {\it Acknowledgements} G.B. and V.M. acknowledge the financial
support of MIUR, PRIN 2008. V.M. gratefully acknowledges also the financial
support from the ERC Starting Grant CoMBoS-239694. The work of P.F. is
supported by the Giorgio and Elena Petronio Fellowship Fund.

\bibliographystyle{unsrt}
\bibliography{Ref_Fermions, Ref_Roma}

\end{document}